\documentclass[11pt]{article}

\usepackage[spanish,activeacute]{babel}
\usepackage[T1]{fontenc}

\usepackage{amsmath, amsfonts, amssymb, amsthm}
\usepackage{bm}
\usepackage{mathrsfs}
\usepackage{dsfont}
\usepackage{scalerel} 

\usepackage{graphicx}
\usepackage{adjustbox}
\usepackage{grffile}
\usepackage{float}
\usepackage{flafter}
\usepackage{subfigure}
\usepackage{pdfpages}
\usepackage{pdflscape}

\usepackage{multirow}
\usepackage{multicol}
\usepackage{rotating}
\usepackage{threeparttable}
\usepackage{array}

\usepackage{comment}
\usepackage{caption}
\usepackage{subcaption}
\usepackage{enumitem}
\usepackage{color, xcolor}
\usepackage{soul}
\usepackage{url}
\usepackage{setspace}
\usepackage[toc]{appendix}
\usepackage[hidelinks]{hyperref}
\usepackage{booktabs}

\usepackage[numbers]{natbib}
\setcitestyle{authoryear,open={(},close={)}}

\hypersetup{
    colorlinks=true,
    linkcolor=black,
    filecolor=black,
    urlcolor=black,
    citecolor=black
}

\def\spacingset#1{\renewcommand{\baselinestretch}{#1}\small\normalsize}
\spacingset{1}

\newcommand{\be}{\begin{eqnarray*}}
\newcommand{\ee}{\end{eqnarray*}}
\newcommand{\bet}{\begin{eqnarray}}
\newcommand{\eet}{\end{eqnarray}}

\def\@roman#1{\romannumeral #1}

\makeatletter
\def\hlinewd#1{%
  \noalign{\ifnum0=`}\fi\hrule \@height #1 %
  \futurelet\reserved@a\@xhline}
\makeatother

\addto\captionsspanish{}

\begin{document}

\title{\textbf{El proceso legislativo colombiano, 2014-2025: \\ redes, tópicos y polarización}}

\author{
    Juan Sosa$^1$\footnote{Contacto: jcsosam@unal.edu.co.}\qquad
    Brayan Riveros$^1$ \qquad
    Emma J. Camargo-Díaz$^2$ \\  
}	 

\date{
    $^1$Universidad Nacional de Colombia, Colombia \\ \vspace{0.3cm}
    $^2$Universidad Externado de Colombia, Colombia
}

\maketitle

\begin{abstract}
Se analiza la producción legislativa de la Cámara de Representantes de Colombia entre 2014 y 2025 a partir de 4{,}083 proyectos de ley. Se construyen redes bipartitas entre partidos y proyectos, y entre representantes y proyectos, junto con sus proyecciones, para caracterizar patrones de copatrocinio, centralidad e influencia y evaluar si la polarización política se refleja en la colaboración legislativa. En paralelo, se estudia el contenido de las iniciativas mediante redes semánticas basadas en coocurrencias obtenidas de descripciones cortas, y se identifican tópicos por partido y periodo con un modelo de bloques estocásticos para redes ponderadas, con contraste adicional usando Asignación Latente de Dirichlet. Además, se aplica un modelo Bayesiano de sociabilidad para detectar términos con conectividad robusta y resumir núcleos discursivos. En conjunto, el enfoque integra estructura relacional y estructura semántica para describir cambios temáticos entre gobiernos, identificar actores y colectividades influyentes, y aportar una síntesis reproducible que favorece la transparencia y el control ciudadano del proceso legislativo.
\end{abstract}

\noindent
{\it Palabras clave: Cámara de Representantes; Redes bipartitas; Redes Sociales; Polarización; Análisis semántico.}

\spacingset{1.1} 

\newpage

\section{Introducción}

En Colombia se observa, desde los inicios de la democracia, una identidad bipartidista que se expresa primero en los conflictos entre federalistas y centralistas y luego en las disputas entre el Partido Liberal y el Partido Conservador. Incluso entre próceres de la independencia se identifican corrientes ideológicas diferenciadas, como las posturas bolivarianas y las santanderistas. Estas tensiones se traducen en periodos complejos de la historia del país, entre ellos las guerras civiles de la década de 1850 \citep{alonso2014ensamblajes,chenou_restrepo2023naciondividida}.

Con el avance tecnológico y los cambios sociales, estas violencias y divisiones no desaparecen, sino que se desplazan hacia nuevos escenarios. En la discusión contemporánea, la polarización deja de formularse exclusivamente en términos de liberalismo y conservatismo y se reconfigura como controversias entre izquierda y derecha, con matices que incorporan ideas del progresismo y del conservatismo. Aunque la Constitución de 1991 fortalece el pluralismo político, estas dinámicas también se expresan en espacios digitales, donde se amplifican debates, identidades y alineamientos políticos \citep{gonzalez2015identidad}.

En este trabajo se aborda este problema mediante análisis de redes sociales y se evalúa si en la Cámara de Representantes de Colombia existen divisiones políticas que afectan el proceso legislativo. La Cámara, junto con el Senado, integra el Congreso de la República y sus integrantes se eligen por voto popular como representantes regionales, con el propósito de tramitar problemáticas específicas de los departamentos. En este marco, se asume que su función principal consiste en proponer y debatir proyectos de ley, al tiempo que se representan los intereses partidistas.

Además, en este trabajo se analiza el discurso legislativo como un componente central de estas dinámicas. En la última década Colombia atraviesa procesos que reordenan prioridades y marcos de discusión, entre ellos la implementación del acuerdo de paz durante el gobierno de Juan Manuel Santos, el periodo de pandemia y el deterioro económico asociado a la pandemia durante el gobierno de Iván Duque, y el impulso de políticas orientadas a enfrentar el cambio climático y a promover reformas estructurales en salud y pensiones durante el gobierno de Gustavo Petro. En este contexto se plantea que los cambios institucionales y sociales se reflejan en los temas promovidos y en los patrones de colaboración entre actores.

Este interés se refuerza por la escasez de literatura que integre, de manera conjunta, análisis de texto y análisis de redes aplicados específicamente a la Cámara de Representantes. Aunque existen iniciativas como Congreso Visible \url{https://congresovisible.uniandes.edu.co/}, que ofrecen una visión general sobre la orientación de los proyectos, aún se carece de un marco sistemático y detallado que describa la colaboración legislativa y la evolución del discurso a lo largo del tiempo, lo que justifica una caracterización empírica basada en redes y en la estructura semántica de los proyectos.

Para abordar de manera integral esta temática, se parte de la información disponible en la página de la Cámara de Representantes de Colombia \url{https://www.camara.gov.co/} y se estudian los periodos presidenciales entre 2014 y 2025. Con estos registros se analizan 4{,}083 proyectos de ley y se construyen redes bipartitas y sus proyecciones, tanto entre partidos políticos y proyectos de ley como entre representantes y proyectos de ley, con el fin de caracterizar patrones de colaboración e identificar actores y colectividades con mayor influencia. Con este diseño se evalúa si la polarización visible en entornos digitales, por ejemplo en \texttt{X}, y en discusiones políticas difundidas por integrantes del Congreso \citep{chenou_restrepo2023naciondividida}, también se reproduce en el trabajo legislativo y de qué manera ocurre.

En este trabajo se ofrece una descripción basada en evidencia cuantitativa sobre los patrones de colaboración legislativa y sobre los representantes y partidos con mayor influencia, y se presenta una síntesis de acceso rápido que facilita el control ciudadano del funcionamiento del Congreso. Este aporte resulta relevante para la democracia porque reduce asimetrías de información entre la ciudadanía y las élites políticas, fortalece la transparencia y la rendición de cuentas, y favorece una participación informada en la deliberación pública.

Además, en este trabajo se aborda un reto técnico central asociado al manejo de los datos disponibles, en particular la falta de estandarización en la presentación de los proyectos y la complejidad de sus formatos. Para mitigar estas dificultades, se trabaja con textos cortos que corresponden a descripciones puntuales de los proyectos, con el fin de reducir ruido y evitar sesgos en los análisis textuales. Asimismo, se contrastan métodos de análisis de texto directo, como la Asignación Latente de Dirichlet \citep{blei2003lda}, con enfoques basados en redes semánticas que explotan la estructura de coocurrencias para identificar comunidades interpretables como tópicos, mediante modelos de bloques estocásticos \citep{abbe2023community} y modelos de sociabilidad \citep{sosa2025bayesian}.

Este artículo se organiza de la siguiente manera. En la Sección 2 se presenta el estado del arte y la base conceptual que sustenta la metodología. En la Sección 3 se describe la metodología asociada al proceso de construcción de redes y a los modelos empleados para el análisis. En la Sección 4 se presentan los resultados y se detallan los principales hallazgos. Finalmente, en la Sección 5 se desarrolla la discusión, se sintetizan los hallazgos y se exponen las limitaciones del trabajo.


\section{Estado del arte}

La popularidad del análisis de redes crece por el aumento de datos digitales y por la necesidad de describir, con evidencia cuantitativa, cómo interactúan individuos y organizaciones en contextos reales. En este trabajo se aprovecha la información pública de la Camara de Representantes de Colombia (\url{https://www.camara.gov.co/}) para construir redes de colaboración entre representantes, partidos y proyectos de ley. En particular, se consideran redes generadas en el ámbito político a partir de la actividad legislativa, mediante interacciones directas entre legisladores y proyectos de ley, y mediante relaciones entre partidos y proyectos que permiten describir patrones de patrocinio, coautoría y coordinación legislativa.

En estudios recientes, como el análisis de dinámicas de colaboración en Corea \citep{kim2025collaboration}, se muestra que en la vigésima legislatura de la Asamblea Nacional, durante 2016 a 2020, la estructura de la red se asocia con afiliaciones políticas, ideológicas y partidistas, y se emplean modelos paramétricos como regresión Probit y regresión Poisson. No obstante, dado que la ventana temporal es corta, se considera que las conclusiones pueden depender de la agenda particular de esa legislatura. Por otro lado, en el estudio de \citet{fowler2006} se examina la dinámica de las relaciones en el Congreso de los Estados Unidos entre 1973 y 2004, y se comparan periodos mediante la densidad de las redes construidas. En ese análisis se observan redes altamente conectadas y se sugiere que, con el paso del tiempo, emergen relaciones que no se explican únicamente por las afiliaciones políticas. En Colombia, se destaca el trabajo de \citet{robles2025dime}, en el que se analizan redes sociales de integrantes del Senado durante el gobierno de Juan Manuel Santos, y se reporta que las alianzas políticas aumentan la probabilidad de que una ley sea sancionada y que métricas de centralidad en medios sociales se asocian con la participación política. En conjunto, estos estudios muestran que el análisis de redes aporta información sustantiva para describir la estructura de poblaciones de interés, identificar liderazgos, caracterizar la conformación de comunidades y evaluar patrones de colaboración y alineamiento político.

Estos enfoques también se aplican a redes construidas a partir de texto, en las que la coocurrencia de palabras permite identificar agrupaciones que reflejan temas de interés en periodos específicos. Por ejemplo, en \citet{hachaj2018bigrams} se examina cómo redes basadas en bigramas se utilizan para sintetizar grandes volúmenes de comentarios, identificar términos recurrentes y reconocer tópicos dominantes dentro de comunidades de seguidores en medios sociales. De manera complementaria, en \citet{luque2024caracterizacion} se caracteriza el discurso de posesión presidencial y, mediante redes construidas desde la perspectiva de palabras y de actores políticos, se identifican patrones de cohesión discursiva y comunidades políticas en Colombia, lo que resulta pertinente en este contexto porque se respalda el uso de redes textuales para describir agendas temáticas y estructuras de relación asociadas al debate público.

En estos métodos se presentan limitaciones asociadas al lenguaje, ya que se observan sesgos derivados de términos de alta frecuencia que pueden distorsionar la identificación de los temas sustantivos \citep{galvez2018copalabras}, por lo que se requiere complementar el preprocesamiento con técnicas de procesamiento de lenguaje natural orientadas a normalización lingüística, desambiguación y representación semántica \citep{manning1999foundations,vaswani2017attention,devlin2019bert}. También se obtiene baja interpretabilidad cuando no se conoce el origen del corpus, lo que obliga a documentar la procedencia y a definir criterios de selección explícitos, y a medida que aumenta el volumen de texto el análisis demanda mayores recursos computacionales. Parte de estas dificultades se atenúa en este caso porque los documentos provienen de una fuente única y siguen una estructura normativa relativamente estable, y porque los textos son breves, lo que reduce el beneficio marginal de aplicar lematización de manera sistemática.

Otra dificultad frecuente en la literatura sobre redes basadas en bigramas corresponde al uso limitado de modelos estadísticos, ya que muchos estudios se apoyan en algoritmos deterministas basados en distancias entre nodos. Sin embargo, estas distancias pueden ser inadecuadas en texto, porque términos aislados introducen ruido y dificultan la identificación de comunidades. Por ello, en este trabajo se emplean modelos de bloques estocásticos para redes ponderadas \citep{abbe2023community,aicher2014learning}, que modelan conteos de coocurrencia y representan de forma natural la intensidad de las relaciones, y al mismo tiempo atenúan coocurrencias esporádicas. Con este enfoque se contrasta el modelo de Asignación Latente de Dirichlet \citep{blei2003lda}, que puede presentar baja interpretabilidad cuando el número de tópicos es grande y ambigüedad en la elección de su número. Además, se aplica un modelo de sociabilidad \citep{sosa2025bayesian} para identificar términos clave con alta conectividad semántica, que actúan como núcleos temáticos o conectores entre comunidades y fortalecen la interpretación de los tópicos estimados.

\subsection{Conceptos fundamentales de redes sociales}

A continuación se presentan definiciones y conceptos básicos de análisis de redes sociales, para lo cual se adapta contenido de textos de referencia ampliamente utilizados, como \cite{al2017python,newman2018networks,menczer2020first,kolaczyk2020statistical}, entre otros.

\subsubsection{Redes sociales}

Se define una red social, o simplemente una red, como un conjunto de relaciones entre individuos que se representa mediante un grafo $G=(\mathcal{V},\mathcal{E})$, donde $\mathcal{V}$ denota el conjunto de vértices, interpretados como individuos u objetos de análisis, y $\mathcal{E}$ denota el conjunto de aristas. Cada arista $e=(i,j)$, con $i,j\in\mathcal{V}$, representa la existencia de una relación entre los vértices $i$ y $j$. Se distinguen dos tipos de redes. En redes dirigidas, la relación tiene orientación, por lo que $e=(i,j)$ es una pareja ordenada y, en general, $(i,j)\neq(j,i)$. En redes no dirigidas, la relación es simétrica y, por tanto, $(i,j)=(j,i)$.

\subsubsection{Matriz de adyacencia}

Se representa una red mediante su matriz de adyacencia $\mathbf{A}=[a_{i,j}]$, que se construye a partir de las relaciones entre pares de vértices. En una red binaria, la entrada $a_{i,j}$ se define como una variable indicadora que toma el valor 1 si existe una relación entre $i$ y $j$, y toma el valor 0 en caso contrario. En este contexto, la diagonal contiene ceros que se interpretan como ceros estructurales, ya que no se admiten relaciones reflexivas. Cuando la red es no dirigida, la matriz es simétrica porque la relación satisface $(i,j)=(j,i)$. En una red dirigida esta igualdad no necesariamente se cumple, por lo que la matriz puede no ser simétrica.

\subsubsection{Red bipartita y proyecciones}

Las redes bipartitas se caracterizan por tener dos grupos disyuntos de vértices. Estos grupos se denotan como $U$ y $P$, y las relaciones solo se definen entre vértices de conjuntos distintos. Un ejemplo corresponde a usuarios y aplicaciones, donde la relación se interpreta como el uso de una aplicación por parte de un usuario, por lo que las aristas pertenecen al producto cartesiano $e \in U \times P$. En este caso, la matriz de adyacencia tiene tamaño $n \times m$, con $n=\lvert U\rvert$ el número de usuarios y $m=\lvert P\rvert$ el número de aplicaciones, y por tanto es rectangular.

Las redes bipartitas son útiles porque permiten relacionar individuos con patrones de conexión similares mediante \textit{proyecciones}. La primera proyección se obtiene mediante $\mathbf{P}_1=\mathbf{A}\mathbf{A}^{\top}$, que corresponde a una matriz de adyacencia ponderada de tamaño $n\times n$, cuyas entradas representan el número de elementos de $P$ compartidos por cada par de vértices de $U$. La segunda proyección se obtiene mediante $\mathbf{P}_2=\mathbf{A}^{\top}\mathbf{A}$, que corresponde a una matriz de adyacencia ponderada de tamaño $m\times m$, cuyas entradas representan el número de elementos de $U$ compartidos por cada par de vértices de $P$. En el ejemplo, las entradas de $\mathbf{P}_1$ corresponden al número de aplicaciones en común utilizadas por dos usuarios, mientras que las entradas de $\mathbf{P}_2$ corresponden al número de usuarios en común entre dos aplicaciones. Estas matrices se denominan proyecciones porque representan relaciones inducidas por la coasociación en el conjunto opuesto.

\subsubsection{Características de los individuos del sistema}

Para determinar la importancia de un individuo dentro de una red no dirigida, se analiza su patrón de conexiones y el nivel de influencia que ejerce sobre la estructura global. Para ello se describen métricas que resumen la conectividad, la intensidad de los vínculos y la posición relativa del individuo dentro de la red.

\textsf{Grado}

Se define el grado como el número de vecinos que tiene un individuo $v_i \in \mathcal{V}$. Para una red representada por una matriz de adyacencia $\mathbf{A}=[a_{i,j}]$ de tamaño $n\times n$, el grado se calcula como la suma de las entradas en la fila correspondiente,
$$
\textsf{deg}(i)=\sum_{j=1}^{n} a_{i,j}.
$$
Un valor alto de $\deg(v_i)$ indica mayor conectividad local, porque el individuo se vincula con un mayor número de nodos.

\textsf{Fuerza}

Se define la fuerza como una extensión del grado para redes ponderadas, en las que las aristas tienen pesos continuos o discretos. Para una red representada por una matriz de adyacencia ponderada $\mathbf{W}=[w_{i,j}]$ de tamaño $n\times n$, la fuerza del nodo $v_i \in \mathcal{V}$ se calcula como la suma de los pesos en la fila correspondiente,
$$
\textsf{str}(i)=\sum_{j=1}^{n} w_{i,j}.
$$
Un valor alto de $\textsf{str}(v_i)$ indica mayor intensidad total de conexión, porque el nodo acumula vínculos más fuertes, lo que permite distinguir nodos con pocas relaciones pero muy intensas de nodos con muchas relaciones pero de baja intensidad.

\textsf{Centralidad de intermediación (\textit{betweenness})}

Se define la centralidad de intermediación como una medida de posición estructural, basada en la proporción de caminos más cortos entre pares de nodos que pasan por el vértice $i$. Para ello se calcula
$$
C_\textsf{B}(i)=\sum_{\substack{s\neq i\neq t}}\frac{\sigma_{s,t}(i)}{\sigma_{s,t}},
$$
donde $\sigma_{s,t}(i)$ denota el número de caminos más cortos entre $s$ y $t$ que pasan por $i$, y $\sigma_{s,t}$ denota el número total de caminos más cortos entre $s$ y $t$. Un valor alto de $C_\textsf{B}(i)$ indica mayor intermediación, porque el nodo opera como puente y puede mediar el flujo de información o coordinación entre distintas partes de la red.

\textsf{Centralidad de cercanía (closeness)}

Se define la centralidad de cercanía como una medida de accesibilidad global, basada en la distancia geodésica promedio entre el vértice $i$ y el resto de la red. Si $d(i,j)$ denota la longitud del camino más corto entre $i$ y $j$, la centralidad de cercanía se calcula como
$$
C_\textsf{C}(i)=\left[\sum_{\, j\neq i} d(i,j)\right]^{-1}.
$$
Un valor alto de $C_\textsf{C}(i)$ indica mayor cercanía global, ya que el nodo se ubica, en promedio, a menor distancia de los demás vértices y, por tanto, puede alcanzar más rápidamente al resto de la red.

\textsf{Centralidad propia (\textit{eigenvector})}

Se define la centralidad propia como una medida que pondera las conexiones de un nodo por la importancia de sus vecinos, de modo que aumenta cuando un vértice se conecta con otros vértices que también presentan alta centralidad. La centralidad propia se calcula como
$$
C_\textsf{E}(i)=\lambda^{-1}\sum_{\{i,j\}\in\mathcal{E}} c(j),
$$
donde el vector $\mathbf{c}$ satisface la ecuación de autovalores $\mathbf{A}\mathbf{c}=\lambda\,\mathbf{c}$, con $\mathbf{A}$ la matriz de adyacencia y $\lambda$ el mayor valor propio de $\mathbf{A}$. En consecuencia, un valor alto de $C_\textsf{E}(i)$ indica influencia por asociación, ya que el nodo adquiere mayor relevancia cuando se conecta con vecinos que también lo son.

\subsubsection{Características estructurales de la red}

Además de caracterizar a los individuos, se considera fundamental estudiar la estructura global de la red en términos de conectividad, con el fin de caracterizar cómo surgen y se organizan las conexiones.

\textsf{Densidad}

Se define la densidad como la proporción entre el número de aristas observadas y el número máximo de aristas posibles,
$$
\textsf{d} = \frac{\lvert \mathcal{E} \rvert}{\lvert \mathcal{V} \rvert(\lvert \mathcal{V} \rvert-1)/2}.
$$
Esta métrica toma valores entre $0$ y $1$ y cuantifica qué tan próxima se encuentra la red a ser completa.

\textsf{Diámetro}

La distancia geodésica $d(i,j)$ se define como la longitud del camino más corto entre dos nodos $i,j\in\mathcal{V}$. Esta distancia permite caracterizar la accesibilidad entre vértices y, por tanto, la facilidad con la que se propaga la información o la interacción en la red. Con base en esta noción, el diámetro se define como la mayor distancia geodésica entre cualquier par de nodos, lo que da cuenta del tamaño efectivo de la red en términos de conectividad.

\textsf{Transitividad}

Se define la transitividad o coeficiente de agrupamiento como una medida de cierre triádico, que cuantifica la frecuencia con la que los vecinos de un nodo también se conectan entre sí y, en consecuencia, la propensión de la red a formar triángulos. Para ello se calcula
\[
\textsf{cl}
=
\frac{3\,\tau_\triangle}{\tau_3},
\]
donde \(\tau_\triangle\) denota el número de triángulos y \(\tau_3\) denota el número de tríadas conectadas. Por lo tanto, un valor alto de \(\textsf{cl}\) indica mayor cohesión local y una estructura comunitaria más densa.

\textsf{Asortatividad}

Se define la asortatividad como una medida de homofilia estructural, que cuantifica la tendencia de los nodos a conectarse con otros nodos que comparten características similares, considerando atributos cualitativos y cuantitativos. Para un atributo cualitativo con \(n\) categorías, se define como
\[
\textsf{r}=
\frac{\sum_{i=1}^{n} f_{i,i}-\sum_{i=1}^{n} f_{i,\bullet}\,f_{\bullet,i}}
{1-\sum_{i=1}^{n} f_{i,\bullet}\,f_{\bullet,i}},
\]
donde \(f_{i,j}\) denota la fracción de aristas que conectan nodos de las categorías \(i\) y \(j\),
$f_{i,\bullet}=\sum_{j=1}^{n} f_{i,j}$ y $f_{\bullet,j}=\sum_{i=1}^{n} f_{i,j}$. Para un atributo numérico, se define la asortatividad como la correlación entre los valores del atributo en los extremos de las aristas, de forma que
\[
\textsf{r}=
\frac{\sum_{e\in\mathcal{E}} (x_e-\bar{x})(y_e-\bar{y})}{s_x\,s_y},
\]
donde \(x_e\) y \(y_e\) denotan los valores del atributo en los dos extremos de la arista \(e\), \(\bar{x}\) y \(\bar{y}\) denotan sus promedios sobre aristas, y \(s_x\) y \(s_y\) denotan las desviaciones estándar correspondientes. En consecuencia, \(\textsf{r}>0\) indica homofilia y \(\textsf{r}<0\) indica heterofilia.

\subsection{Modelamiento}

Aquí se introducen los elementos esenciales de los modelos de bloques estocásticos \citep{aicher2014learning,abbe2023community}, de la Asignación Latente de Dirichlet \citep{silge2017text} y de los modelos de sociabilidad \citep{sosa2025bayesian}, los cuales se utilizan como herramientas centrales para el análisis de las redes objeto de estudio.

\subsubsection{Modelo de bloques estocásticos}\label{sec_modelo_bloques}

Para identificar tópicos como comunidades de palabras que tienden a coocurrir en el corpus, se ajusta un modelo de bloques estocásticos (SBM, por sus siglas en inglés; e.g., \citealt{lee2019review}) a una red semántica ponderada. Sea $G=(\mathcal{V},\mathcal{E})$ una red no dirigida, donde $\mathcal{V}$ denota el vocabulario y donde los pesos $y_{i,j}\in\mathbb{N}_0$ representan el número de coocurrencias observadas entre las palabras $i$ y $j$ (con $y_{i,i}=0$ y $y_{i,j}=y_{j,i}$). El modelo asume una partición latente $\mathcal{P}=\{C_1,\ldots,C_Q\}$ de $\mathcal{V}$ en $Q$ comunidades, interpretadas como tópicos, y satisface un supuesto de equivalencia estocástica: condicional en las asignaciones, la distribución de $y_{i,j}$ depende únicamente de las comunidades a las que pertenecen $i$ y $j$.

Bajo este modelo, para cada nodo $i$ se introduce una variable latente $\xi_i\in\{1,\ldots,Q\}$ tal que $\xi_i=q$ indica que el vértice $i$ pertenece al grupo $C_q$. Esta asignación se modela mediante $\xi_i \mid \boldsymbol{\alpha} \overset{\text{ind}}{\sim} \textsf{Categorical}(\boldsymbol{\alpha})$, con $\boldsymbol{\alpha}\in\Delta^{Q-1}$, donde $\Delta^{Q-1}=\{\boldsymbol{\alpha}\in\mathbb{R}^Q:\ \alpha_q\ge 0\,\,\text{y}\, \sum_{q=1}^{Q}\alpha_q=1\}$ denota el $(Q-1)$-simplex. Condicional en $\boldsymbol{\xi}=(\xi_1,\ldots,\xi_n)$, para $i<j$ se asume que
\[
y_{i,j} \mid \xi_i,\xi_j \overset{\text{ind}}{\sim} \textsf{Poisson}\!\big(\lambda_{\xi_i,\xi_j}\big),
\qquad \lambda_{q,r}>0,\qquad \lambda_{q,r}=\lambda_{r,q}.
\]
En esta especificación, $\lambda_{q,r}$ representa la intensidad promedio de coocurrencia entre palabras de las comunidades $q$ y $r$, lo que permite distinguir tópicos con vínculos semánticos densos o persistentes de tópicos con vínculos débiles.

El modelo se ajusta con la librería \texttt{blockmodels} de \texttt{R}, en la cual se estiman los parámetros del modelo mediante un algoritmo EM variacional que maximiza de forma aproximada la verosimilitud para modelos de bloques, incluida la especificación Poisson para redes ponderadas \citep{leger2016blockmodels}. La selección del número de grupos $Q$ se realiza con el criterio de verosimilitud integrada completada, ICL por sus siglas en inglés, evaluando distintos valores de $Q$ y eligiendo el que maximiza dicho criterio, de modo que se balancea calidad de ajuste y complejidad del modelo de forma análoga al BIC en problemas de agrupamiento \citep{biernacki2002assessing}.

\subsubsection{Asignación latente de Dirichlet}\label{sec_LDA}

Como método de contraste se utiliza la Asignación Latente de Dirichlet (LDA, por sus siglas en inglés; e.g., \citealt{chauhan2021topic}), uno de los modelos más difundidos para inferir tópicos latentes en un corpus. En LDA, cada documento se modela como una mezcla de $K$ tópicos y cada tópico se modela como una distribución de probabilidad sobre el vocabulario. El ajuste se implementa con la librería \texttt{topicmodels} de \texttt{R}, cuya inferencia se realiza mediante muestreo de Gibbs colapsado (e.g., \citealt{gamerman2006markov}), lo que permite aproximar la posterior de las asignaciones latentes y recuperar estimadores para las distribuciones de tópicos y de términos \citep{grun2011topicmodels}.

Sea $D$ un corpus con $M$ documentos y un vocabulario de tamaño $V$. Para cada documento $d$, con longitud $N_d$, se denota la secuencia de términos por $\boldsymbol{w}_d=(w_{d,1},\ldots,w_{d,N_d})$, con $w_{d,i}\in\{1,\ldots,V\}$. El modelo asume un número de tópicos $K$ fijado de antemano y, para cada documento, introduce proporciones latentes de tópicos $\boldsymbol{\theta}_d\in\Delta^{K-1}$, así como, para cada tópico, distribuciones latentes de términos $\boldsymbol{\beta}_k\in\Delta^{V-1}$, además de asignaciones $z_{d,i}\in\{1,\ldots,K\}$ para cada posición $i$ del documento $d$. 

Específicamente, el proceso generativo establece que $\boldsymbol{\beta}_k \sim \textsf{Dirichlet}(\delta\,\mathbf{1})$, para $k=1,\ldots,K$, $\boldsymbol{\theta}_d \sim \textsf{Dirichlet}(\alpha\,\mathbf{1})$, para $d=1,\ldots,M$, dados hiperparámetros $\alpha>0$ y $\delta>0$, y condicional en $\boldsymbol{\theta}_d$ y $\{\boldsymbol{\beta}_k\}$, para $i=1,\ldots,N_d$,
\[
z_{d,i}\mid \boldsymbol{\theta}_d \sim \textsf{Categorica}(\boldsymbol{\theta}_d),
\qquad
w_{d,i}\mid z_{d,i},\{\boldsymbol{\beta}_k\} \sim \textsf{Categorica}(\boldsymbol{\beta}_{z_{d,i}}),
\]
donde $z_{d,i}$ indica el tópico asignado a la palabra $w_{d,i}$, $\boldsymbol{\beta}_k$ determina la distribución de términos del tópico $k$ y $\boldsymbol{\theta}_d$ determina la mezcla de tópicos del documento $d$.

Dado que $K$ debe fijarse a priori, su selección se guía con base en la perplejidad \citep{blei2003lda}, calculada sobre un conjunto de evaluación como medida de desempeño predictivo:
\[
\textsf{p}(\boldsymbol{w})
=
\exp\left\{
-\frac{\log p(\boldsymbol{w})}{\sum_{d=1}^{M} N_d}
\right\},
\]
donde $p(\boldsymbol{w})$ denota la probabilidad predictiva del conjunto de evaluación bajo el modelo ajustado. Un valor más bajo de perplejidad indica mejor capacidad predictiva. Sin embargo, este criterio no evalúa de forma directa la coherencia semántica de los tópicos, por lo que puede favorecer valores grandes de $K$ o soluciones con tópicos solapados y de interpretación débil. En consecuencia, su uso se complementa con una revisión sustantiva de los tópicos estimados \citep{grun2011topicmodels}.

\subsubsection{Modelo de sociabilidad}\label{sec_modelo_sociabilidad}

Con el fin de identificar palabras relevantes en el discurso político de los partidos, se utiliza el modelo Bayesiano de sociabilidad propuesto por \citet{sosa2025bayesian}. En este contexto, los nodos se interpretan como palabras y la sociabilidad se entiende como una propensión a coocurrir con múltiples términos en la red, lo que sugiere mayor presencia y centralidad temática en el corpus. Este modelo permite capturar heterogeneidad en el número de relaciones que forman los nodos y, para su aplicación, se trabaja con una red no dirigida y no ponderada. Por tanto, se utiliza una versión binaria de la matriz de adyacencia construida a partir de coocurrencias.

Sea $\mathbf{Y}=[y_{i,j}]$ la matriz de adyacencia binaria de una red no dirigida, con $y_{i,j}\in\{0,1\}$ para $i<j$. Se denota con $\mu$ el efecto global de conectividad y con $\delta_i$ el efecto de sociabilidad del nodo $i\in\mathcal{V}$, y se denota con $\sigma^2$ y $\tau^2$ las varianzas asociadas a $\mu$ y a los $\delta_i$, respectivamente. Las entradas de la matriz se modelan como
\[
y_{i,j}\mid \theta_{i,j}\ \overset{\text{ind}}{\sim}\ \textsf{Ber}(\theta_{i,j}),
\qquad i<j,
\]
donde $\theta_{i,j}=\Phi(\mu+\delta_i+\delta_j)$ y $\Phi(\cdot)$ es la función de distribución acumulada de una normal estándar, lo que induce una formulación tipo probit.

El modelo jerárquico se especifica mediante previas para los parámetros de conectividad y sus componentes de variabilidad. En particular, se asigna al efecto global $\mu\mid\sigma^2\sim \textsf{N}(0,\sigma^2)$. Para los efectos de sociabilidad individual se asume $\delta_i\mid\tau^2 \overset{\text{iid}}{\sim}\textsf{N}(0,\tau^2)$, para $i=1,\ldots,n$, y se garantiza identificabilidad imponiendo la restricción $\sum_{i=1}^{n}\delta_i=0$. Finalmente, las varianzas se modelan con previas Gamma Inversa, de forma que $\sigma^2\sim \textsf{IG}(a_\sigma,b_\sigma)$ para la variabilidad global y $\tau^2\sim \textsf{IG}(a_\tau,b_\tau)$ para la variabilidad de sociabilidad. Siguiendo \citet{sosa2025bayesian}, se fijan $a_\sigma=a_\tau=2$ y $b_\sigma=b_\tau=\tfrac{1}{3}$, con el fin de no favorecer a priori valores extremos de la probabilidad de conexión.

\section{Metodología}

Para el lector que desee replicar los resultados de este análisis, se ponen a disposición los pasos correspondientes en un repositorio de \texttt{GitHub}, disponible en el siguiente enlace \url{https://github.com/briveros23/Tesis_analisis_de_congreso}.

\subsection{Datos de estudio}

Con el objetivo de fortalecer la cercanía entre la ciudadanía y el proceso legislativo, en la Cámara de Representantes se dispone una página web que concentra información pública relevante en \url{https://www.camara.gov.co/}. En particular, en el apartado dedicado a los proyectos de ley se proporciona información esencial para construir las redes y el corpus de texto analizado. Esta información incluye el periodo legislativo en el cual se propone el proyecto de ley, el título y una breve descripción, el código de identificación asignado por la Cámara y el Senado, los representantes e individuos externos que lo proponen y la comisión ante la cual se tramita y se sustenta.

En este trabajo se analizan los periodos presidenciales comprendidos entre 2014 y 2025 con el fin de comparar patrones de colaboración y cambios temáticos a lo largo de tres gobiernos. En primer lugar se estudia el gobierno de Juan Manuel Santos durante el periodo legislativo 2014 a 2018, en el que el proceso de paz ocupa un lugar central en la agenda pública. Posteriormente se considera el gobierno de Iván Duque durante el periodo 2018 a 2022, que se desarrolla bajo el impacto de la pandemia de COVID 19 y sus efectos sociales y económicos. Finalmente se examina el gobierno de Gustavo Petro durante el periodo legislativo 2022 a 2025, cuyo enfoque se orienta al fortalecimiento de una democracia incluyente y a la participación de grupos históricamente marginados. El número total de proyectos de ley registrados durante estos periodos asciende a \(4{,}083\), de acuerdo con la información disponible hasta el 9 de marzo, fecha en la que se inicia el proceso de minería de datos.

\subsection{Recolección de los datos}

En este proyecto se parte de la base de datos suministrada por la página web de la Cámara de Representantes. En esta fuente se observan problemas de codificación y falta de estandarización, por ejemplo nombres incompletos y denominaciones de cargo inconsistentes. Para normalizar esta información se utiliza un enfoque híbrido con componentes automatizados y manuales. En una primera etapa se aplica el modelo generativo \texttt{o3-mini} de \texttt{OpenAI},, mediante el siguiente \textit{prompt}.

\begin{quote}
``Extrae y separa los cargos y nombres de las personas en un texto dado. Los resultados deben presentarse en formato de pares entre paréntesis, donde el primer elemento es el cargo y el segundo el nombre completo, separados por una coma. Mantén la correspondencia de filas del texto original. Cada conjunto de nombres en una línea debe mantenerse en una línea separada en la salida.''
\end{quote}

Con esta instrucción el modelo genera una lista de tuplas \((\texttt{rol},\texttt{nombre})\) que permite automatizar la extracción y separación inicial de los datos. Posteriormente se realiza una revisión manual para verificar que los cargos y los nombres asociados a cada proyecto concuerdan con los registros originales. En esta revisión se corrigen ambigüedades, se completan nombres incompletos y se unifica la nomenclatura de los cargos con el fin de garantizar consistencia y trazabilidad sin alterar el contenido de los registros.

La base de datos inicial no incluye información detallada sobre cada actor legislativo. Por ello se implementa un proceso de minería web para recolectar variables adicionales como el partido político de cada representante, la comisión en la que trabaja y el departamento al que se asocia. Esta recolección se realiza mediante \textit{scripts} en \texttt{Python} empleando las librerías \texttt{BeautifulSoup} y \texttt{requests}.

\subsection{Generación de las redes}

En esta sección se describe la construcción de redes bipartitas, sus proyecciones y redes semánticas basadas en coocurrencias, junto con el preprocesamiento textual requerido, para comparar patrones de colaboración y temas entre los tres periodos presidenciales.

\subsubsection{Redes bipartitas}

Se generan redes bipartitas a partir de dos conjuntos de nodos disyuntos y se definen sus aristas a partir de la incidencia observada en los proyectos de ley, de modo que cada conexión representa participación o patrocinio en una iniciativa legislativa. La primera red se conforma por partidos políticos y proyectos de ley propuestos por sus representantes, lo que permite representar de manera explícita la relación entre colectividades y agendas legislativas. Con esta estructura se caracterizan dinámicas internas y patrones de colaboración entre partidos, al identificar proyectos impulsados de forma individual y proyectos promovidos de manera conjunta, y se busca evidencia de alianzas, afinidades o fragmentaciones que se reflejen en la coautoría, el acompañamiento y la coordinación legislativa.

La segunda red se construye con representantes, sin distinguir afiliación partidista, y proyectos de ley presentados, con el fin de describir la actividad legislativa a nivel individual. Con esta estructura se identifican actores con alta productividad y con posiciones estructuralmente relevantes, al tiempo que se evalúa si la producción de proyectos se concentra en pocos nodos o si se distribuye de manera más homogénea entre los representantes, lo que aporta una medida descriptiva de concentración y liderazgo legislativo. Este procedimiento se implementa con la librería \texttt{igraph} en \texttt{R} y \texttt{Python}.

\subsubsection{Proyecciones}

Las proyecciones de las redes bipartitas permiten inducir redes de un solo modo y resumir patrones de colaboración entre actores del mismo tipo. En la proyección de partidos, se obtiene una red no dirigida y ponderada en la que cada nodo representa un partido político y en la que el peso \(y_{i,j}\) corresponde al número de proyectos de ley en los que participan conjuntamente los partidos \(i\) y \(j\), entendido como copatrocinio o coautoría entre sus representantes. Esta construcción cuantifica la intensidad de colaboración entre colectividades y permite describir la contribución relativa de cada partido dentro de la producción legislativa conjunta.

De manera análoga, la proyección de representantes define una red no dirigida y ponderada en la que los nodos corresponden a congresistas y el peso \(y_{i,j}\) denota el número de proyectos en los que los representantes \(i\) y \(j\) figuran simultáneamente como proponentes. Esta representación permite identificar actores con alta participación, detectar pares o subgrupos con trabajo conjunto persistente y describir patrones de asociación recurrentes dentro del proceso legislativo.

Sobre estas redes de patrocinio y cocreación se calculan métricas descriptivas y se ajustan los modelos propuestos para caracterizar la estructura de colaboración. Aunque en redes proyectadas suele aplicarse un umbral y conservar únicamente aristas con peso superior a un valor prefijado, en este estudio no se realiza filtrado con el fin de evitar pérdida de información, dado el número limitado de partidos con representación efectiva y la relevancia analítica de vínculos de baja frecuencia. Finalmente, el análisis se lleva a cabo por separado para los tres periodos legislativos considerados, con el propósito de evaluar cambios en los patrones de colaboración y en la participación relativa bajo enfoques presidenciales distintos, tanto en términos de alianzas como de prioridades legislativas.

\subsubsection{Limpieza de texto}

La limpieza del texto se utiliza como una etapa de preprocesamiento orientada a reducir ruido y a estandarizar el corpus, con el fin de mejorar la validez de las relaciones de coocurrencia que se emplean para construir las redes semánticas. Para ello, las descripciones de cada proyecto de ley se transforman a minúsculas y se eliminan tildes, de modo que variantes ortográficas del mismo término no induzcan nodos distintos. También se eliminan números, fechas y caracteres no alfabéticos, ya que suelen aparecer en contextos heterogéneos sin aportar contenido temático y pueden generar aristas espurias. Adicionalmente, se excluyen palabras vacías y términos de alta frecuencia propios del lenguaje legislativo que aportan poca información semántica y tienden a dominar la coocurrencia.

No se aplica lematización ni tokenización avanzada debido a que los textos son breves y una normalización morfológica agresiva puede reducir en exceso la diversidad léxica, afectando la identificación de asociaciones entre términos y, en consecuencia, la estructura de las redes. Finalmente, los nombres propios se tratan bajo el mismo esquema de depuración y se remueven cuando funcionan principalmente como marcadores contextuales, como ocurre con topónimos y nombres institucionales, para evitar que concentren conectividad y desplacen términos que representan de manera más directa los tópicos legislativos.

\subsubsection{Redes semánticas}

Una vez se cuenta con los textos limpios, se generan coocurrencias mediante \textit{skip-gramas} de orden uno para construir la matriz de adyacencia ponderada. En esta matriz, los nodos corresponden a palabras y las entradas \(y_{i,j}\) representan el número de coocurrencias observadas entre las palabras \(i\) y \(j\) a lo largo del corpus bajo el esquema de \textit{skip-gramas}. Los \textit{skip-gramas} se forman al considerar pares de palabras dentro de una ventana definida en cada oración, de modo que se capturan asociaciones locales del discurso.

Para fijar ideas se considera un ejemplo simple. Dada la oración \textit{``programa educación superior gratuita''} se obtiene el conjunto de pares
\begin{gather*}
(\text{programa, educación}),
(\text{programa, superior}),
(\text{educación, superior}),\\
(\text{educación, gratuita}),
(\text{superior, gratuita}),    
\end{gather*}
lo que permite representar relaciones semánticas entre términos cercanos, incluso cuando no son estrictamente consecutivos. La matriz de adyacencia asociada puede escribirse como
\[
\mathbf{Y}=
\begin{bmatrix}
0 & 1 & 1 & 0 \\
1 & 0 & 1 & 1 \\
1 & 1 & 0 & 1 \\
0 & 1 & 1 & 0
\end{bmatrix},
\]
donde las filas y columnas corresponden, en ese orden, a \textit{programa}, \textit{educación}, \textit{superior} y \textit{gratuita}, y \(y_{i,j}=1\) indica al menos una coocurrencia dentro de la ventana considerada. Para el análisis se emplean dos representaciones, una binaria donde \(y_{i,j}=1\) si existe al menos una coocurrencia, y una ponderada donde \(y_{i,j}=n\) es el número total de coocurrencias observadas, lo que induce una red ponderada. En ambos casos, \(\mathbf{Y}\) es simétrica porque la coocurrencia se trata como una relación bidireccional.

\section{Resultados}

Esta sección resume los resultados para \(4{,}083\) proyectos en tres gobiernos. Se describe la distribución de iniciativas por partido (Tabla~\ref{tab:comparativa-subcolumnas}), se analizan redes bipartitas y sus proyecciones para caracterizar colaboración y centralidad, y se estudian redes semánticas por colectividad. Finalmente, se identifican hubs con el modelo de sociabilidad y se estiman tópicos mediante un modelo de bloques estocásticos, con contraste adicional usando LDA.

\subsection{Generalidades}

En total se analizan \(4{,}083\) proyectos de ley distribuidos en tres periodos presidenciales. En el gobierno de Juan Manuel Santos se incluyen \(992\) proyectos con sus respectivos representantes. En el gobierno de Iván Duque se incluyen \(1{,}779\) proyectos. En el gobierno de Gustavo Petro se incluyen \(1{,}312\) proyectos.

\begin{table}[!htb]
\centering
\footnotesize
\begin{tabular}{l r l r l r}
\toprule
\multicolumn{2}{c}{\textbf{Santos}} & \multicolumn{2}{c}{\textbf{Duque}} & \multicolumn{2}{c}{\textbf{Petro}} \\
\cmidrule(lr){1-2} \cmidrule(lr){3-4} \cmidrule(lr){5-6}
\textbf{Partido} & \textbf{No.} & \textbf{Partido} & \textbf{No.} & \textbf{Partido} & \textbf{No.} \\
\midrule
Otros & 182 & CD & 215 & PC & 140 \\
PL    & 115 & PL & 175 & PL &  78 \\
CD    & 102 & PC & 163 & PH &  58 \\
PC    & 101 & CR & 144 & CD &  52 \\
U     &  86 & Otros & 97 & Otros & 48 \\
\bottomrule
\end{tabular}
\caption{{\footnotesize Distribución de proyectos propuestos por los cinco partidos con mayor número de iniciativas en los gobiernos de Santos, Duque y Petro. Se usan las abreviaciones PC para Partido Conservador, PL para Partido Liberal, PH para Pacto Histórico, CD para Centro Democrático, CR para Cambio Radical y U para Partido de la U.}}
\label{tab:comparativa-subcolumnas}
\end{table}

En la Tabla~\ref{tab:comparativa-subcolumnas} se observan cambios en la distribución de proyectos propuestos por los partidos con mayor actividad en cada periodo. En el gobierno de Santos destaca la categoría Otros, lo que sugiere una agenda con participación de múltiples actores y presencia de coaliciones en el contexto del proceso de paz. En el gobierno de Duque se observa alta actividad del Centro Democrático, en línea con la filiación del presidente y con un mayor alineamiento legislativo. En el gobierno de Petro se registra un número elevado de proyectos aun con el periodo incompleto, pero sin predominio del partido presidencial, lo que sugiere una iniciativa más dispersa. Estos patrones motivan el análisis de redes y el análisis semántico para evaluar cambios en colaboración, centralidad e identificación de tópicos.

\subsection{Redes Bipartitas}

\begin{figure}[H]
    \centering
    \subfigure[Santos.]{
        \includegraphics[scale=0.2]{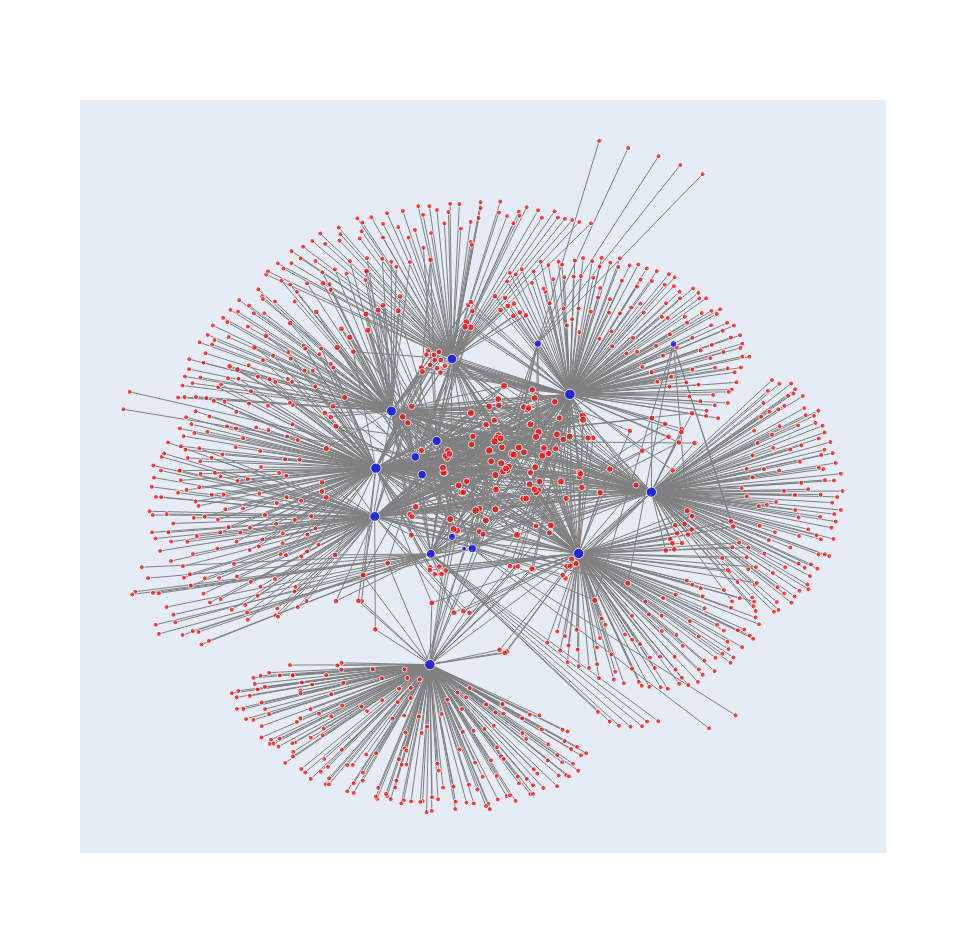}
    }
    \subfigure[Duque.]{
        \includegraphics[scale=0.2]{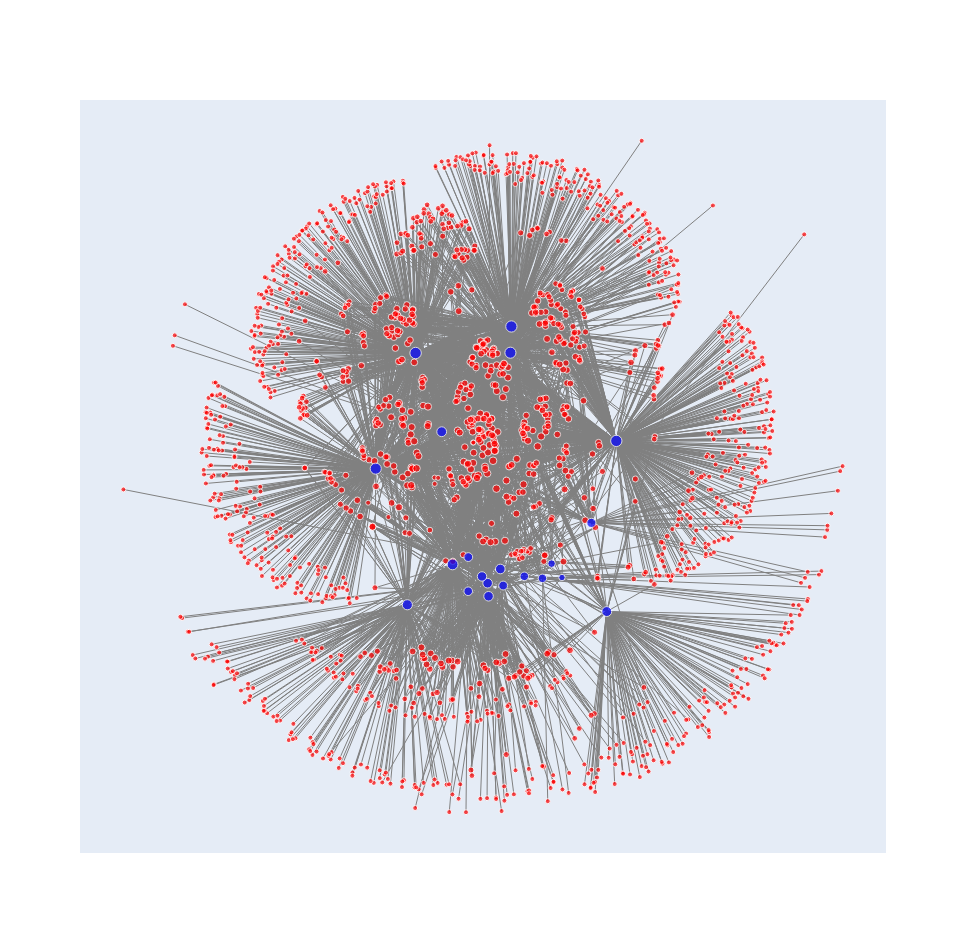}
    }
    \subfigure[Petro.]{
        \includegraphics[scale=0.2]{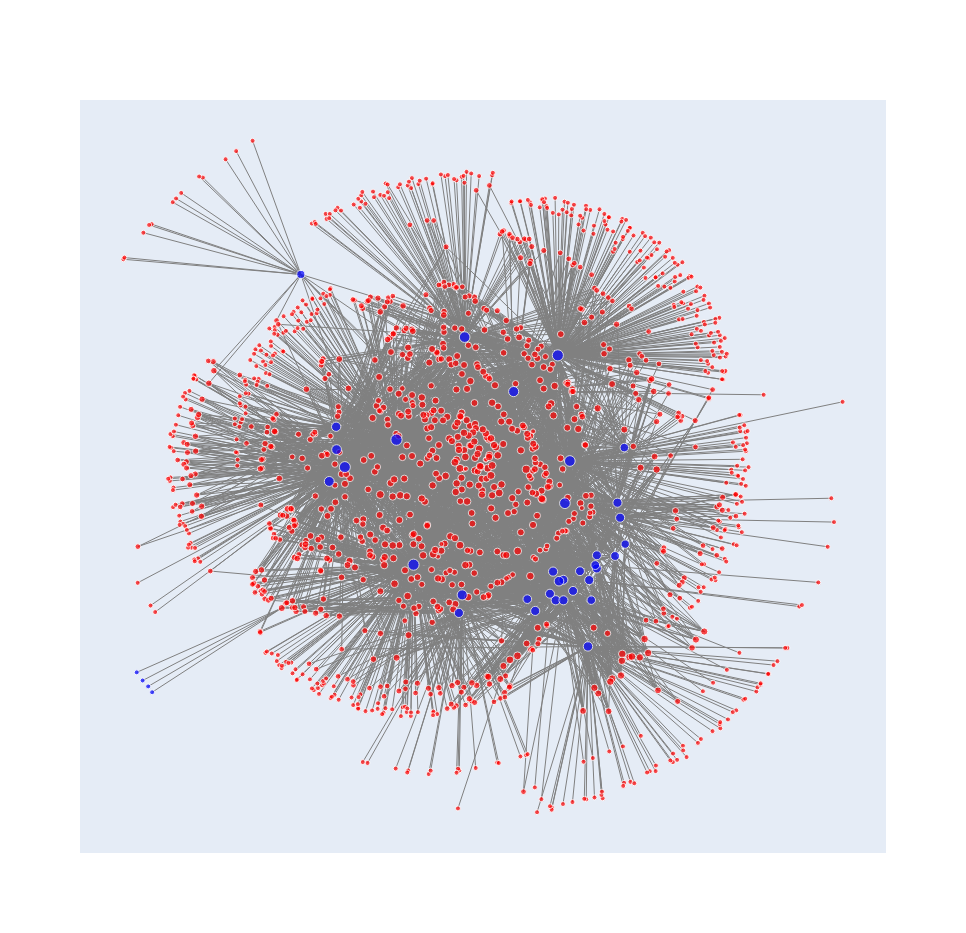}
    }
    \caption{{\footnotesize Redes bipartitas partidos políticos vs. proyectos de ley.}}
    \label{fig:Bipartitas_pl}
\end{figure}

\begin{figure}[H]
    \centering
    \subfigure[Santos.]{
        \includegraphics[scale=0.23]{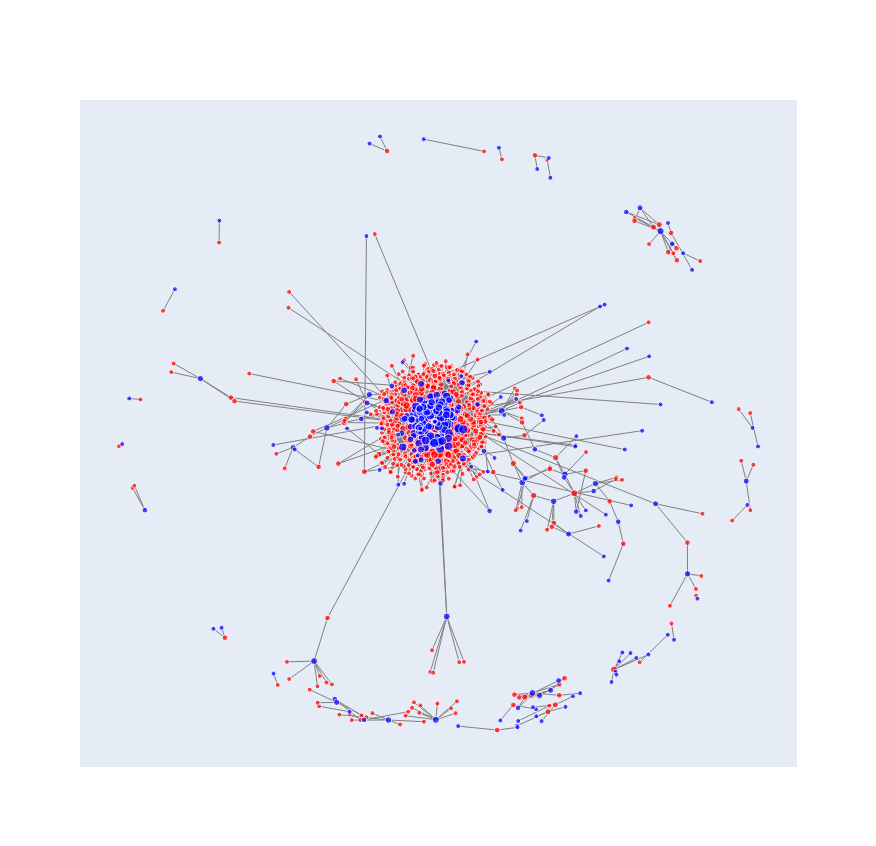}
    }
    \subfigure[Duque.]{
        \includegraphics[scale=0.2]{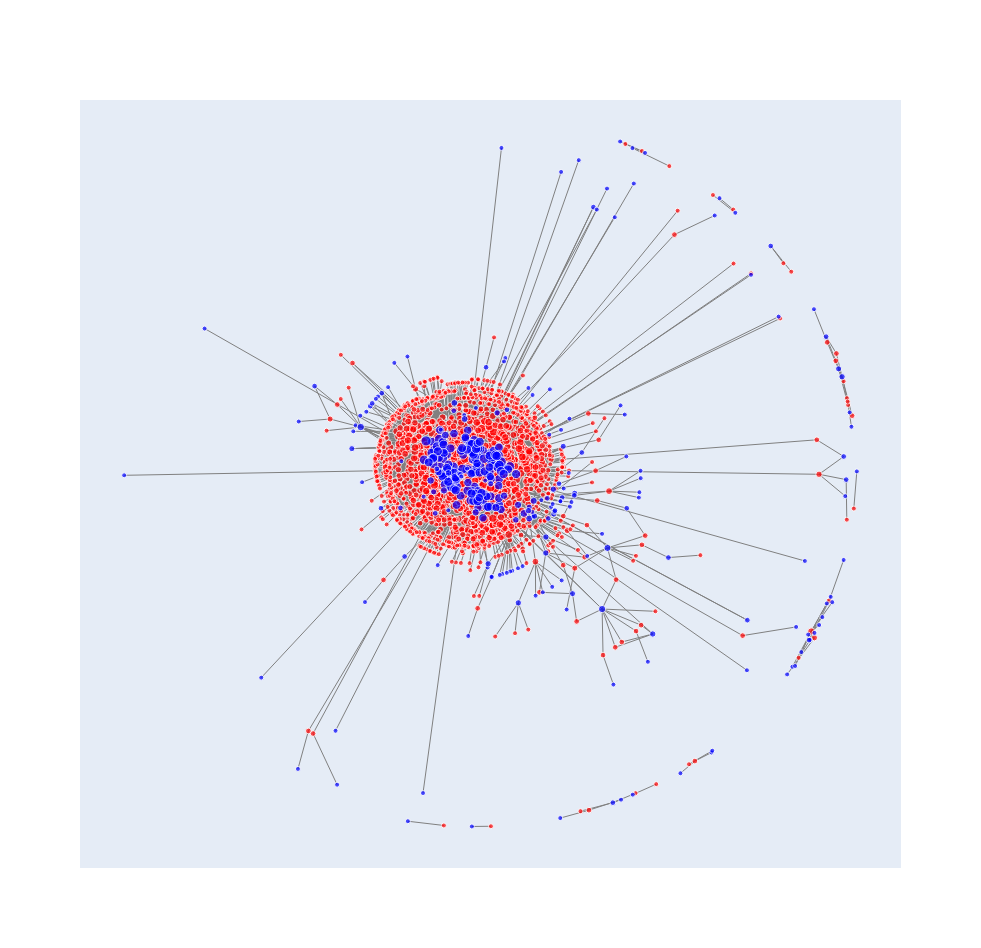}
    }
    \subfigure[Petro.]{
        \includegraphics[scale=0.2]{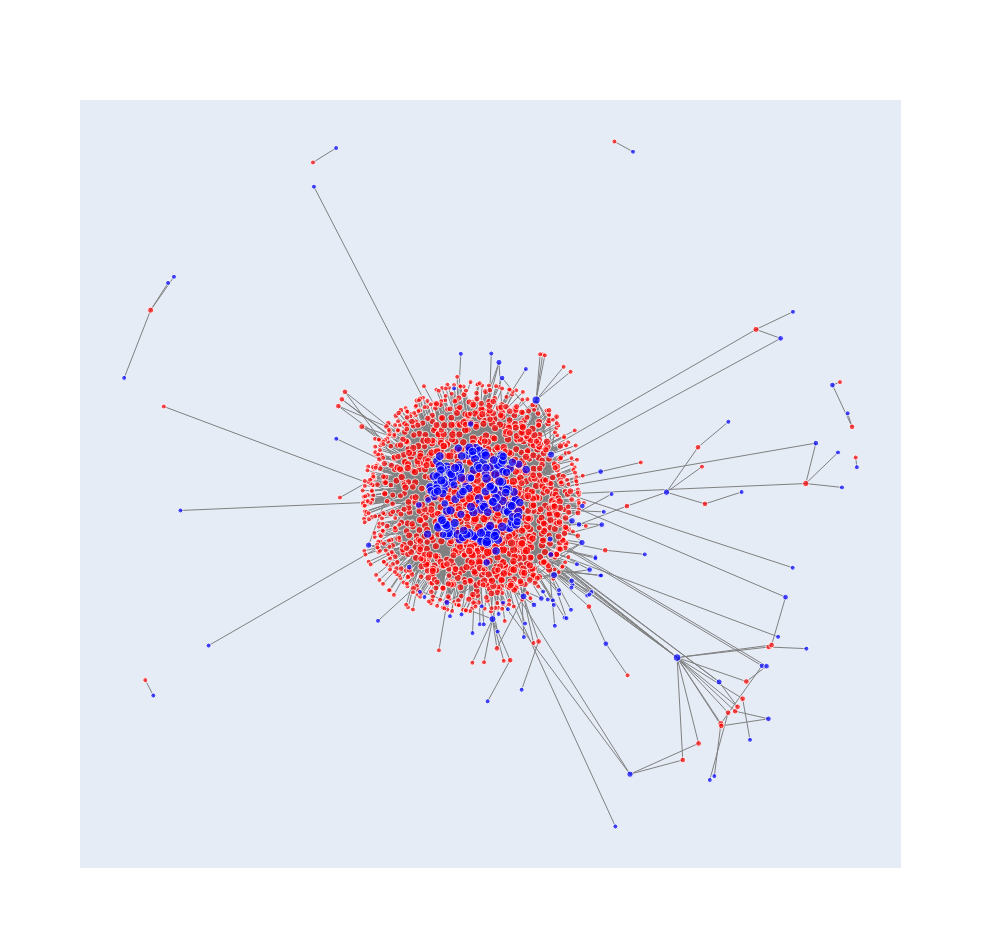}
    }
   \caption{{\footnotesize Redes bipartitas representantes vs. proyectos de ley.}}
    \label{fig:Bipartitas_rl}
\end{figure}

Se construyen dos redes bipartitas a partir de relaciones de incidencia. La primera vincula proyectos de ley con los partidos cuyos representantes participan en su elaboración, lo que da lugar a las redes presentadas en la Figura~\ref{fig:Bipartitas_pl}. La segunda vincula directamente representantes con los proyectos de ley en los que participan, lo que permite describir patrones de actividad y participación individual que se presentan en la Figura~\ref{fig:Bipartitas_rl}.

En la Tabla~\ref{tab:metricas_redes_bipartitas} se resume la estructura de las redes bipartitas y de sus proyecciones para los tres gobiernos. En la proyección Partidos--Leyes se observa una presencia dominante y persistente del Partido Liberal, que aparece como el nodo de mayor grado en los tres periodos y además concentra la mayor centralidad propia en Duque y Petro, lo que indica influencia por asociación al conectarse con partidos que también son centrales. En términos de conectividad agregada, el grado promedio de Partidos--Leyes aumenta de 3.21 en Santos a 4.54 en Duque y a 7.05 en Petro, lo que sugiere un incremento de la colaboración interpartidista medida por proyectos compartidos, mientras que el diámetro se mantiene en 6, consistente con una estructura donde los partidos quedan a pocas etapas unos de otros.

En la proyección Individuos--Leyes se identifican representantes con alta actividad y posiciones estructurales destacadas. Carlos Guevara Villabón en Santos, Fabián Díaz Plata en Duque y Gabriel Parrado Durán en Petro aparecen con el mayor grado y además con alta centralidad propia o intermediación, lo que sugiere un rol relevante en la articulación de coautorías y en la conectividad global. De manera complementaria, los valores máximos de intermediación en Individuos--Leyes se asocian con proyectos específicos, lo que indica que ciertas iniciativas concentran conexiones entre grupos y actúan como puntos de articulación del trabajo legislativo dentro de cada periodo.

\subsection{Proyecciones}

\begin{figure}[H]
    \centering
    \subfigure[Santos.]{
        \includegraphics[scale=0.2]{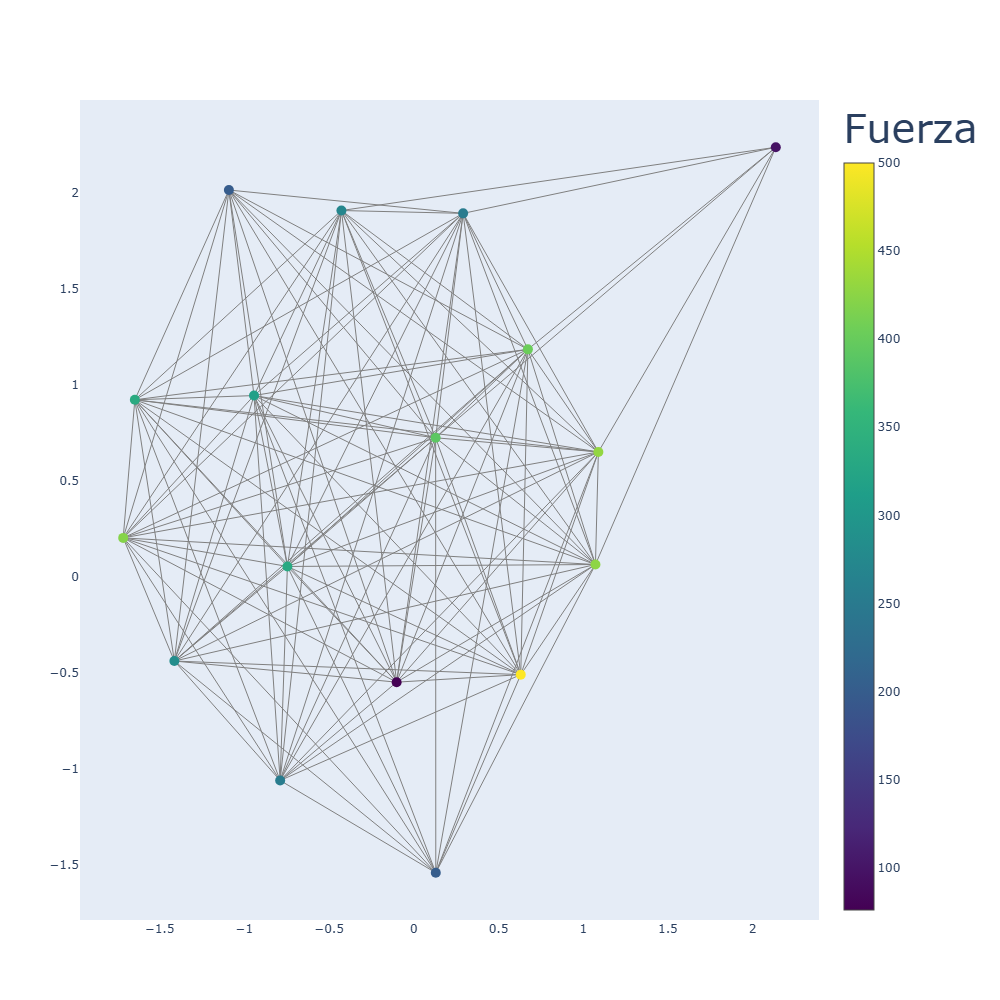}
    }
    \subfigure[Duque.]{
        \includegraphics[scale=0.2]{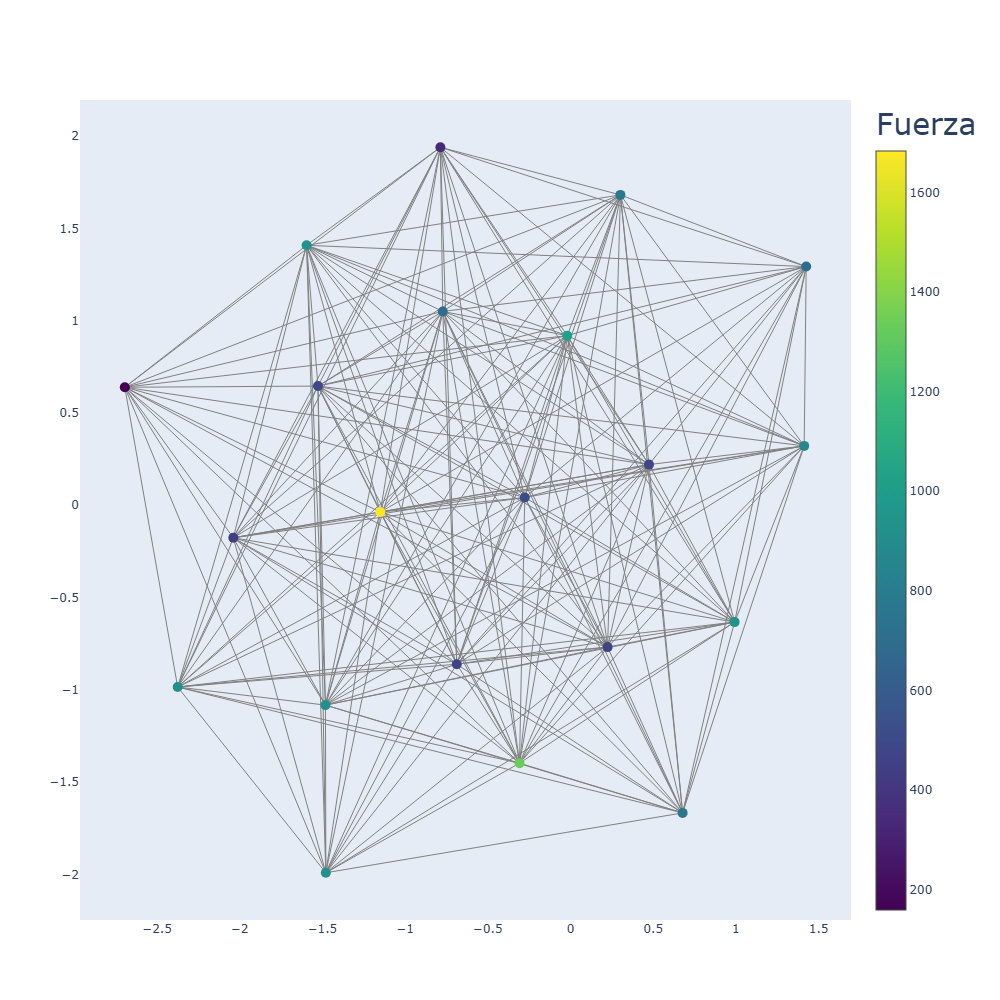}
    }
    \subfigure[Petro.]{
        \includegraphics[scale=0.2]{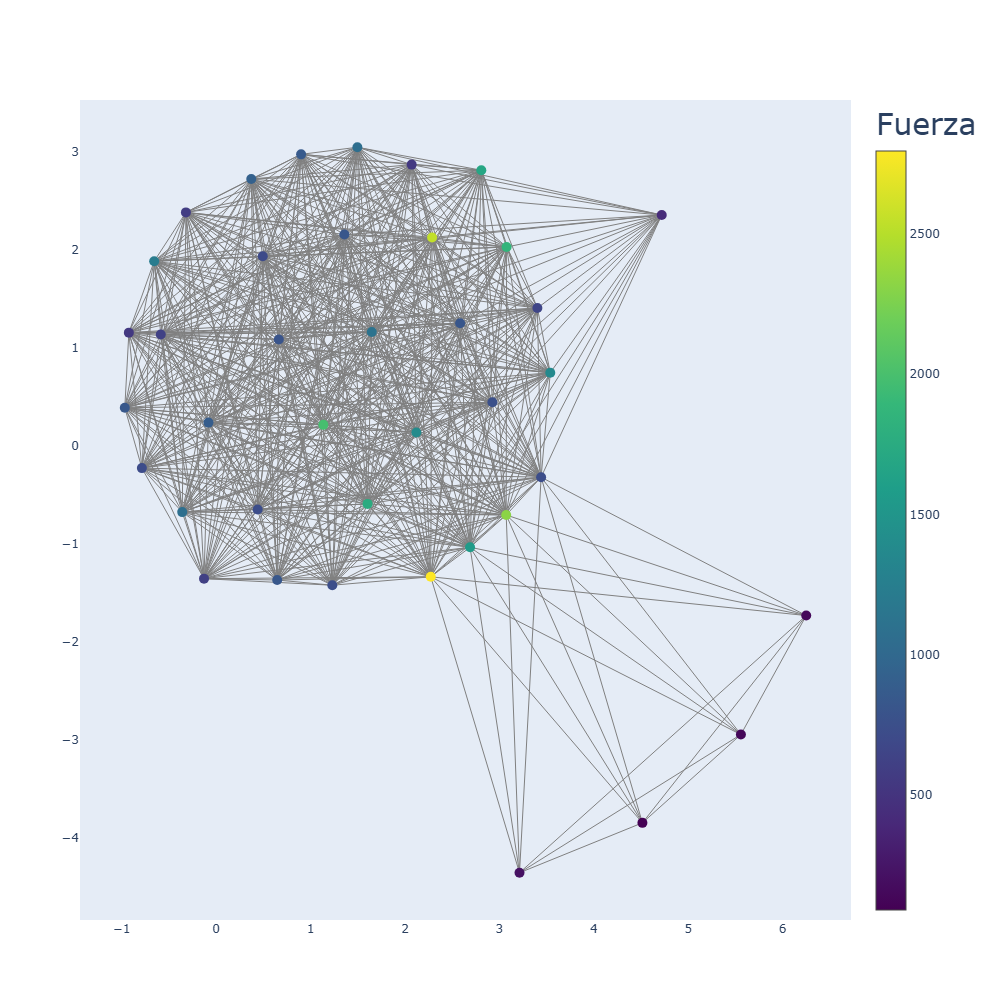}
    }
    \caption{{\footnotesize Proyección de las colaboraciones de partidos políticos.}}
    \label{fig:Proyeccion_leyes}
\end{figure}

\begin{figure}[H]
    \centering
    \subfigure[Santos.]{
        \includegraphics[scale=0.2]{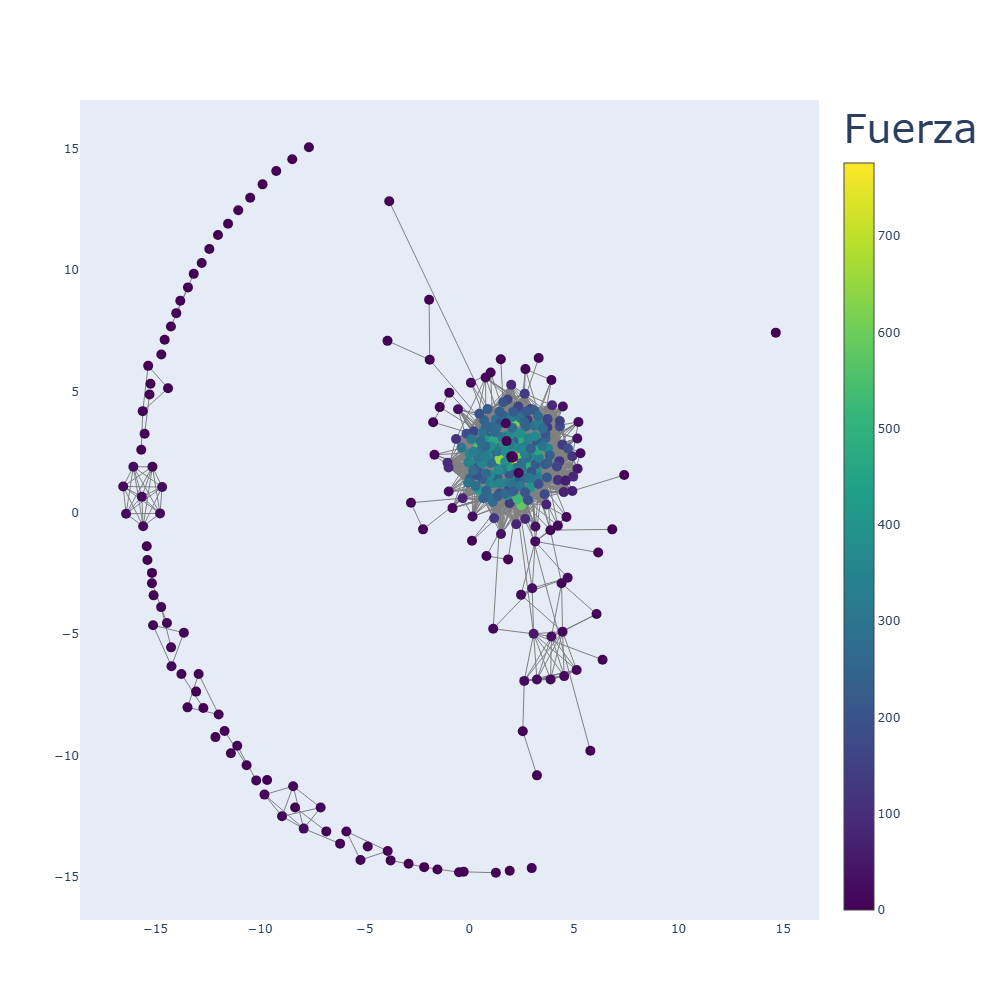}
    }
    \subfigure[Duque.]{
        \includegraphics[scale=0.2]{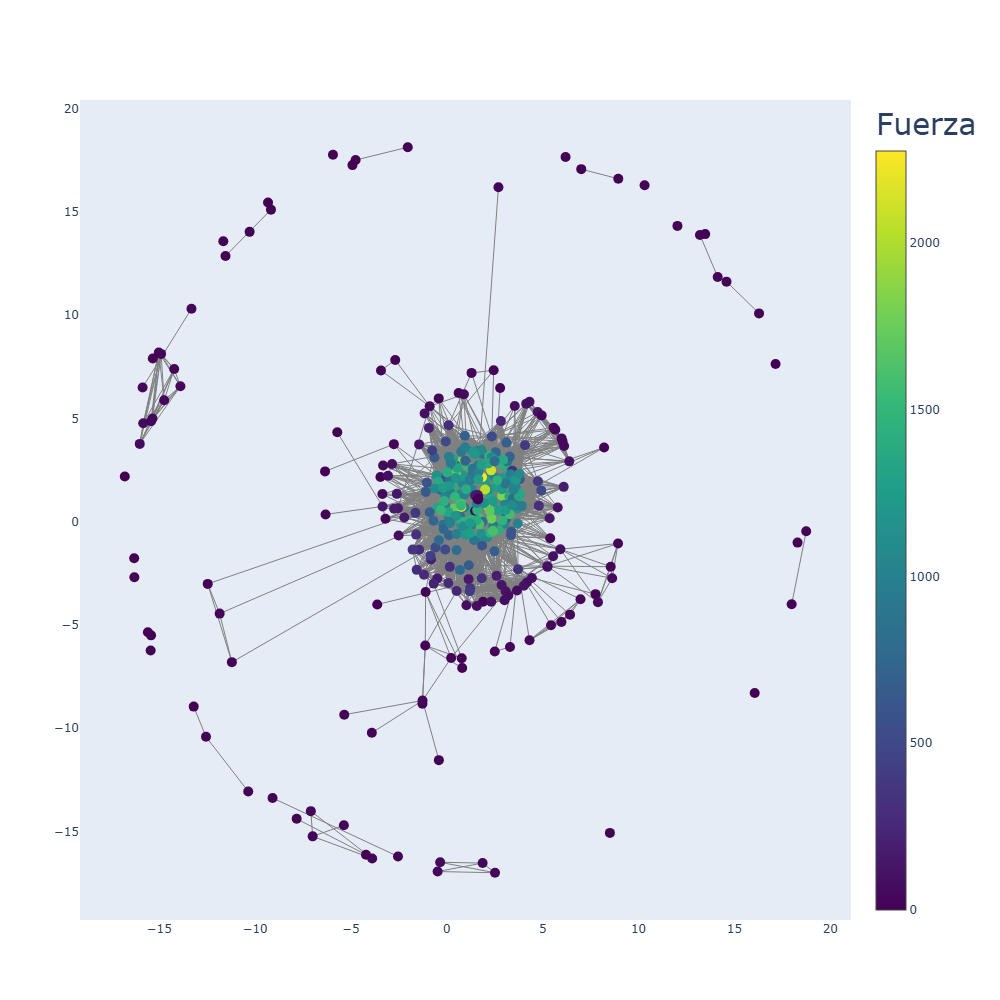}
    }
    \subfigure[Petro.]{
        \includegraphics[scale=0.2]{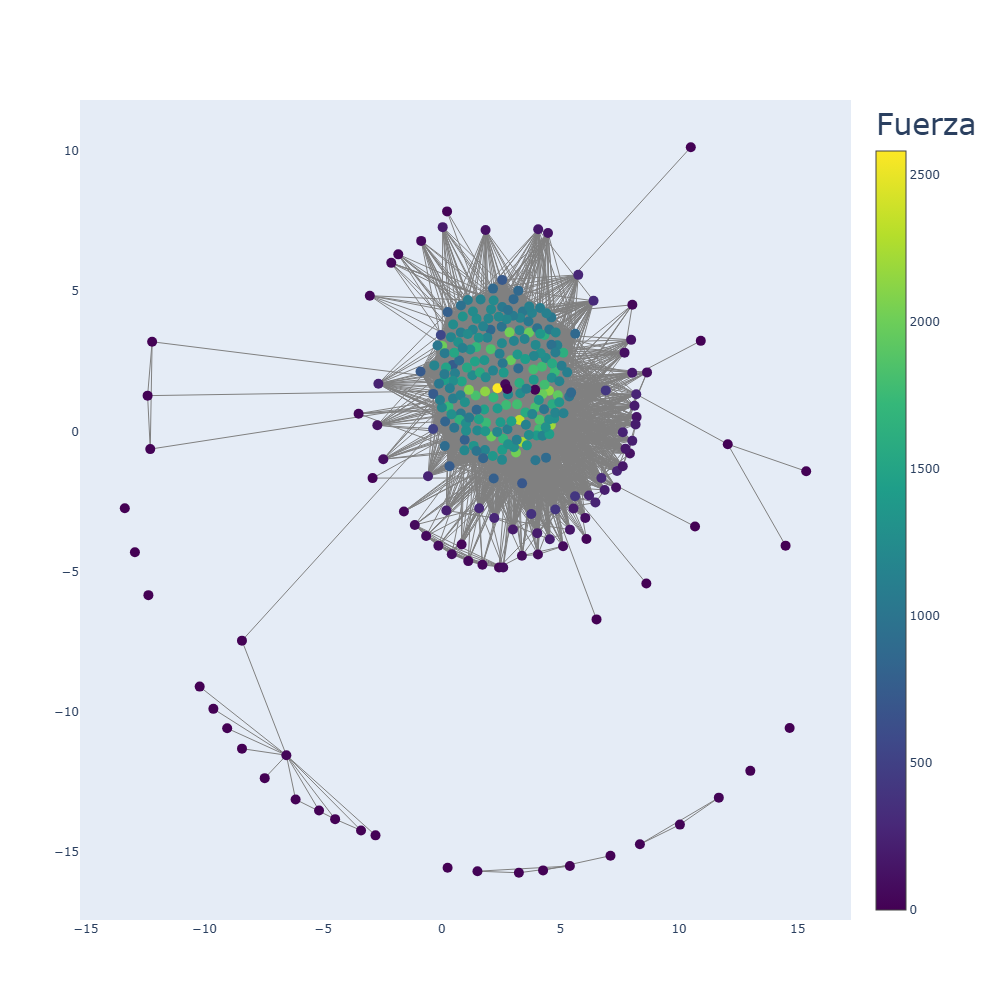}
    }
    \caption{{\footnotesize Proyección de las colaboraciones de representantes.}}
    \label{fig:Proyeccion_autores}
\end{figure}

A partir de las redes bipartitas se construyen proyecciones que resumen la colaboración entre actores del mismo tipo. En la proyección de partidos cada nodo representa una colectividad y cada arista resume el copatrocinio entre dos partidos, ponderado por el número de iniciativas compartidas, como se ilustra en la Figura~\ref{fig:Proyeccion_leyes}. En la proyección de representantes cada nodo corresponde a un congresista y las aristas ponderadas indican cuántos proyectos comparte con otros representantes, como se muestra en la Figura~\ref{fig:Proyeccion_autores}. El color de los nodos refleja la fuerza, de modo que tonos más intensos identifican actores con mayor volumen de coautorías.

La Tabla~\ref{tab:metricas_proyeccion_gobiernos} indica que las redes de partidos son muy densas en los tres gobiernos, con diámetro \(2\) y densidades superiores a \(0.8\), lo que sugiere un copatrocinio cercano a completo y conexiones a pocos pasos. El grado promedio aumenta de \(14.12\) en Santos a \(18.57\) en Duque y a \(30.87\) en Petro, con alta transitividad, en concordancia con el núcleo interconectado observado en la Figura~\ref{fig:Proyeccion_leyes}. En este contexto, la mayor fuerza se concentra en el Partido Liberal en Santos, el Centro Democrático en Duque y la coalición Pacto Histórico en Petro, mientras que el Partido Conservador Colombiano y el Partido Alianza Verde destacan por cercanía e intermediación en periodos puntuales.

Las proyecciones de representantes son más dispersas, pero muestran núcleos cada vez más cohesivos. El diámetro cae de \(9\) a \(7\) y a \(5\) y la densidad sube de \(0.18\) a \(0.22\) y a \(0.41\), lo que sugiere mayor coautoría y formación de clústeres, como se aprecia en la Figura~\ref{fig:Proyeccion_autores}. Sobresalen por grado o fuerza Víctor Javier Correa Vélez en Santos, Jennifer Arias Falla y Eloy Quintero Romero en Duque, y Eduard Sarmiento Hidalgo y Carolina Giraldo Botero en Petro, mientras que \textit{closeness}, \textit{betweenness} y \textit{eigenvector} resaltan a Gonzálo Araújo Muñoz, Alicia Arango Olmos, David Racero y Gabriel Parrado Durán como actores con capacidad de articulación entre subgrupos.

\subsection{Redes semánticas}

\begin{figure}[H]
    \centering
    \subfigure[Otros.]{
        \includegraphics[scale=0.24]{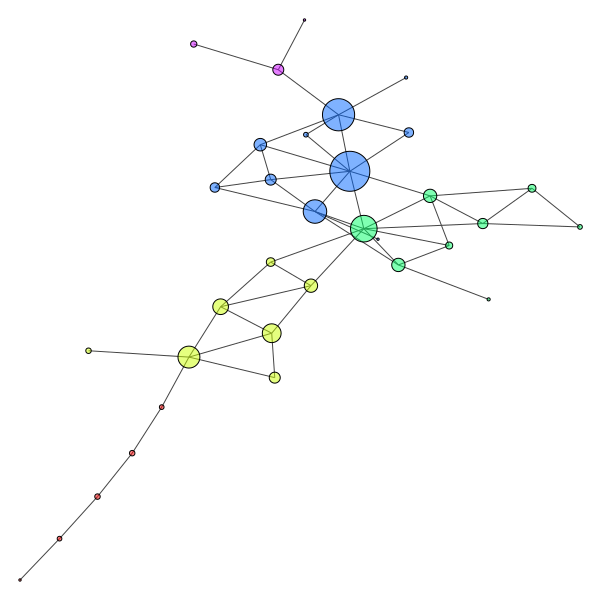}
        \label{fig:otros}
    }
    \subfigure[Partido Liberal.]{
        \includegraphics[scale=0.24]{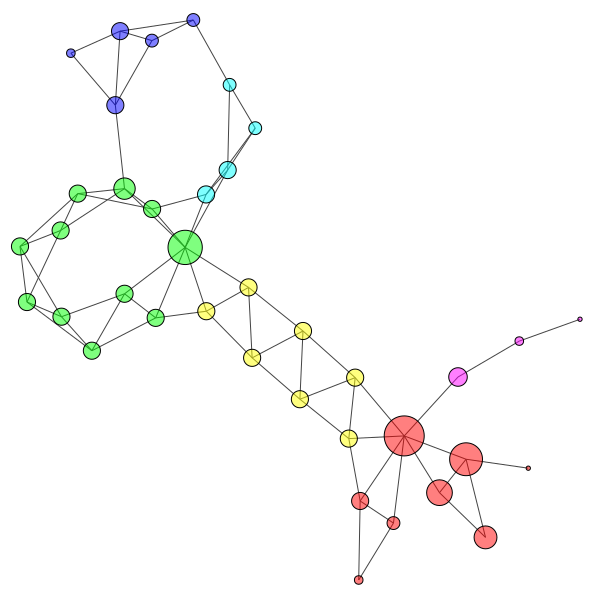}
        \label{fig:liberal}
    }

    \subfigure[Centro Democrático.]{
        \includegraphics[scale=0.24]{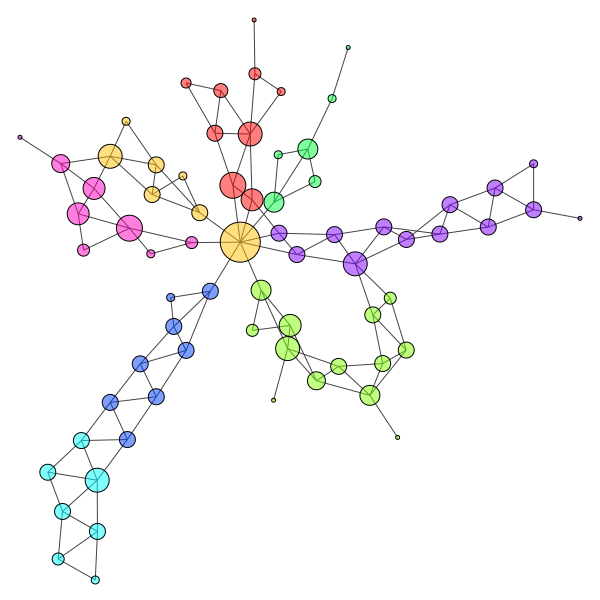}
        \label{fig:centrodemocratico}
    }
    \subfigure[Partido Conservador.]{
        \includegraphics[scale=0.24]{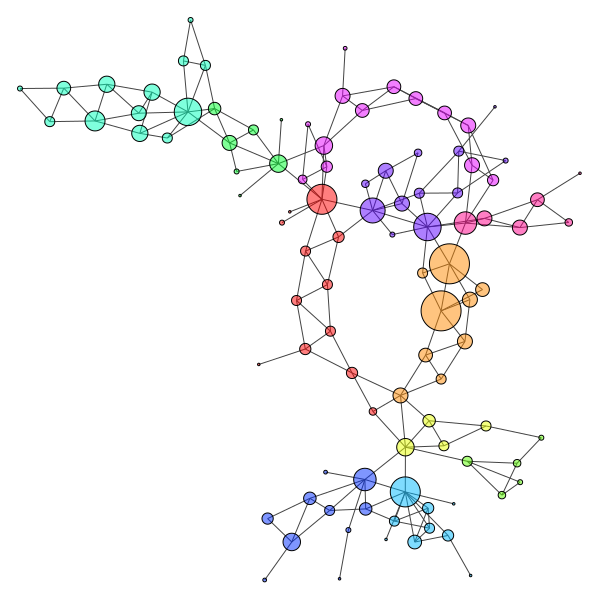}
        \label{fig:conservador}
    }

    \subfigure[Partido de la U.]{
        \includegraphics[scale=0.24]{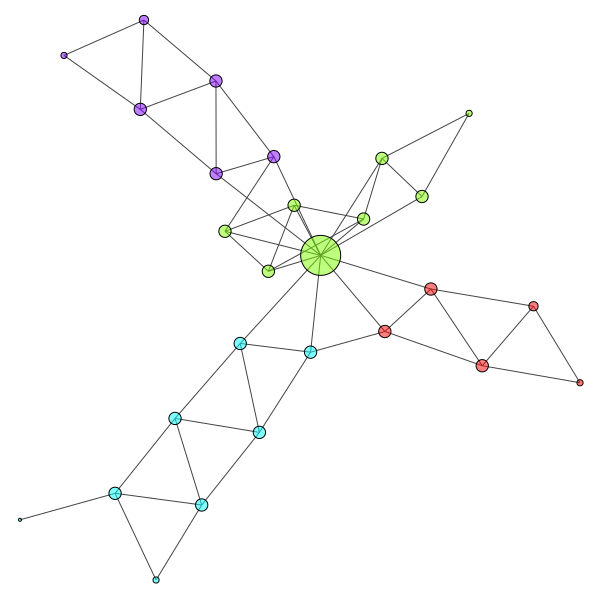}
        \label{fig:partidodelau}
    }
    \caption{{\footnotesize Comunidades semánticas por partido político durante el gobierno de Santos.}}
    \label{fig:comunidades_partidos_santos}
\end{figure}

\begin{figure}[H]
    \centering
    \subfigure[Centro Democrático.]{
        \includegraphics[scale=0.24]{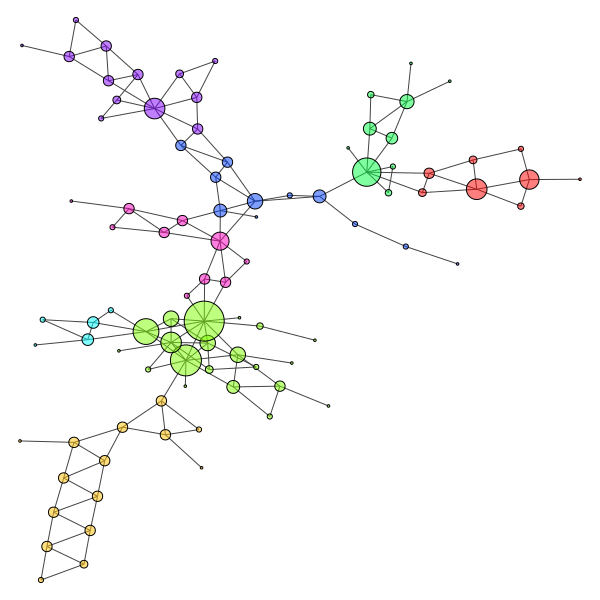}
        \label{fig:comunidades_cc_duque}
    }
    \subfigure[Partido Liberal.]{
        \includegraphics[scale=0.24]{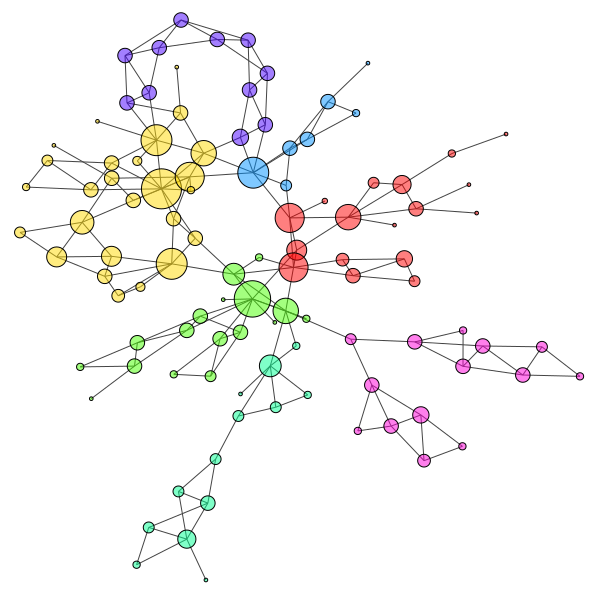}
        \label{fig:comunidades_pl_duque}
    }

    \subfigure[Partido Conservador.]{
        \includegraphics[scale=0.24]{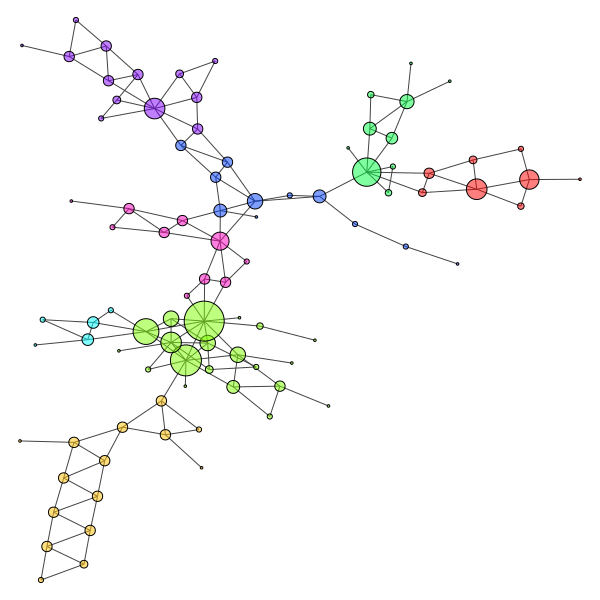}
        \label{fig:comunidades_pcc_duque}
    }
    \subfigure[Cambio Radical.]{
        \includegraphics[scale=0.24]{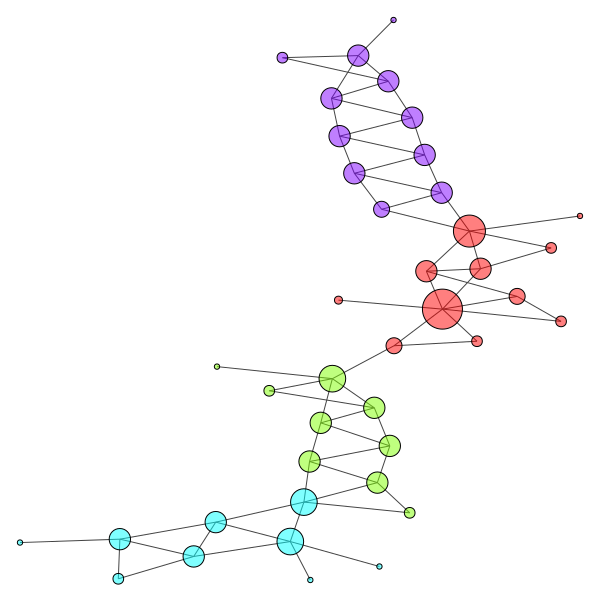}
        \label{fig:comunidades_cr_duque}
    }

    \subfigure[Otros.]{
        \includegraphics[scale=0.24]{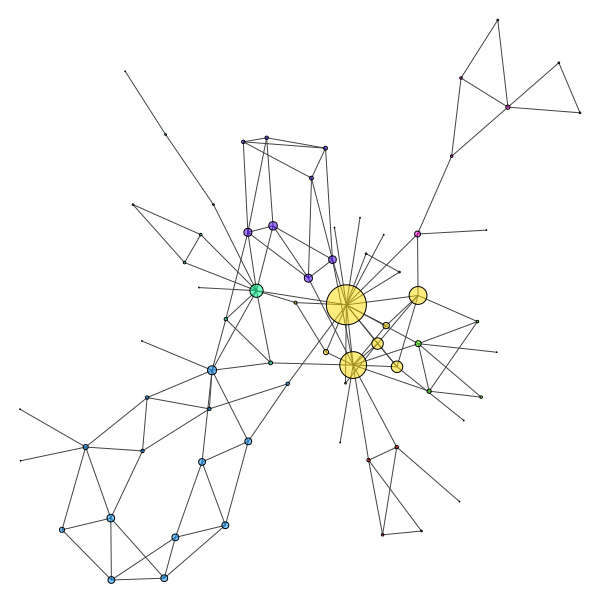}
        \label{fig:comunidades_otros_duque}
    }
    \caption{{\footnotesize Comunidades semánticas por partido político durante el gobierno de Duque.}}
    \label{fig:comunidades_partidos_duque}
\end{figure}

\begin{figure}[H]
    \centering
    \subfigure[Partido Conservador.]{
        \includegraphics[scale=0.24]{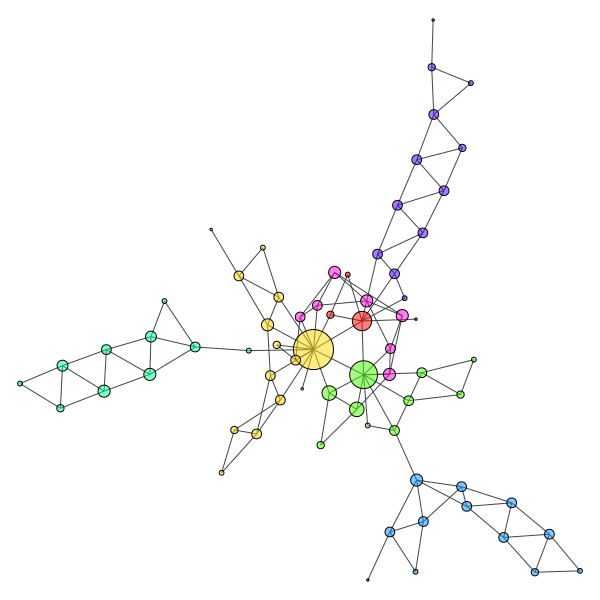}
        \label{fig:comunidades_pcc_petro}
    }
    \subfigure[Partido Liberal.]{
        \includegraphics[scale=0.24]{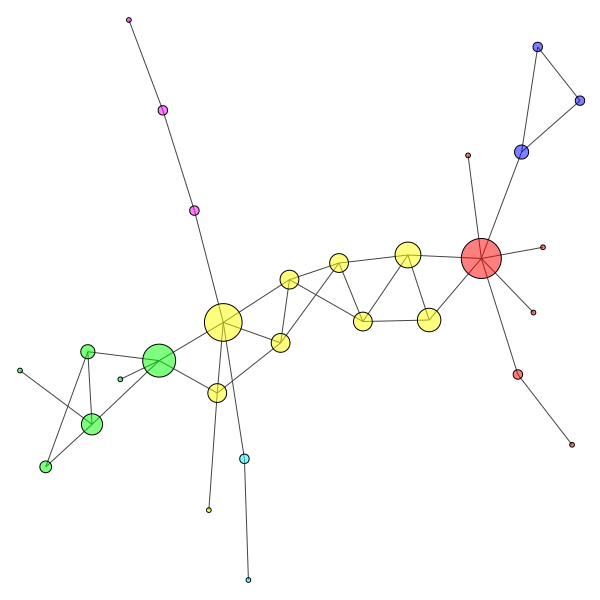}
        \label{fig:comunidades_pl_petro}
    }

    \subfigure[Coalición Pacto Histórico.]{
        \includegraphics[scale=0.24]{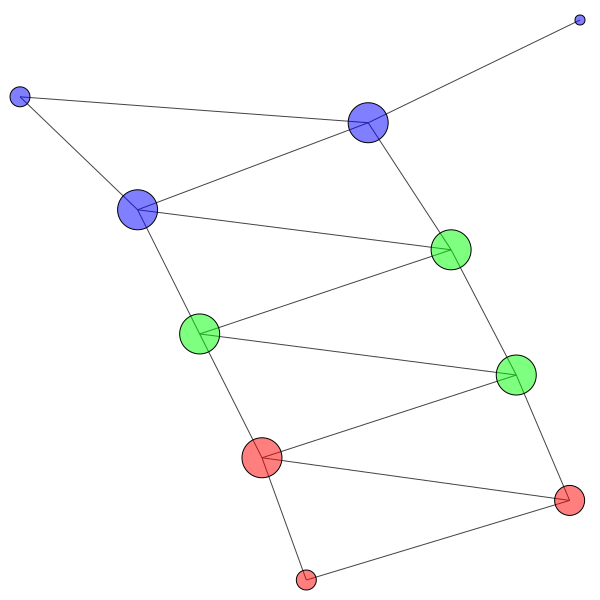}
        \label{fig:comunidades_cph_petro}
    }
    \subfigure[Centro Democrático.]{
        \includegraphics[scale=0.24]{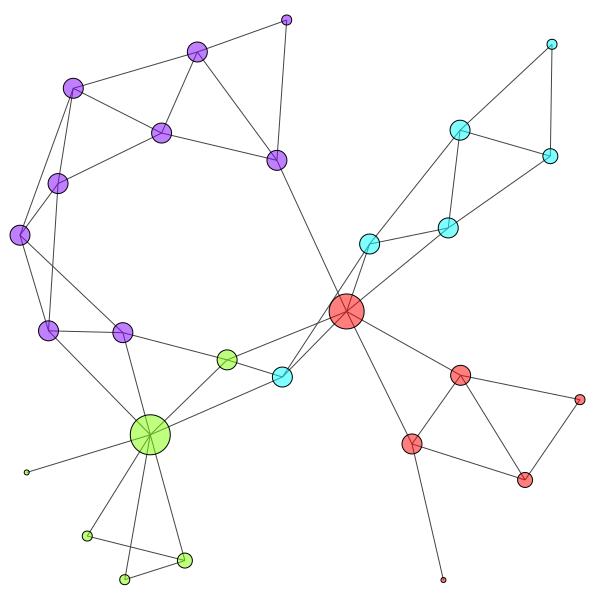}
        \label{fig:comunidades_cd_petro}
    }

    \subfigure[Otros.]{
        \includegraphics[scale=0.24]{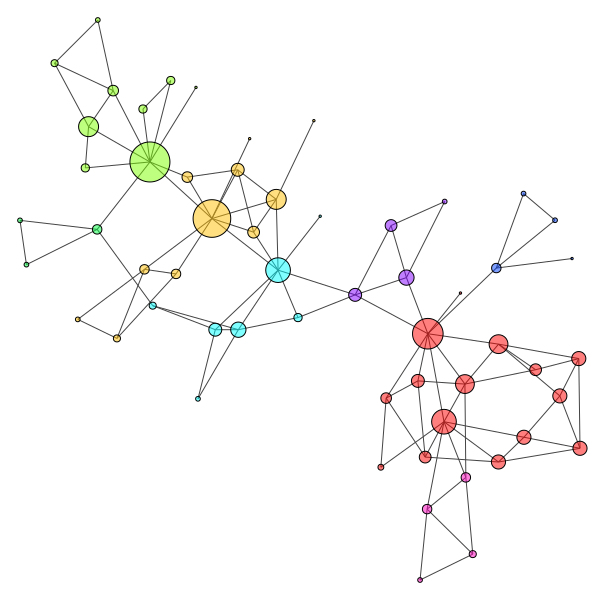}
        \label{fig:comunidades_otros_petro}
    }
    \caption{{\footnotesize Comunidades semánticas por partido político durante el gobierno de Petro.}}
    \label{fig:comunidades_partidos_petro}
\end{figure}

Con base en los resultados de la sección anterior y en la cohesión observada en las redes bipartitas y sus proyecciones, se aplica un análisis semántico a los proyectos de ley asociados con los partidos más influyentes y con las iniciativas promovidas de forma independiente. Este análisis busca identificar tópicos que sinteticen el contenido temático y caractericen el contenido de los proyectos de ley impulsados o apoyados por cada colectividad. Las componentes conexas de las redes semánticas resultantes se presentan por gobierno y por partido en las Figuras~\ref{fig:comunidades_partidos_santos}, \ref{fig:comunidades_partidos_duque} y \ref{fig:comunidades_partidos_petro}.

\subsubsection{Gobierno Santos}

La Tabla~\ref{tab:metricas_semanticas_santos_partidos} sugiere que en el gobierno de Santos los discursos legislativos se organizan en comunidades claras en un contexto marcado por el proceso de paz, el fortalecimiento institucional y agendas territoriales. El Partido de la U presenta una red más compacta, compatible con pocos ejes dominantes, mientras que el Partido Conservador exhibe una estructura más extendida, con mayor diámetro y menor densidad, consistente con una agenda más diversificada. En la misma línea, Otros y el Partido Liberal tienden a redes relativamente más compactas, mientras que Centro Democrático y el Partido Conservador muestran comunidades más separadas.

El panel (b) resume los términos que anclan cada discurso. En Otros, \textit{república} concentra coocurrencias y \textit{acuerdo} actúa como articulador, coherente con un lenguaje institucional asociado con negociación e implementación. En el Partido Liberal, \textit{liberal} organiza el discurso y \textit{municipio} concentra asociaciones ligadas a énfasis territorial. En Centro Democrático, \textit{colombia} estructura gran parte de las conexiones y \textit{cuerpos} agrupa asociaciones locales. En el Partido Conservador, \textit{aceites}, \textit{sostenible}, \textit{servicios} y \textit{territorio} sugieren un perfil multifocal. En el Partido de la U, \textit{vivienda} domina centralidad e intensidad, reforzando un foco temático concentrado.

\begin{table}[!htb]
\centering
\footnotesize

\textit{Panel (a). Métricas globales por partido para redes semánticas en el gobierno de Santos.}

\medskip
\begin{tabular}{lp{1.5cm}p{1.5cm}p{1.5cm}p{1.5cm}p{1.5cm}p{1.5cm}}
\toprule
Partido & Diám. & G.\ Prom. & G.\ Máx. & F.\ Máx. & Den. & Tran. \\
\midrule
Otros               & 12 & 2.95 &  8 & 51 & 0.05 & 0.38 \\
Partido Liberal     & 10 & 2.67 &  8 & 28 & 0.04 & 0.44 \\
Centro Democrático  & 14 & 3.09 &  9 & 22 & 0.02 & 0.42 \\
Partido Conservador & 17 & 3.24 & 10 & 32 & 0.02 & 0.38 \\
Partido de la U     &  9 & 2.88 & 12 & 26 & 0.03 & 0.45 \\
\bottomrule
\end{tabular}

\bigskip
\textit{Panel (b). Palabras más importantes según la métrica indicada.}

\medskip
\begin{tabular}{lp{1.9cm}p{1.9cm}p{1.9cm}p{1.9cm}p{1.9cm}}
\toprule
Partido & Grado & Fuerza & \textit{Closeness} & \textit{Betweenness} & \textit{Eigenvector} \\
\midrule
Otros              & acuerdo, república & república & adopta     & acuerdo   & república \\
Partido Liberal    & liberal            & municipio & frente     & liberal   & liberal \\
Centro Democrático & colombia           & cuerpos   & adoptan    & colombia  & colombia \\
Partido Conservador& años               & aceites   & sostenible & servicios & territorio \\
Partido de la U    & vivienda           & vivienda  & cultural   & vivienda  & vivienda \\
\bottomrule
\end{tabular}

\caption{{\footnotesize Resumen de métricas globales y palabras más importantes por partido en el gobierno de Santos. En el panel (a), Diám.\ denota el diámetro, G.\ Prom.\ denota el grado promedio, G.\ Máx.\ denota el grado máximo, F.\ Máx.\ denota la fuerza máxima, Den.\ denota la densidad y Tran.\ denota la transitividad.}}
\label{tab:metricas_semanticas_santos_partidos}
\end{table}

\subsubsection{Gobierno Duque}

La Tabla~\ref{tab:metricas_semanticas_duque_partidos} sugiere que, en el gobierno de Duque, las redes semánticas capturan un entorno legislativo marcado por prioridades de orden público, agenda social y coordinación territorial, en un contexto de choques económicos y sanitarios. En ese marco, el Partido Conservador presenta la red más extendida, con diámetro \(26\), lo que es compatible con una agenda más diversificada, mientras que Otros muestra una estructura más compacta, con mayor densidad y fuerza máxima, lo que sugiere un discurso más reiterativo y estandarizado.

El panel (b) permite interpretar esos patrones. En Centro Democrático, \textit{seguridad} organiza el discurso y \textit{hacinamiento} aparece como término cercano, consistente con énfasis en política criminal y capacidad institucional. En el Partido Liberal, \textit{instituciones}, \textit{niños} y \textit{funcionamiento} se asocian con fortalecimiento institucional y protección social. En el Partido Conservador, \textit{municipio}, \textit{familia} y \textit{personas} sugieren foco territorial y social. En Cambio Radical, \textit{mecanismos}, \textit{asamblea} y \textit{laboral} apuntan a reglas de participación y mercado de trabajo. En Otros, \textit{república} y \textit{derechos} reflejan un lenguaje más institucional y transversal.

\begin{table}[!htb]
\centering
\footnotesize

\textit{Panel (a). Métricas globales por partido para redes semánticas en el gobierno de Duque.}

\medskip
\begin{tabular}{lp{1.5cm}p{1.5cm}p{1.5cm}p{1.5cm}p{1.5cm}p{1.5cm}}
\toprule
Partido & Diám. & G.\ Prom. & G.\ Máx. & F.\ Máx. & Den. & Tran. \\
\midrule
Centro Democrático & 17 & 2.67 & 11 & 31 & 0.01 & 0.37 \\
Partido Liberal    & 13 & 2.97 & 10 & 22 & 0.01 & 0.39 \\
Partido Conservador& 26 & 3.32 & 15 & 34 & 0.02 & 0.40 \\
Cambio Radical     & 16 & 2.74 &  8 & 16 & 0.02 & 0.43 \\
Otros              &  9 & 3.18 & 18 & 94 & 0.03 & 0.31 \\
\bottomrule
\end{tabular}

\bigskip
\textit{Panel (b). Palabras más importantes según la métrica indicada.}

\medskip
\begin{tabular}{lp{1.9cm}p{1.9cm}p{1.9cm}p{1.9cm}p{1.9cm}}
\toprule
Partido & Grado & Fuerza & \textit{Closeness} & \textit{Betweenness} & \textit{Eigenvector} \\
\midrule
Centro Democrático & seguridad & seguridad & hacinamiento & programa & seguridad \\
Partido Liberal    & instituciones, programa & instituciones & niños & funcionamiento & instituciones \\
Partido Conservador& municipio & municipio & familia & personas & municipio \\
Cambio Radical     & mecanismos & mecanismos & asamblea & laboral & mecanismos \\
Otros              & república & república & derechos & república & república \\
\bottomrule
\end{tabular}

\caption{{\footnotesize Resumen de métricas globales y palabras más importantes por partido en el gobierno de Duque. En el panel (a), Diám.\ denota el diámetro, G.\ Prom.\ denota el grado promedio, G.\ Máx.\ denota el grado máximo, F.\ Máx.\ denota la fuerza máxima, Den.\ denota la densidad y Tran.\ denota la transitividad.}}
\label{tab:metricas_semanticas_duque_partidos}
\end{table}

\subsubsection{Gobierno Petro}

La Tabla~\ref{tab:metricas_semanticas_petro_partidos} sugiere que, en el gobierno de Petro, las redes semánticas reflejan un contexto de reformas sociales y debates sobre modelo productivo y seguridad, con discursos de distinta concentración temática por partido. Pacto Histórico muestra la estructura más compacta, con diámetro \(5\) y la mayor transitividad, \(0.54\), consistente con pocos ejes bien cohesionados. En contraste, el Partido Conservador presenta una red más extendida, con diámetro \(13\), compatible con una agenda más dispersa. La densidad es mayor en Centro Democrático y Otros, con \(0.04\), lo que indica coocurrencias más recurrentes y vocabulario más redundante.

El panel (b) identifica los anclajes del discurso. En el Partido Conservador, \textit{salud} y \textit{niñas} apuntan a prioridades de bienestar y protección. En el Partido Liberal, \textit{cultural} y \textit{municipio} sugieren énfasis territorial e institucional. En Centro Democrático, \textit{seguridad} domina y \textit{protección} articula, coherente con una agenda de orden público y garantías. En Pacto Histórico, \textit{explotación}, \textit{recursos} e \textit{inmaterial}, junto con \textit{públicos} y \textit{debates}, reflejan discusiones sobre regulación, bienes comunes y reforma del Estado. En Otros, \textit{república}, \textit{aprueba}, \textit{adopta} y \textit{general} describen un léxico más procedimental e institucional.

\begin{table}[!htb]
\centering
\footnotesize

\textit{Panel (a). Métricas globales por partido para redes semánticas en el gobierno de Petro.}

\medskip
\begin{tabular}{lp{1.5cm}p{1.5cm}p{1.5cm}p{1.5cm}p{1.5cm}p{1.5cm}}
\toprule
Partido & Diám. & G.\ Prom. & G.\ Máx. & F.\ Máx. & Den. & Tran. \\
\midrule
Partido Conservador   & 13 & 2.56 & 14 & 33 & 0.01 & 0.39 \\
Partido Liberal       &  9 & 2.53 &  7 & 17 & 0.03 & 0.45 \\
Centro Democrático    &  8 & 3.05 &  8 & 16 & 0.04 & 0.48 \\
Pacto Histórico       &  5 & 2.16 &  4 & 16 & 0.03 & 0.54 \\
Otros                 &  9 & 3.08 &  9 & 34 & 0.04 & 0.36 \\
\bottomrule
\end{tabular}

\bigskip
\textit{Panel (b). Palabras más importantes según la métrica indicada.}

\medskip
\begin{tabular}{lp{1.9cm}p{1.9cm}p{1.9cm}p{1.9cm}p{1.9cm}}
\toprule
Partido & Grado & Fuerza & \textit{Closeness} & \textit{Betweenness} & \textit{Eigenvector} \\
\midrule
Partido Conservador & \textit{salud} & \textit{salud} & \textit{niñas} & \textit{salud} & \textit{salud} \\
Partido Liberal     & \textit{cultural} & \textit{cultural} & \textit{quindiuense} & \textit{municipio} & \textit{municipio} \\
Centro Democrático  & \textit{seguridad} & \textit{seguridad} & \textit{años} & \textit{protección} & \textit{seguridad} \\
Pacto Histórico     & \textit{explotación}, \textit{recursos} & \textit{inmaterial} & \textit{afectada} & \textit{públicos} & \textit{debates} \\
Otros               & \textit{aprueba}, \textit{república} & \textit{república} & \textit{adopta} & \textit{general} & \textit{general} \\
\bottomrule
\end{tabular}

\caption{{\footnotesize Resumen de métricas globales y palabras más importantes por partido en el gobierno de Petro. En el panel (a), Diám.\ denota el diámetro, G.\ Prom.\ denota el grado promedio, G.\ Máx.\ denota el grado máximo, F.\ Máx.\ denota la fuerza máxima, Den.\ denota la densidad y Tran.\ denota la transitividad.}}
\label{tab:metricas_semanticas_petro_partidos}
\end{table}

\subsubsection{Comparación}

Las Tablas~\ref{tab:metricas_semanticas_santos_partidos}--\ref{tab:metricas_semanticas_petro_partidos} sugieren un desplazamiento del discurso legislativo que acompaña los cambios de coyuntura. En Santos, el léxico se organiza alrededor de un marco institucional asociado con implementación y territorialidad, con anclas como \textit{acuerdo}, \textit{república} y \textit{municipio}, y con mayor dispersión en el Partido Conservador. En Duque, el énfasis se reordena hacia orden público y gestión social en un contexto de choques económicos y sanitarios, reflejado en la centralidad de \textit{seguridad}, la aparición de \textit{hacinamiento} y el peso de \textit{instituciones} y \textit{funcionamiento}, mientras el Partido Conservador mantiene el mayor alcance temático. En Petro, Pacto Histórico muestra la mayor concentración y cohesión, coherente con pocos ejes ligados a regulación y bienes públicos mediante \textit{explotación}, \textit{recursos}, \textit{públicos} y \textit{debates}, mientras persisten continuidades como el eje territorial del Partido Liberal vía \textit{municipio}, el foco de Centro Democrático en \textit{seguridad} y \textit{protección}, y un lenguaje procedimental transversal en Otros con \textit{república}, \textit{aprueba} y \textit{general}.


\subsection{Modelo de sociabilidad}

A continuación, se ajusta el modelo de sociabilidad (véase la Sección \ref{sec_modelo_sociabilidad}) para identificar, en cada red semántica completa (no solo la componente conexa) por partido, los términos con parámetros de sociabilidad (popularidad) $\delta_i$ significativos al \(95\%\), lo que permite detectar hubs discursivos y distinguirlos de palabras meramente frecuentes. Las Tablas~\ref{tab:delta_periodo1}--\ref{tab:delta_periodo3} reportan los términos con $\delta_i>0$ y con evidencia suficiente. Cuando un partido no aparece en estas tablas, se debe a que ningún término de su red semántica resulta significativo bajo este criterio. En el gobierno de Santos, el modelo destaca pocos términos, \textit{vivienda} en el Partido de la U y \textit{años} en el Partido Conservador, en línea con redes donde el Partido de la U se concentra en un eje dominante y el Partido Conservador exhibe una agenda más dispersa. En un contexto marcado por implementación institucional y territorialidad, \textit{vivienda} se asocia con política social y provisión de bienes públicos, mientras \textit{años} sugiere un registro normativo ligado a vigencias y trayectorias regulatorias que conecta múltiples temas.

\begin{table}[!htb]
\centering
\footnotesize
\begin{tabular}{l p{2cm} p{1.3cm} p{1.3cm} p{1.3cm}}
\toprule
\textbf{Partido Político} & \textbf{Palabra} & \textbf{$\delta_i$} & \textbf{LI} & \textbf{LS} \\
\midrule
Partido Conservador & años     & 0.271 & 0.027 & 0.528 \\
Partido de la U     & vivienda & 0.398 & 0.123 & 0.666 \\
\bottomrule
\end{tabular}
\caption{{\footnotesize Estimaciones e intervalos de credibilidad al \(95\%\) para los parámetros de sociabilidad que resultan significativos en el gobierno de Santos, por partido político.}}
\label{tab:delta_periodo1}
\end{table}

En el gobierno de Duque, el modelo identifica más términos significativos y en varias colectividades, coherente con un contexto legislativo atravesado por orden público, tensiones sociales y choques económicos y sanitarios. En Centro Democrático, \textit{seguridad} y \textit{municipio} sobresalen, en línea con la lectura semántica donde \textit{seguridad} estructura el discurso y \textit{hacinamiento} aparece cercano como señal de capacidad institucional. En el Partido Liberal, \textit{instituciones} y \textit{programa} apuntan a énfasis en diseño y ejecución de política pública, consistente con la centralidad de \textit{instituciones} y con términos próximos como \textit{funcionamiento} y \textit{niños}. En Otros, \textit{republica}, \textit{colombia} y \textit{suscrito} reflejan un registro procedimental e institucional que conecta comunidades, también visible en el léxico transversal de sus redes semánticas.

\begin{table}[!htb]
\centering
\footnotesize
\begin{tabular}{l p{2cm} p{1.3cm} p{1.3cm} p{1.3cm}}
\toprule
\textbf{Partido Político} & \textbf{Palabra} & \textbf{$\delta_i$} & \textbf{LI} & \textbf{LS} \\
\midrule
Centro Democrático & social        & 0.277 & 0.036 & 0.531 \\
Centro Democrático & seguridad     & 0.344 & 0.099 & 0.589 \\
Centro Democrático & municipio     & 0.370 & 0.147 & 0.585 \\
\midrule
Otros              & suscrito      & 0.414 & 0.133 & 0.709 \\
Otros              & colombia      & 0.499 & 0.214 & 0.788 \\
Otros              & republica     & 0.674 & 0.386 & 0.946 \\
\midrule
Partido Liberal    & instituciones & 0.256 & 0.024 & 0.496 \\
Partido Liberal    & programa      & 0.257 & 0.039 & 0.488 \\
\bottomrule
\end{tabular}
\caption{{\footnotesize Estimaciones e intervalos de credibilidad al \(95\%\) para los parámetros de sociabilidad que resultan significativos en el gobierno de Duque, por partido político.}}
\label{tab:delta_periodo2}
\end{table}

En el gobierno de Petro, el modelo concentra la señal en pocos hubs, principalmente en Otros y en el Partido Conservador. En Otros, \textit{general}, \textit{aprueba} y \textit{republica} reflejan un léxico procedimental e institucional que se repite y conecta tópicos, coherente con la lectura de redes semánticas. En el Partido Conservador, \textit{salud} y \textit{proteccion} apuntan a una agenda de bienestar y garantías, plausible en un contexto de reformas sociales, mientras \textit{sector} muestra evidencia más débil al incluir valores cercanos a cero en su intervalo. Frente a la sección de redes semánticas, términos como \textit{seguridad} o \textit{cultural} y \textit{municipio} no aparecen como significativos aquí, lo que sugiere que su centralidad descriptiva no se traduce en popularidad claramente separable una vez se incorpora incertidumbre.

\begin{table}[!htb]
\centering
\footnotesize
\begin{tabular}{l p{2cm} p{1.3cm} p{1.3cm} p{1.3cm}}
\toprule
\textbf{Partido Político} & \textbf{Palabra} & \textbf{$\delta_i$} & \textbf{LI} & \textbf{LS} \\
\midrule
Otros & general   & 0.349 & 0.031 & 0.662 \\
Otros & aprueba   & 0.351 & 0.021 & 0.683 \\
Otros & republica & 0.364 & 0.056 & 0.680 \\
\midrule
Partido Conservador & sector     & 0.237 & -0.021 & 0.497 \\
Partido Conservador & proteccion & 0.312 & 0.066  & 0.564 \\
Partido Conservador & salud      & 0.446 & 0.205  & 0.691 \\
\bottomrule
\end{tabular}
\caption{{\footnotesize Estimaciones e intervalos de credibilidad al \(95\%\) para los parámetros de sociabilidad que resultan significativos en el gobierno de Petro, por partido político.}}
\label{tab:delta_periodo3}
\end{table}

En conjunto, los resultados sugieren un tránsito desde Santos, donde pocos términos exhiben popularidad claramente distinguible, hacia Duque, donde la coyuntura y la estandarización institucional amplían el conjunto de términos con sociabilidad positiva, y hacia Petro, donde la señal vuelve a concentrarse en vocabularios procedimentales y de bienestar mientras otras colectividades mantienen discursos más repartidos. Esto resalta el valor agregado del modelo de sociabilidad frente al análisis semántico descriptivo, porque no solo identifica palabras centrales por estructura, sino cuáles destacan de forma robusta al incorporar incertidumbre, de modo que la ausencia de un partido en las tablas de $\delta_i$ se interpreta como falta de hubs estadísticamente dominantes en su red y no como ausencia de temas o de actividad legislativa.


\begin{table}[!htb]
\centering
\scriptsize
\begin{tabular}{cc p{8.5cm}}
\toprule
\textbf{Tópico} & \textbf{$\lambda$} & \textbf{Interpretación} \\
\midrule
\multicolumn{3}{c}{Centro Democrático} \\
\midrule
1 & 0.23 & Regulación sobre maternidad y derechos digitales. \\
2 & 0.53 & Regulación médica y cirugías estéticas. \\
3 & 0.34 & Procesos de paz y amnistía. \\
4 & 0.40 & Pesca, acuicultura y regulación política. \\
5 & 0.55 & Programas sociales y salud pública. \\
6 & 0.59 & Turismo histórico y patrimonio local. \\
7 & 0.40 & Seguridad educativa y monitoreo. \\
8 & 0.18 & Regulaciones profesionales y transporte. \\
\midrule
\multicolumn{3}{c}{Partido Liberal} \\
\midrule
1 & 1.14 & Patrimonio cultural y reconocimientos locales. \\
2 & 1.53 & Bicentenario y conmemoraciones. \\
3 & 0.82 & Trayectoria política e intelectual del Partido Liberal. \\
4 & 2.01 & Obras de infraestructura y tecnología. \\
5 & 0.16 & Agrupamiento de palabras poco usadas. \\
\midrule
\multicolumn{3}{c}{Partido Conservador} \\
\midrule
1 & 0.42 & Regulación ambiental y protección animal. \\
2 & 1.98 & Manejo y disposición de aceites y residuos. \\
3 & 0.59 & Servicios ambientales y normas de infraestructura de vivienda. \\
4 & 0.32 & Patrimonio cultural y medidas anticorrupción. \\
5 & 0.49 & Régimen sancionatorio y deporte y educación física. \\
6 & 0.44 & Conmemoraciones y homenajes municipales. \\
7 & 0.14 & Protección social, producción sostenible y ordenamiento territorial. \\
8 & 0.58 & Derechos de la niñez y conflictos socioambientales. \\
9 & 0.61 & Declaración de distritos portuarios y agroindustriales de Turbo Antioquia. \\
\midrule
\multicolumn{3}{c}{Partido de la U} \\
\midrule
1 & 0.59 & Reconocimientos municipales y servicios públicos. \\
2 & 0.15 & Regulación cultural, educativa y fiscal. \\
3 & 0.61 & Vivienda y créditos habitacionales. \\
4 & 0.51 & Deporte y reconocimiento a logros deportivos. \\
5 & 0.88 & Ordenamiento territorial y áreas marinas. \\
6 & 0.61 & Conmemoraciones universitarias y presupuestos regionales. \\
7 & 0.70 & Protección de datos y delitos sexuales. \\
\midrule
\multicolumn{3}{c}{Otros} \\
\midrule
1 & 1.29 & Presupuesto y recursos fiscales. \\
2 & 2.59 & Paz y posconflicto. \\
3 & 6.13 & Acuerdos institucionales. \\
4 & 1.48 & Elecciones locales y gobierno distrital. \\
5 & 0.35 & Tratados internacionales y comercio exterior. \\
6 & 0.26 & Reforma legislativa y lucha contra la corrupción. \\
\bottomrule
\end{tabular}
\caption{{\footnotesize Interpretación de tópicos de las leyes propuestas en el gobierno de Santos.}}
\label{tab:sbm_topicos_santos}
\end{table}

\begin{table}[!htb]
\centering
\scriptsize
\begin{tabular}{cc p{8.5cm}}
\toprule
\textbf{Tópico} & \textbf{$\lambda$} & \textbf{Interpretación} \\
\midrule
\multicolumn{3}{c}{Centro Democrático} \\
\midrule
1 & 0.31 & Patrimonio cultural y conmemoraciones históricas. \\
2 & 0.50 & Radiodifusión y políticas de comunicación. \\
3 & 0.24 & Programas rurales y fomento al café. \\
4 & 2.44 & Seguridad social para trabajadores independientes. \\
5 & 0.30 & Regulación del transporte y derechos en salud. \\
6 & 0.34 & Régimen especial y normas ético-médicas. \\
7 & 0.05 & Salud pública, ambiente y derechos humanos. \\
8 & 0.18 & Fortalecimiento económico y salud preventiva. \\
\midrule
\multicolumn{3}{c}{Partido Liberal} \\
\midrule
1 & 0.35 & Patrimonio cultural del Caribe colombiano. \\
2 & 0.27 & Protección al usuario y gestión ambiental agropecuaria. \\
3 & 0.43 & Bienestar animal y envejecimiento digno. \\
4 & 0.32 & Educación ambiental y prohibición de plásticos. \\
5 & 0.53 & Seguridad vial y seguros obligatorios. \\
6 & 0.40 & Educación inclusiva y formación de alto rendimiento. \\
7 & 0.30 & Programas sociales y apoyo a mujeres. \\
8 & 0.40 & Conmemoraciones institucionales y gestión pública. \\
9 & 0.06 & Salud, educación y desarrollo territorial. \\
\midrule
\multicolumn{3}{c}{Partido Conservador} \\
\midrule
1 & 0.26 & Bienestar y protección animal. \\
2 & 0.35 & Gobernanza institucional y equidad de género. \\
3 & 0.68 & Protección animal y procesos penales. \\
4 & 0.41 & Acceso a vivienda, educación y pensiones. \\
5 & 0.52 & Gestión integral de residuos y responsabilidad ambiental. \\
6 & 0.51 & Patrimonio cultural y festividades del Caribe. \\
7 & 0.36 & Reconocimiento institucional y obras públicas. \\
8 & 0.46 & Regulación profesional y ética laboral. \\
9 & 0.08 & Inclusión social, fortalecimiento institucional y leyes sueltas. \\
\midrule
\multicolumn{3}{c}{Cambio Radical} \\
\midrule
1 & 0.36 & Registro de vivienda y administración urbana. \\
2 & 0.22 & Regulación educativa, penitenciaria y económica. \\
3 & 1.14 & Participación ciudadana y revocatoria del mandato. \\
4 & 0.20 & Reforma tributaria y descentralización administrativa. \\
5 & 0.63 & Protección de recursos hídricos y derechos de la infancia. \\
6 & 0.31 & Inclusión laboral y derechos de personas con discapacidad. \\
7 & 0.07 & Asamblea, servicios públicos y fomento cultural. \\
8 & 0.70 & Patrimonio cultural y conmemoraciones locales. \\
\midrule
\multicolumn{3}{c}{Otros} \\
\midrule
1 & 0.14 & Acuerdos y convenios internacionales. \\
2 & 11.26 & Aprobaciones oficiales y relaciones internacionales. \\
3 & 0.95 & Presupuesto nacional y recursos fiscales. \\
4 & 0.06 & Relaciones multilaterales y cooperación internacional. \\
5 & 0.44 & Acuerdos tributarios y eliminación de la doble tributación. \\
6 & 1.67 & Instituciones internacionales y cooperación verde. \\
7 & 0.59 & Protección ambiental y cooperación internacional. \\
8 & 0.11 & Reforma administrativa y cooperación europea. \\
\bottomrule
\end{tabular}
\caption{{\footnotesize Interpretación de tópicos de las leyes propuestas en el gobierno de Duque.}}
\label{tab:sbm_topicos_duque}
\end{table}

\begin{table}[!htb]
\centering
\scriptsize
\begin{tabular}{cc p{8.5cm}}
\toprule
\textbf{Tópico} & \textbf{$\lambda$} & \textbf{Interpretación} \\
\midrule
\multicolumn{3}{c}{Partido Conservador} \\
\midrule
1 & 0.42 & Fortalecimiento energético y asistencia agropecuaria. \\
2 & 0.38 & Derechos territoriales y ambientales locales. \\
3 & 0.62 & Conmemoraciones y homenajes municipales. \\
4 & 0.60 & Educación constitucional en instituciones públicas. \\
5 & 0.41 & Regulaciones laborales y procesos administrativos. \\
6 & 0.04 & Protección infantil, bienestar animal y sostenibilidad. \\
7 & 0.57 & Políticas ambientales y gestión de recursos hídricos. \\
8 & 0.47 & Salud mental y bienestar laboral. \\
\midrule
\multicolumn{3}{c}{Partido Liberal} \\
\midrule
1 & 0.68 & Patrimonio cultural y celebraciones locales. \\
2 & 0.54 & Conservación ambiental y regulación médica. \\
3 & 0.45 & Legislación laboral y fortalecimiento institucional. \\
4 & 0.31 & Subsidios y beneficios sociales estudiantiles. \\
5 & 0.64 & Espacios de reflexión y pluralidad religiosa. \\
6 & 0.08 & Inclusión, salud mental y reconocimiento cultural. \\
\midrule
\multicolumn{3}{c}{Centro Democrático} \\
\midrule
1 & 0.38 & Políticas de salud pública y protección ambiental. \\
2 & 0.54 & Incentivos económicos y conmemoraciones locales. \\
3 & 0.82 & Protección de líderes sociales y derechos humanos. \\
4 & 0.56 & Conservación ambiental y restauración de ecosistemas rurales. \\
5 & 0.46 & Regulación de cargos públicos y gestión administrativa estatal. \\
6 & 0.26 & Reformas legislativas y políticas territoriales. \\
\midrule
\multicolumn{3}{c}{Pacto Histórico} \\
\midrule
1 & 0.70 & Participación política y debates presidenciales. \\
2 & 0.34 & Protección ambiental y acceso al agua. \\
3 & 2.58 & Patrimonio cultural e inmaterial. \\
4 & 0.52 & Explotación de recursos naturales y homenajes públicos. \\
5 & 0.12 & Promoción cultural y celebración religiosa. \\
\midrule
\multicolumn{3}{c}{Otros} \\
\midrule
1 & 0.53 & Acuerdos internacionales y convenios de traslado de personas condenadas. \\
2 & 1.17 & Relaciones diplomáticas y cooperación entre países de América Latina. \\
3 & 0.91 & Presupuestos y gestión de recursos públicos nacionales. \\
4 & 0.83 & Participación en organismos internacionales y control institucional. \\
5 & 0.13 & Reformas y acuerdos internacionales en educación, trabajo y participación social. \\
\bottomrule
\end{tabular}
\caption{{\footnotesize Interpretación de tópicos de las leyes propuestas en el gobierno de Petro.}}
\label{tab:sbm_topicos_petro}
\end{table}

\subsection{Modelo de bloques estocásticos}

\begin{figure}[H]
    \centering
    \includegraphics[width=0.3\textwidth]{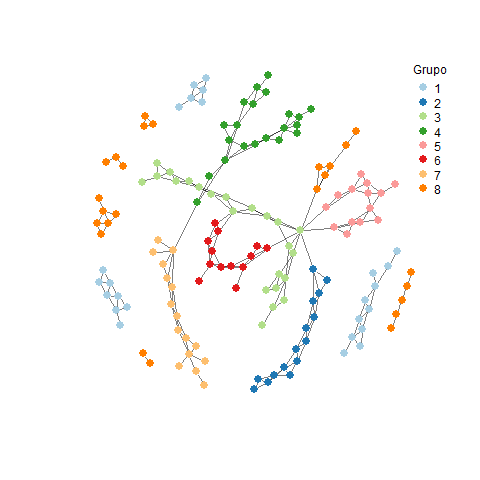}
    \quad
    \includegraphics[width=0.3\textwidth]{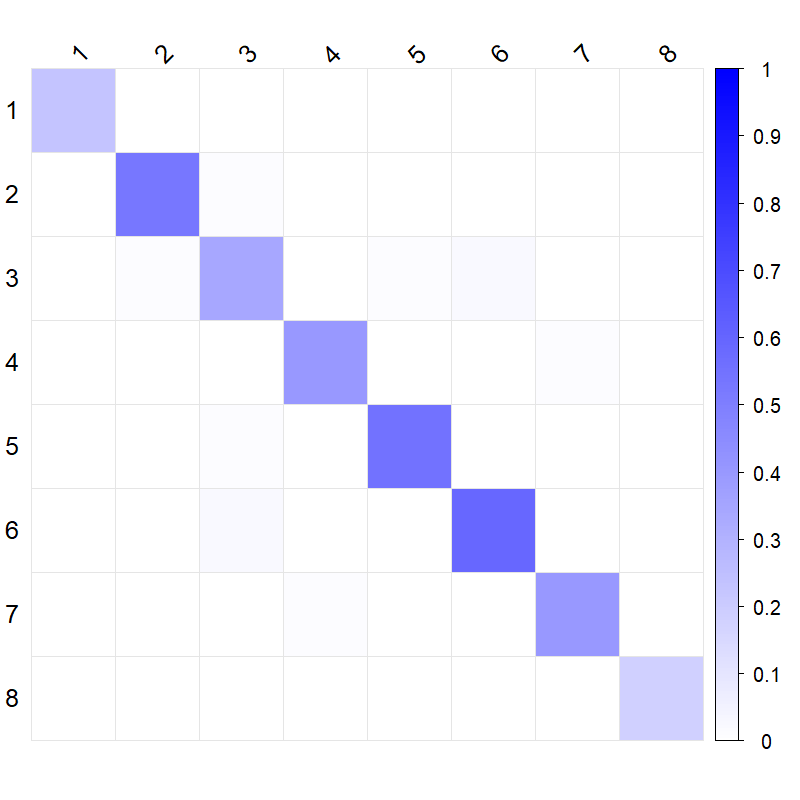}
    \caption{{\footnotesize Comunidades semánticas de Centro Democrático, gobierno Santos.}}
    \label{fig:sbm_santos_cc}
\end{figure}

\begin{figure}[H]
    \centering
    \includegraphics[width=0.3\textwidth]{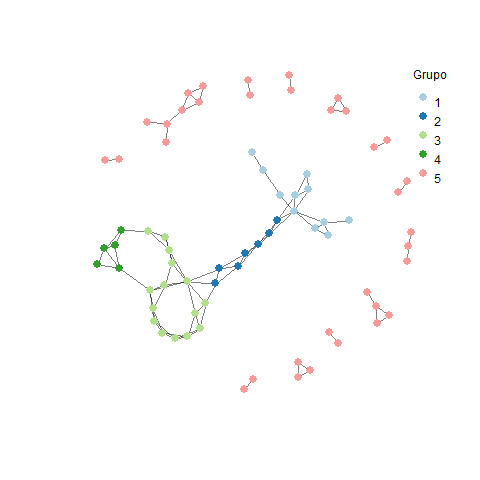}
    \quad
    \includegraphics[width=0.3\textwidth]{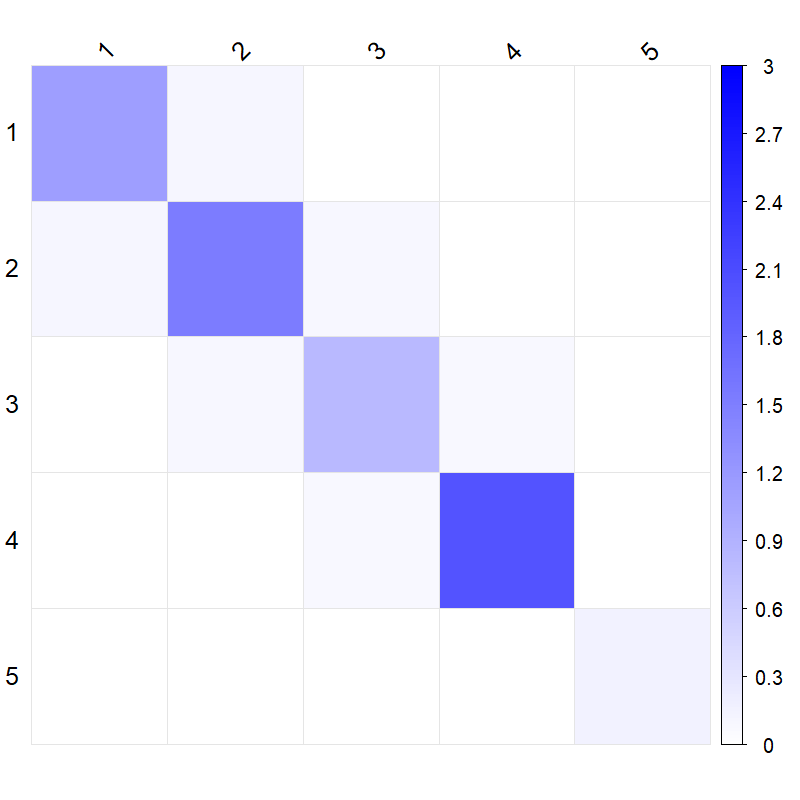}
    \caption{{\footnotesize Comunidades semánticas de Partido Liberal, gobierno Santos.}}
    \label{fig:sbm_santos_pl}
\end{figure}

\begin{figure}[H]
    \centering
    \includegraphics[width=0.3\textwidth]{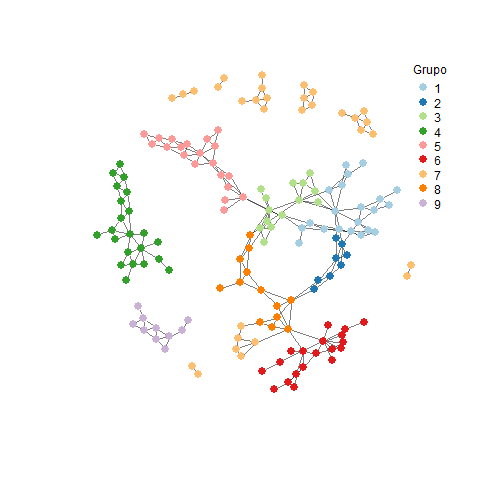}
    \quad
    \includegraphics[width=0.3\textwidth]{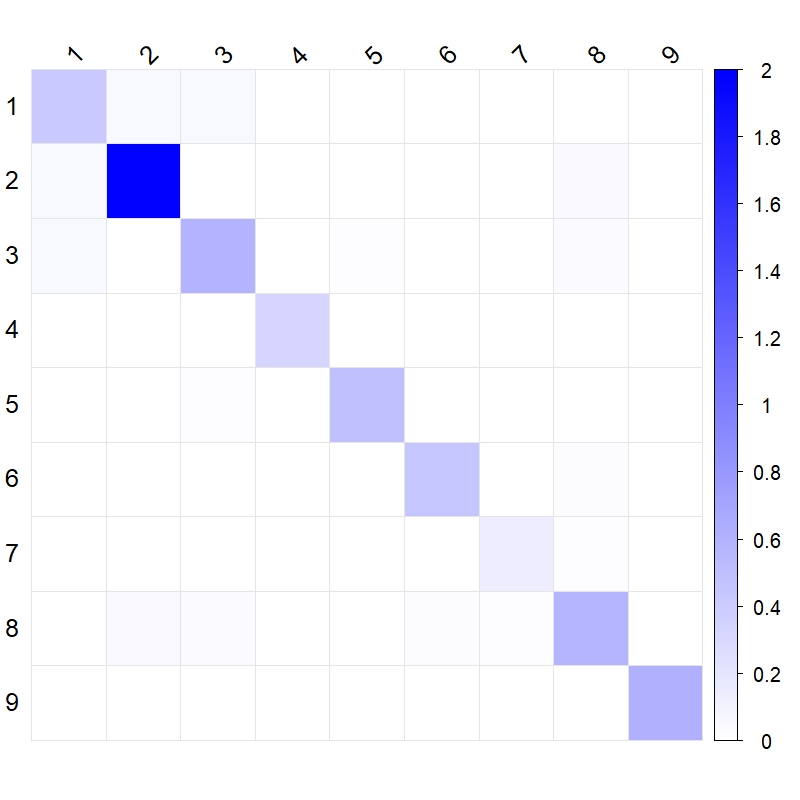}
    \caption{{\footnotesize Comunidades semánticas de Partido Conservador, gobierno Santos.}}
    \label{fig:sbm_santos_pcc}
\end{figure}

\begin{figure}[H]
    \centering
    \includegraphics[width=0.3\textwidth]{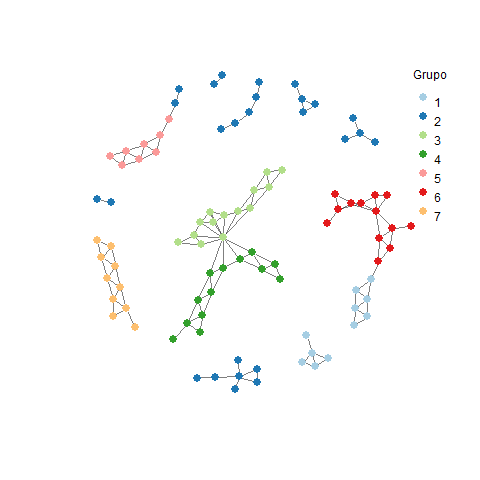}
    \quad
    \includegraphics[width=0.3\textwidth]{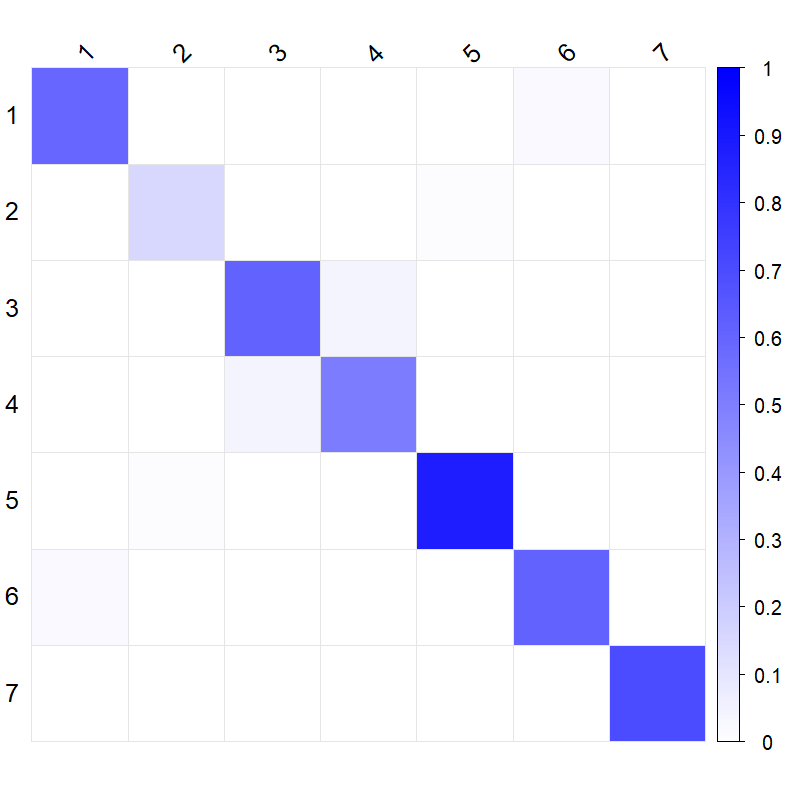}
    \caption{{\footnotesize Comunidades semánticas de Partido de la U, gobierno Santos.}}
    \label{fig:sbm_santos_pu}
\end{figure}

\begin{figure}[H]
    \centering
    \includegraphics[width=0.3\textwidth]{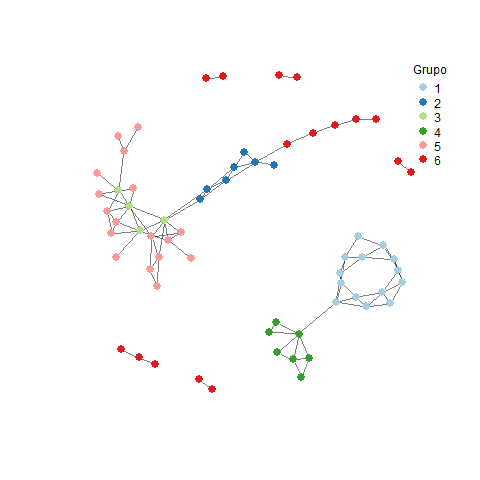}
    \quad
    \includegraphics[width=0.3\textwidth]{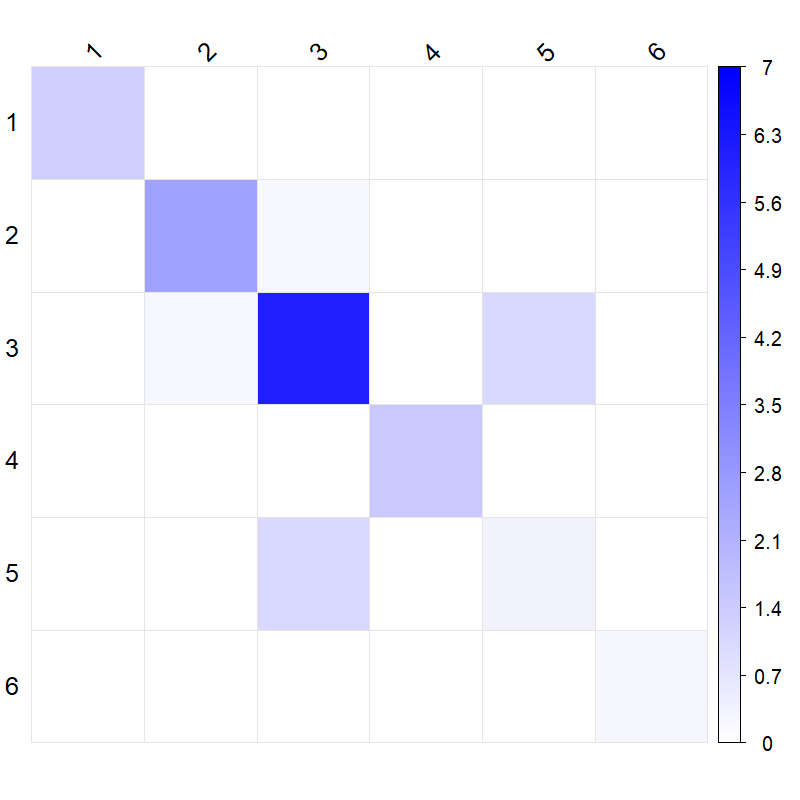}
    \caption{{\footnotesize Comunidades semánticas de Otros, gobierno Santos.}}
    \label{fig:sbm_santos_na}
\end{figure}

\begin{figure}[H]
    \centering
    \includegraphics[width=0.3\textwidth]{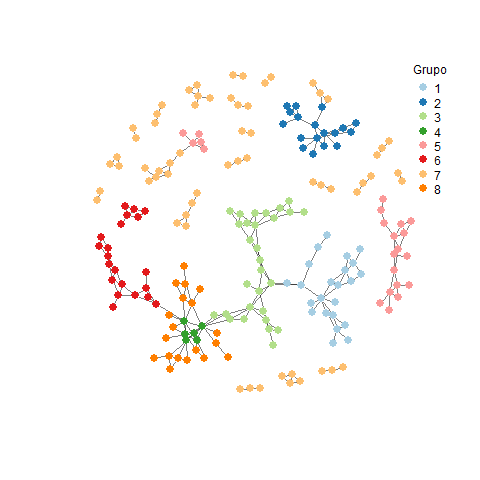}
    \quad
    \includegraphics[width=0.3\textwidth]{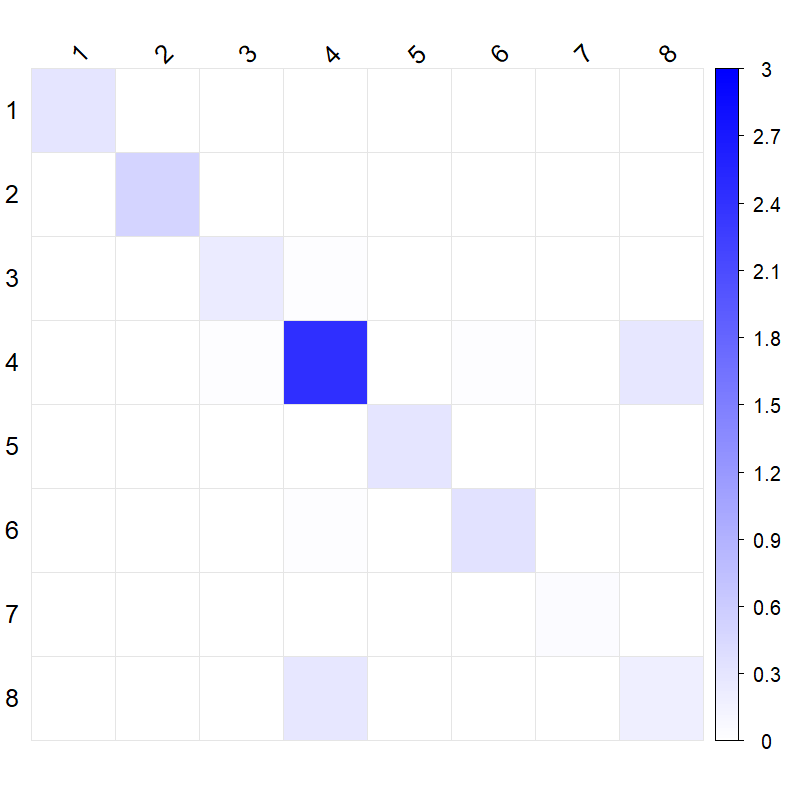}
    \caption{{\footnotesize Comunidades semánticas de Centro Democrático, gobierno Duque.}}
    \label{fig:sbm_duque_cc}
\end{figure}

\begin{figure}[H]
    \centering
    \includegraphics[width=0.3\textwidth]{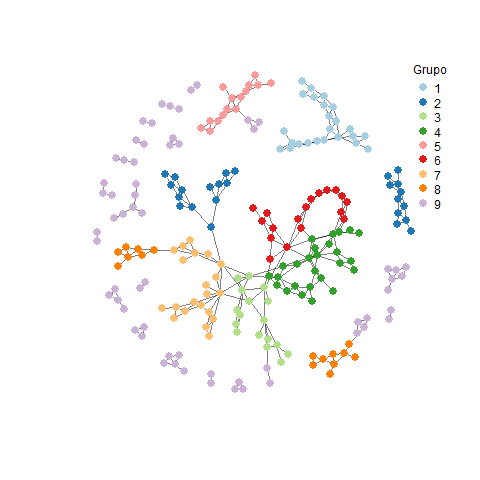}
    \quad
    \includegraphics[width=0.3\textwidth]{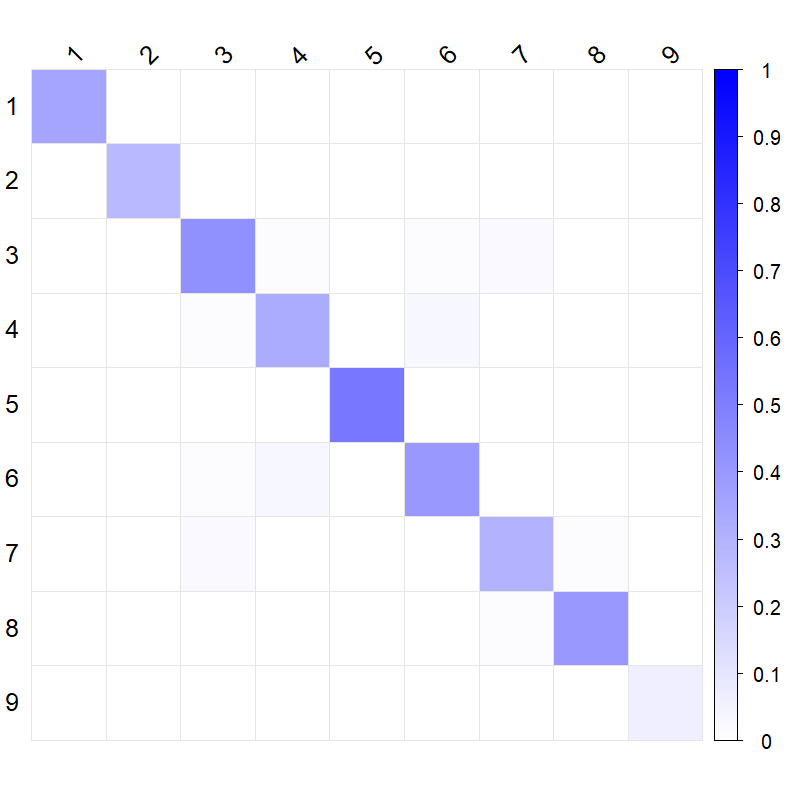}
    \caption{{\footnotesize Comunidades semánticas de Partido Liberal, gobierno Duque.}}
    \label{fig:sbm_duque_pl}
\end{figure}

\begin{figure}[H]
    \centering
    \includegraphics[width=0.3\textwidth]{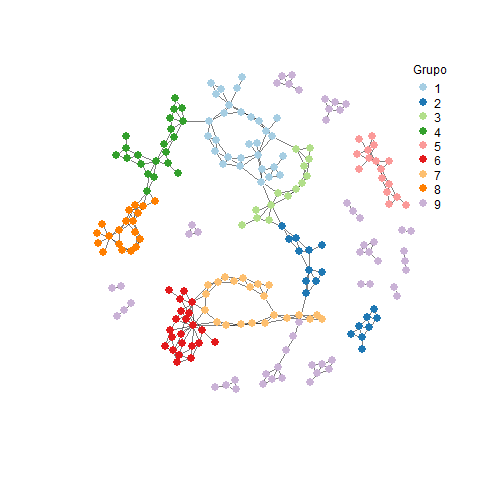}
    \quad
    \includegraphics[width=0.3\textwidth]{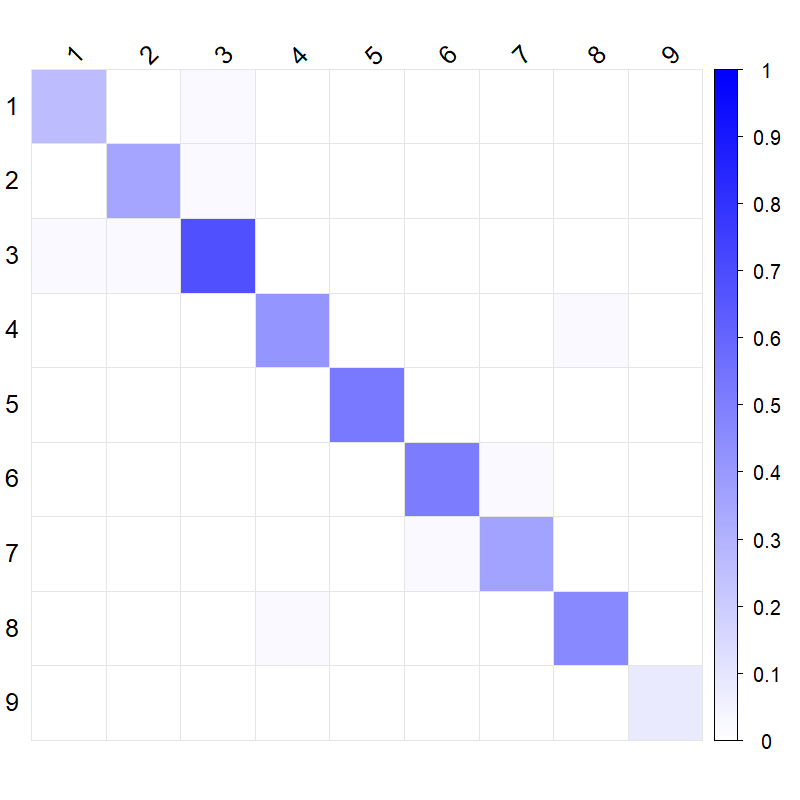}
    \caption{{\footnotesize Comunidades semánticas de Partido Conservador, gobierno Duque.}}
    \label{fig:sbm_duque_pcc}
\end{figure}

\begin{figure}[H]
    \centering
    \includegraphics[width=0.3\textwidth]{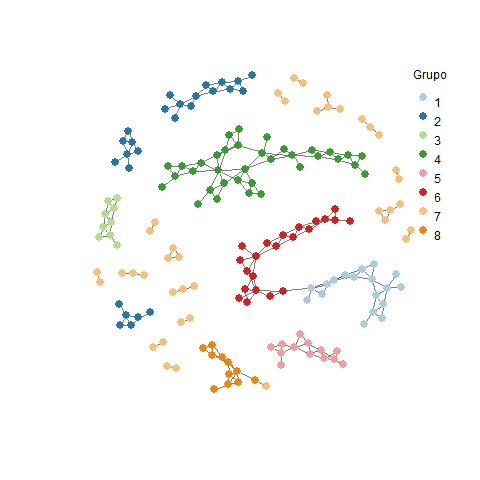}
    \quad
    \includegraphics[width=0.3\textwidth]{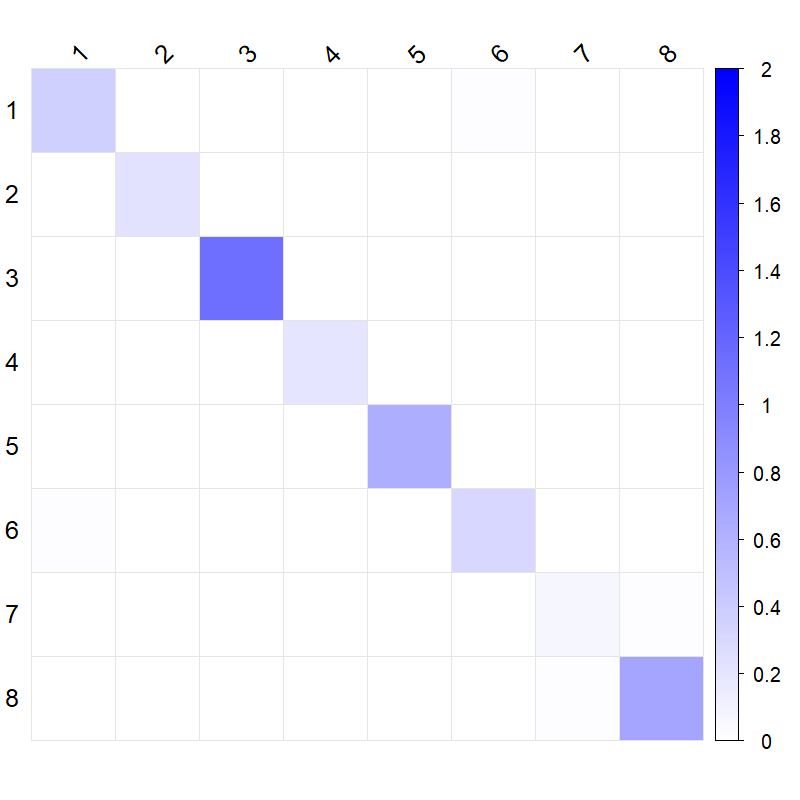}
    \caption{{\footnotesize Comunidades semánticas de Cambio Radical, gobierno Duque.}}
    \label{fig:sbm_duque_cr}
\end{figure}

\begin{figure}[H]
    \centering
    \includegraphics[width=0.3\textwidth]{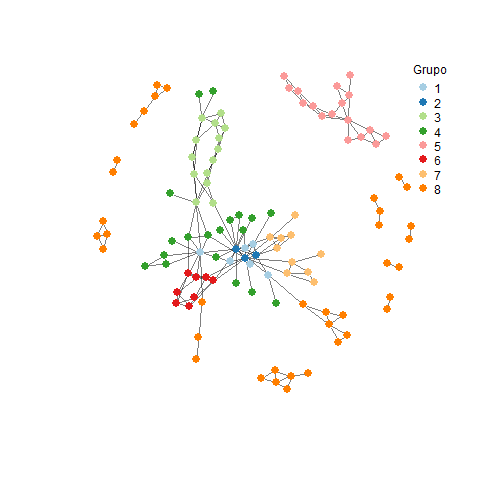}
    \quad
    \includegraphics[width=0.3\textwidth]{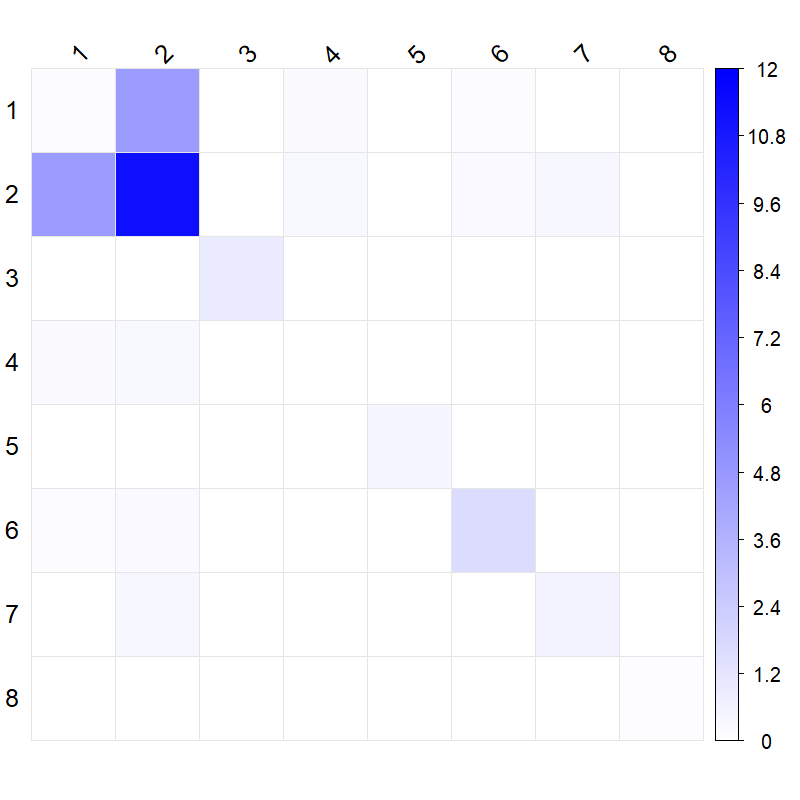}
    \caption{{\footnotesize Comunidades semánticas de Otros, gobierno Duque.}}
    \label{fig:sbm_duque_na}
\end{figure}

\begin{figure}[H]
    \centering
    \includegraphics[width=0.3\textwidth]{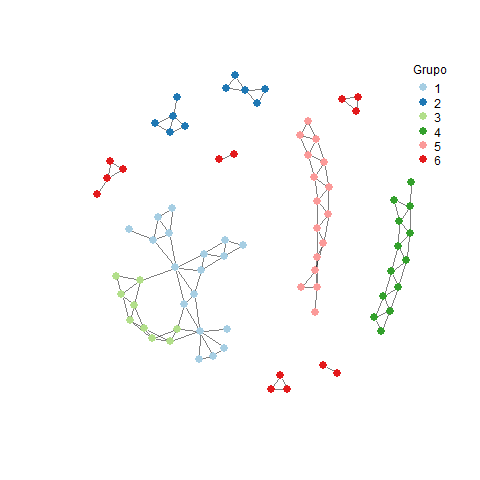}
    \quad
    \includegraphics[width=0.3\textwidth]{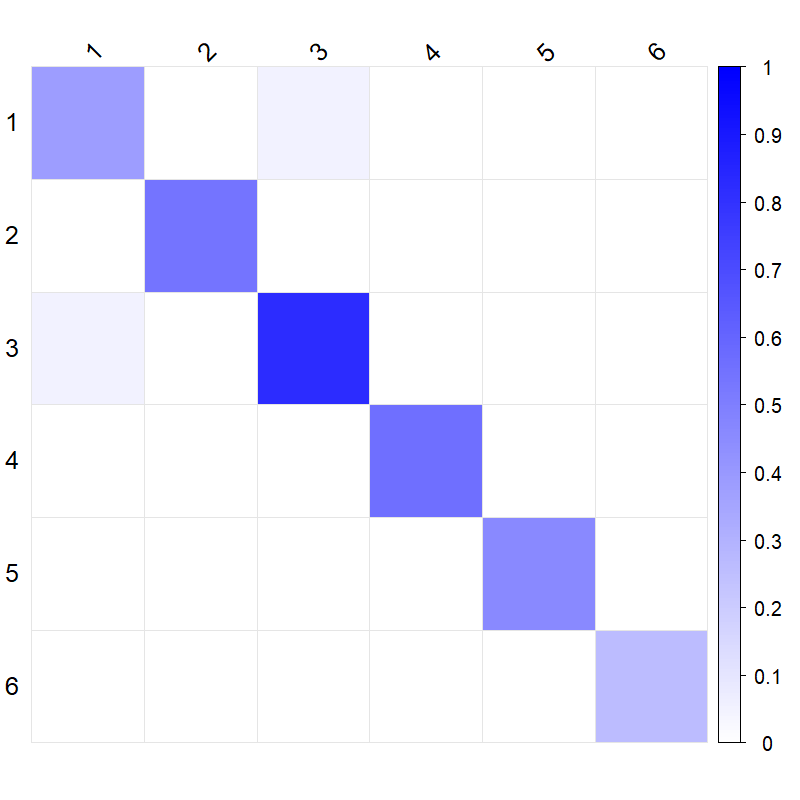}
    \caption{{\footnotesize Comunidades semánticas de Centro Democrático, gobierno Petro.}}
    \label{fig:sbm_petro_cc}
\end{figure}

\begin{figure}[H]
    \centering
    \includegraphics[width=0.3\textwidth]{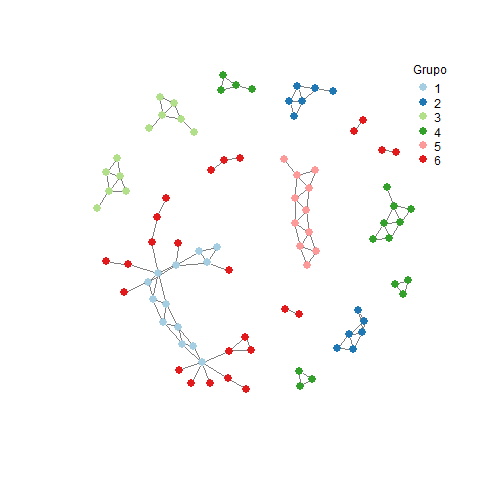}
    \quad
    \includegraphics[width=0.3\textwidth]{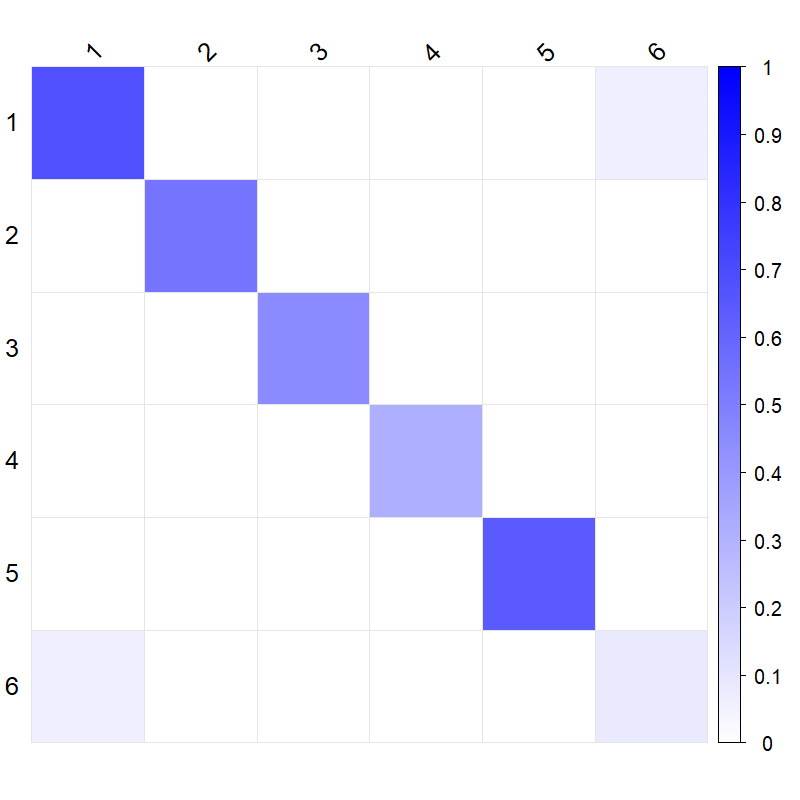}
    \caption{{\footnotesize Comunidades semánticas de Partido Liberal, gobierno Petro.}}
    \label{fig:sbm_petro_pl}
\end{figure}

\begin{figure}[H]
    \centering
    \includegraphics[width=0.3\textwidth]{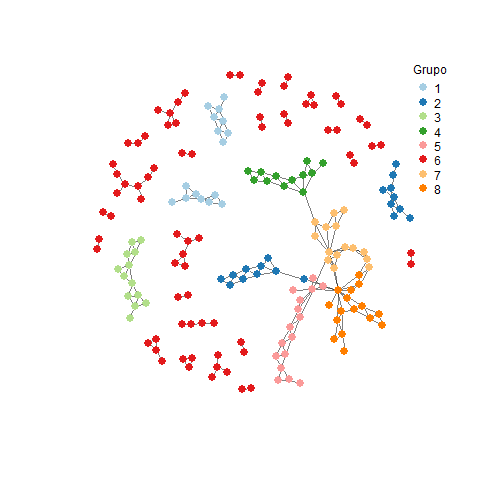}
    \quad
    \includegraphics[width=0.3\textwidth]{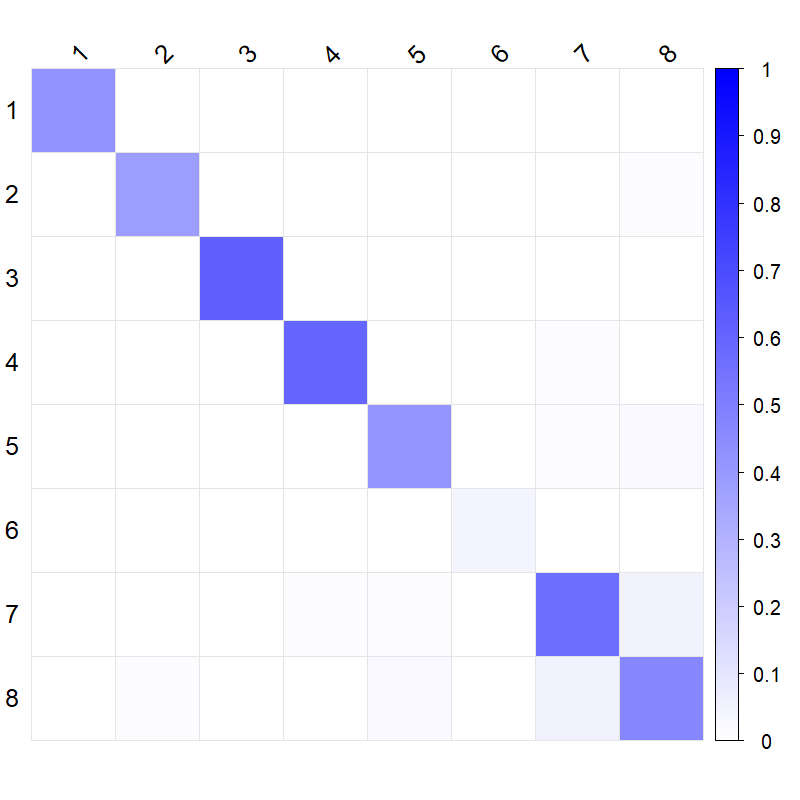}
    \caption{{\footnotesize Comunidades semánticas de Partido Conservador, gobierno Petro.}}
    \label{fig:sbm_petro_pcc}
\end{figure}

\begin{figure}[H]
    \centering
    \includegraphics[width=0.3\textwidth]{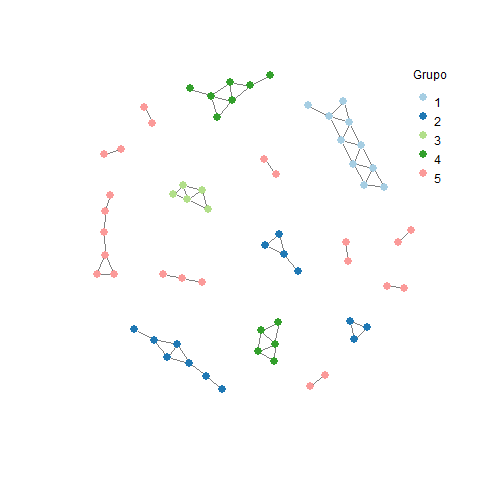}
    \quad
    \includegraphics[width=0.3\textwidth]{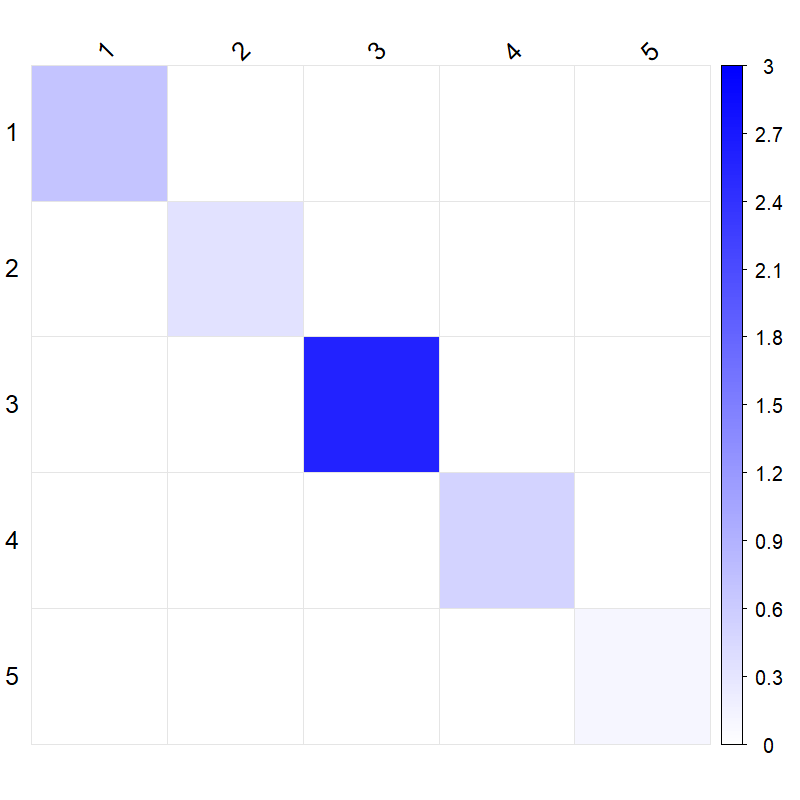}
    \caption{{\footnotesize Comunidades semánticas de Pacto Histórico, gobierno Petro.}}
    \label{fig:sbm_petro_cph}
\end{figure}

\begin{figure}[H]
    \centering
    \includegraphics[width=0.3\textwidth]{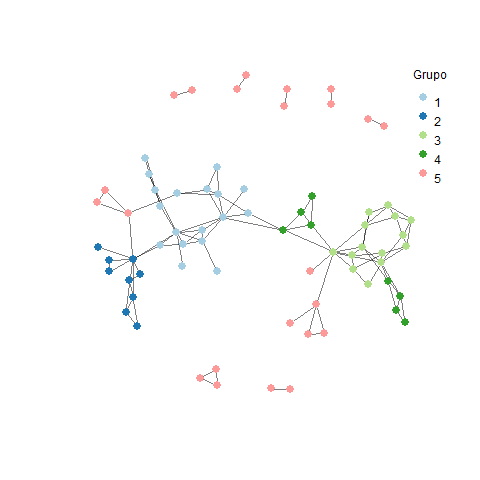}
    \quad
    \includegraphics[width=0.3\textwidth]{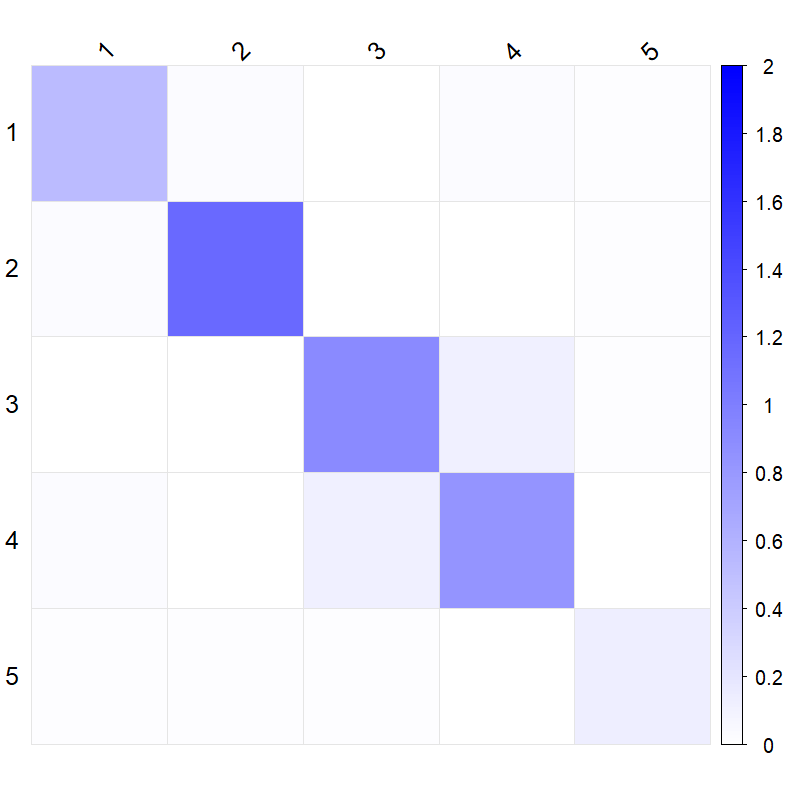}
    \caption{{\footnotesize Comunidades semánticas de Otros, gobierno Petro.}}
    \label{fig:sbm_petro_na}
\end{figure}

La estructura y el orden con que se articulan las palabras en un texto son determinantes para identificar tópicos y describir cómo se transmite el conocimiento. Para complementar el modelo de sociabilidad, se ajusta un modelo de bloques estocásticos (véase la Sección~\ref{sec_modelo_bloques}) sobre la red semántica completa, no solo sobre su componente conexa, con el fin de detectar estructuras latentes de coocurrencia y definir agrupamientos interpretables como tópicos. Al ajustar el modelo a cada red semántica por partido y gobierno, las matrices de probabilidades de interacción entre bloques muestran una señal marcada en la diagonal, lo que sugiere que la interacción se concentra dentro de cada agrupamiento temático. En este marco, el parámetro $\lambda$ resume el número medio de interacciones entre pares de palabras del mismo bloque, entendido como la frecuencia de coocurrencia dentro del tópico. En contraste, los valores fuera de la diagonal tienden a ser bajos, lo que indica una interacción limitada entre bloques y una organización relativamente lineal del discurso, con pocos puentes entre temas. En lo que sigue se presentan únicamente los bloques con mayor interacción interna y algunos ejemplos representativos de interacción entre agrupamientos. Las redes semánticas y sus matrices de interacción se reportan en las Figuras~\ref{fig:sbm_santos_cc}--\ref{fig:sbm_petro_na}, y la interpretación de los tópicos junto con su respectivo $\lambda$ se resume en las Tablas~\ref{tab:sbm_topicos_santos}--\ref{tab:sbm_topicos_petro}.

En el gobierno de Santos, los tópicos son coherentes con una agenda centrada en implementación institucional, territorialidad y proceso de paz. En Otros, acuerdos institucionales concentra la mayor cohesión, \(\lambda=6.13\), y se articula con paz y posconflicto, \(\lambda=2.59\), gobierno local, \(\lambda=1.48\), y agenda fiscal, \(\lambda=1.29\), lo que sugiere un discurso procedimental e institucional dominante. En el Partido Liberal, infraestructura y tecnología, \(\lambda=2.01\), junto con conmemoraciones y patrimonio, \(\lambda=1.53\) y \(\lambda=1.14\), indican una mezcla de inversión pública e identidad simbólica. En el Partido de la U, ordenamiento territorial, \(\lambda=0.88\), vivienda, \(\lambda=0.61\), y servicios públicos, \(\lambda=0.59\), apuntan a provisión de bienes públicos, con un eje regulatorio adicional en protección de datos y delitos sexuales, \(\lambda=0.70\). El Partido Conservador muestra un perfil sectorial donde residuos y aceites es el núcleo, \(\lambda=1.98\), acompañado por subagendas ambientales y sociales con cohesión moderada, mientras Centro Democrático presenta bloques con \(\lambda\) bajos o moderados, consistente con mayor dispersión temática.

En el gobierno de Duque, los tópicos capturan un entorno de presión social y económica, intensificado por el choque sanitario y la disputa institucional. En Otros, se observa una concentración dominante en aprobaciones oficiales y relaciones internacionales, $\lambda=11.26$, acompañada por cooperación verde, $\lambda=1.67$, y presupuesto y recursos fiscales, $\lambda=0.95$, lo que sugiere un léxico procedimental altamente estandarizado. En Centro Democrático, seguridad social para trabajadores independientes, $\lambda=2.44$, destaca como núcleo, coherente con tensiones laborales y ampliación de coberturas, mientras el resto se distribuye entre comunicación, ruralidad y regulación sanitaria. En Cambio Radical, participación ciudadana y revocatoria del mandato, $\lambda=1.14$, señala un eje de reglas de participación y control político, complementado por tópicos fiscales y de descentralización. En el Partido Conservador y el Partido Liberal, los bloques se reparten entre ambiente, protección, cultura, obras públicas y regulación, con $\lambda$ moderados, consistente con agendas amplias y menos concentradas que en Otros.

En el gobierno de Petro, los tópicos reflejan reformas sociales, debate distributivo, transición productiva y una agenda ambiental más visible. Pacto Histórico concentra su mayor cohesión en patrimonio cultural e inmaterial, $\lambda=2.58$, y lo articula con participación política, $\lambda=0.70$, y explotación de recursos naturales, $\lambda=0.52$, lo que sugiere un discurso que combina legitimidad política, cultura y disputa por el modelo de desarrollo. En el Partido Conservador predominan núcleos moderados sobre energía y asistencia agropecuaria, gestión hídrica, regulación laboral y salud mental, coherentes con tensiones territoriales y sectoriales. Centro Democrático enfatiza protección de líderes sociales y derechos humanos, $\lambda=0.82$, junto con gestión administrativa y política territorial, mientras el Partido Liberal mantiene un perfil mixto entre cultura, ambiente, trabajo y apoyos sociales, con $\lambda$ moderados. En Otros, destacan diplomacia regional, acuerdos internacionales y control institucional con $\lambda$ intermedios, lo que indica una agenda exterior relevante pero menos concentrada que en el periodo anterior.

En conjunto, el modelo de bloques muestra un desplazamiento claro del núcleo temático entre gobiernos. En Santos, la mayor cohesión se concentra en acuerdos institucionales y paz, con apoyos en gobierno local, fiscalidad y bienes públicos, mientras los partidos tradicionales reparten su agenda entre infraestructura, ordenamiento territorial y regulación sectorial. En Duque, el énfasis se mueve hacia un repertorio procedimental e internacional muy recurrente en “Otros”, acompañado por tópicos de protección social, participación y control político en las colectividades, consistente con el choque sanitario y la presión socioeconómica. En Petro, la señal se reequilibra hacia cultura, ambiente, recursos y reformas, con un bloque cultural muy cohesionado en el Pacto Histórico y núcleos intermedios en los demás partidos ligados a transición energética, bienestar, derechos y proyección internacional menos concentrada que en el periodo anterior.


\subsection{Asignación latente de Dirichlet}

Para contrastar la detección de tópicos mediante bloques estocásticos, se ajusta también un modelo de Asignación Latente de Dirichlet, LDA (véase la Sección~\ref{sec_LDA}), una metodología ampliamente usada en minería de texto que no depende de una representación en redes. En este caso, el LDA exhibe limitaciones esperables dada la naturaleza del corpus, que está compuesto por descripciones cortas de proyectos de ley. Primero, el modelo tiende a producir un número elevado de tópicos con alta variabilidad léxica, pues la escasa longitud de los textos reduce la evidencia para estimar mezclas temáticas estables y, además, el supuesto de mezcla de tópicos por documento induce asignaciones fragmentadas cuando la señal es débil.

\begin{table}[!htb]
\centering
\footnotesize
\begin{tabular}{l l r}
\toprule
\textbf{Periodo presidencial} & \textbf{Partido político} & \textbf{Perplejidad} \\
\midrule
\multirow{5}{*}{Juan Santos}
& Otros & 508.37 \\
& Partido Liberal & 403.46 \\
& Centro Democrático & 415.31 \\
& Partido Conservador Colombiano & 329.63 \\
& Partido de la U & 327.10 \\
\midrule
\multirow{5}{*}{Iván Duque}
& Centro Democrático & 591.53 \\
& Partido Liberal & 551.12 \\
& Partido Conservador Colombiano & 499.13 \\
& Cambio Radical & 465.46 \\
& Otros & 313.42 \\
\midrule
\multirow{5}{*}{Gustavo Petro}
& Partido Conservador Colombiano & 538.63 \\
& Partido Liberal & 361.86 \\
& Coalición Pacto Histórico & 323.95 \\
& Centro Democrático & 249.71 \\
& Otros & 194.67 \\
\bottomrule
\end{tabular}
\caption{{\footnotesize Perplejidad del modelo LDA por partido político y periodo presidencial.}}
\label{tab:perplejidad_periodos}
\end{table}

Segundo, la comparación entre partidos y periodos presidenciales es intrínsecamente difícil porque los corpus difieren en tamaño, vocabulario y solapamiento léxico, de modo que métricas como la perplejidad no son estrictamente comparables entre grupos y deben interpretarse con cautela, ya que capturan tanto rasgos del discurso como propiedades del conjunto de textos. Aun así, en la Tabla~\ref{tab:perplejidad_periodos} se observa que la perplejidad tiende a ser más alta durante el gobierno de Duque, lo que es consistente con textos menos predecibles y con redes semánticas más dispersas, mientras que en el gobierno de Petro aparecen valores más bajos en varios partidos, lo que sugiere mayor consistencia interna y una estructura temática relativamente más homogénea, posiblemente asociada con menor volumen de iniciativas por colectividad o con discursos más cohesionados.

\begin{figure}[!htb]
\centering
\includegraphics[scale=0.55]{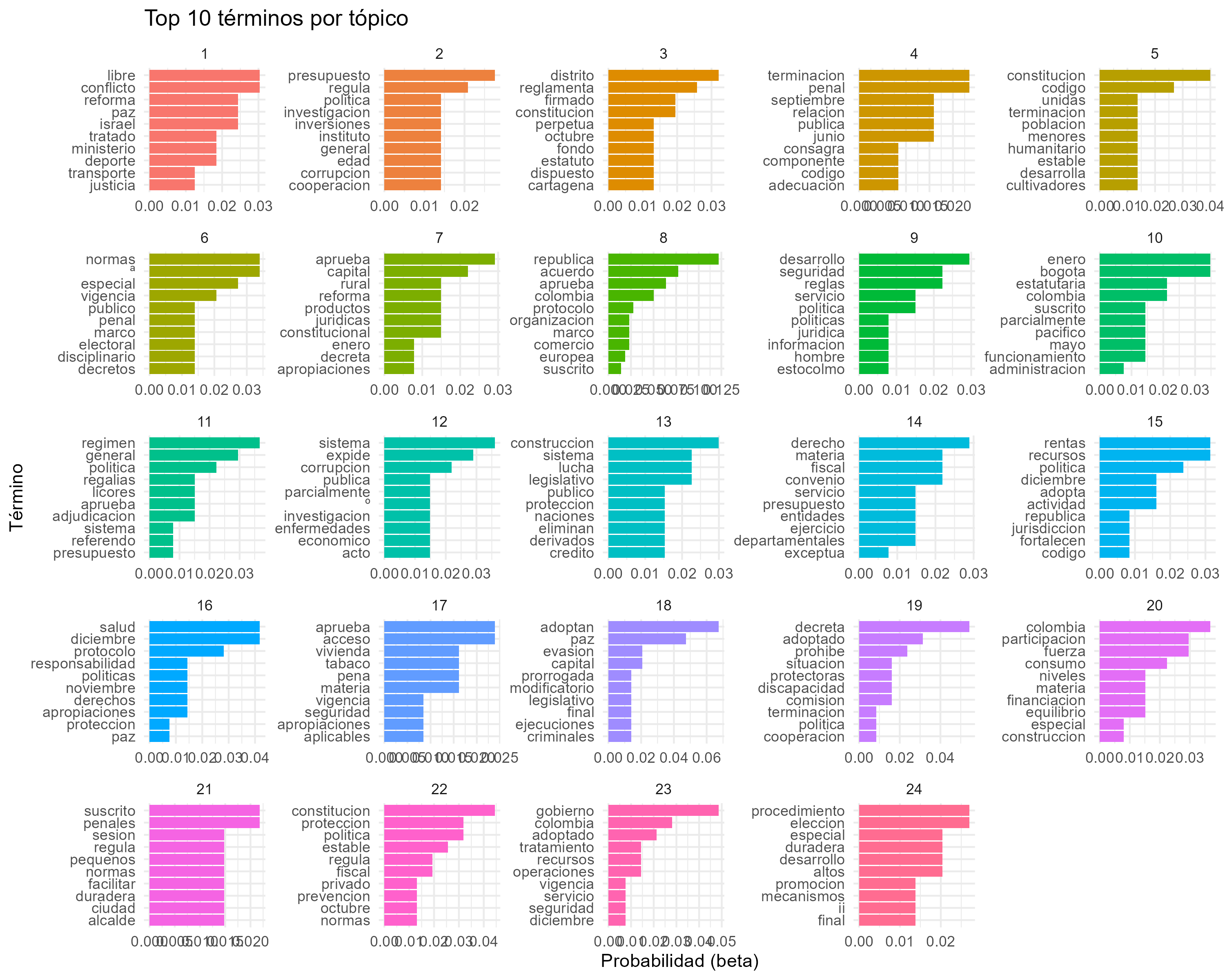}
\caption{Distribución de $\boldsymbol{\beta}$ y tópicos estimados mediante LDA para Otros partidos en el gobierno de Santos.}
\label{fig:lda_santos_na}
\end{figure}

Como ilustración, para los individuos externos en el gobierno de Santos el modelo de bloques estocásticos identifica seis tópicos coherentes, mientras que el LDA genera un número considerablemente mayor y menos interpretable, con distribuciones $\beta$ muy planas y pesos bajos debido a la baja frecuencia de términos, lo que fuerza al modelo a agrupar palabras que coocurren de manera esporádica en los resúmenes. Incluso cuando la cohesión es mayor, pueden aparecer mezclas poco consistentes dentro de un mismo tópico, como la asociación de vocabulario de convenios internacionales con términos ajenos al tema en algunos tópicos del gobierno de Petro. En conjunto, estos resultados respaldan el valor agregado del enfoque basado en redes semánticas, ya que produce agrupamientos más estables e interpretables bajo textos cortos y facilita la lectura temática al explotar explícitamente la estructura de coocurrencia.


\section{Discusión}

Este trabajo estudia la Cámara de Representantes de Colombia en el periodo 2014 a 2025 con el fin de evaluar si la polarización observada en otros escenarios se reproduce en el trabajo legislativo, y de caracterizar de manera sistemática los patrones de colaboración y el contenido temático de los proyectos de ley. El análisis integra redes bipartitas de incidencia, sus proyecciones para describir copatrocinio y coautoría, y redes semánticas basadas en coocurrencias para extraer tópicos y términos centrales con modelos estadísticos. En particular, el modelo de sociabilidad permite identificar hubs discursivos con incertidumbre explícita, y el modelo de bloques estocásticos para redes ponderadas permite recuperar comunidades interpretables como tópicos a partir de intensidades de interacción.

Los resultados sobre colaboración sugieren redes de partidos densas y con distancias pequeñas, lo que indica un patrón de copatrocinio extendido y ausencia de una separación estructural tajante entre colectividades. En los tres gobiernos se observa persistencia de actores con alta centralidad, con el Partido Liberal como nodo influyente de forma sostenida, lo que respalda la idea de que la dinámica legislativa se organiza alrededor de núcleos de articulación más que de bloques aislados. A nivel individual, se identifican representantes con posiciones relevantes por grado, intermediación y centralidad propia, lo que sugiere que ciertas figuras actúan como conectores entre subgrupos y que proyectos puntuales concentran articulación dentro de cada periodo.

En el componente semántico, la baja conectividad observada en varias redes, producto de la construcción por coocurrencias y del filtrado implícito que introducen los modelos, se interpreta como una ventaja para separar vocabulario procedimental de señales temáticas sustantivas. En este marco, las comunidades con palabras muy frecuentes tienden a exhibir mayores niveles medios de interacción dentro de bloque, lo que coincide con una mayor cohesión temática, mientras que los grupos con baja interacción media se asocian con tópicos periféricos o tratados de manera superficial por las colectividades. Además, la evidencia sugiere desplazamientos temáticos coherentes con la coyuntura, con un énfasis institucional y territorial en Santos, una agenda marcada por orden público y gestión social en Duque, y un reequilibrio hacia reformas sociales, ambiente, cultura y recursos en Petro. El contraste con Asignación Latente de Dirichlet respalda el uso del enfoque basado en redes en textos cortos, dado que el modelo tiende a fragmentar tópicos y a reducir interpretabilidad cuando la evidencia por documento es limitada.

La caracterización partidista complementa esta lectura metodológica. El Partido Liberal mantiene una presencia relevante a lo largo de los periodos y muestra cambios de énfasis, con señales de diversidad cultural y territorial y un desplazamiento hacia educación y programas sociales en el gobierno de Duque. Centro Democrático concentra de manera consistente una agenda asociada con seguridad y protección en sentido amplio. El Partido Conservador exhibe una diversidad temática elevada y, al mismo tiempo, una agenda más dispersa fuera de algunos ejes, lo que es compatible con mayor heterogeneidad interna. Pacto Histórico adquiere relevancia en el periodo reciente con una dinámica que depende en buena medida de articulación y colaboración, y con tópicos asociados con participación, patrimonio, ambiente y acceso al agua. En conjunto, la evidencia permite responder la hipótesis inicial, no se observa una división política clara en las redes de coautoría, pero sí se identifica influencia diferencial de algunas colectividades y actores.

Este trabajo no se encuentra excento de limitaciones. Primero, las descripciones legislativas son breves, lo que restringe la inferencia temática y puede ocultar beneficiarios, afectados y alcances reales de las iniciativas. Segundo, la estandarización incompleta de los registros y la presencia de vocabulario procedimental introducen sesgos que, aunque se mitigan con limpieza y con modelos de red, no desaparecen por completo. Tercero, en redes pequeñas o muy dispersas el modelo de bloques puede agrupar leyes con patrones de coocurrencia similares por la fuerza de asociaciones locales, sin garantizar coherencia temática plena. Cuarto, el análisis no incorpora resultados legislativos, como tránsito, aprobación o sanción, por lo que la influencia se interpreta en términos de estructura de colaboración y no de efectividad normativa.

A partir de estas limitaciones, se proponen varias extensiones. Resulta natural incorporar textos más largos cuando estén disponibles y combinar coocurrencias con representaciones semánticas modernas para mejorar resolución temática. También se sugiere extender el análisis hacia modelos dinámicos que capten cambios dentro de cada legislatura, así como integrar covariables de comisión, región y trayectorias políticas para explicar centralidad y colaboración. En la dimensión sustantiva, conviene vincular redes y tópicos con desenlaces legislativos para evaluar si ciertos patrones de copatrocinio o ciertos temas se asocian con mayor probabilidad de avance. Finalmente, la construcción de herramientas de visualización y consulta pública basadas en estos resultados puede fortalecer la rendición de cuentas al ofrecer síntesis interpretables sobre quién colabora con quién y qué temas prioriza cada colectividad, con cautela explícita sobre el alcance que impone la brevedad del texto disponible.


\section*{Declaración}

Los autores declaran que no tienen conflictos de interés financieros conocidos ni relaciones personales que pudieran haber influido, o parecer influir, en el trabajo reportado en este artículo. Durante la preparación de este trabajo, los autores utilizaron ChatGPT-5.1-turbo con el fin de mejorar el lenguaje y la legibilidad. Tras utilizar esta herramienta, los autores revisaron y editaron el contenido según fue necesario y asumen plena responsabilidad por el contenido de la publicación.

\bibliographystyle{apalike}
\bibliography{references.bib}

@article{lee2019review,
  title={A review of stochastic block models and extensions for graph clustering},
  author={Lee, Clement and Wilkinson, Darren J},
  journal={Applied Network Science},
  volume={4},
  number={1},
  pages={1--50},
  year={2019},
  publisher={Springer}
}

@book{gamerman2006markov,
  title={Markov chain Monte Carlo: stochastic simulation for Bayesian inference},
  author={Gamerman, Dani and Lopes, Hedibert F},
  year={2006},
  publisher={Chapman and Hall/CRC}
}

@article{chauhan2021topic,
  title={Topic modeling using latent Dirichlet allocation: A survey},
  author={Chauhan, Uttam and Shah, Apurva},
  journal={ACM Computing Surveys (CSUR)},
  volume={54},
  number={7},
  pages={1--35},
  year={2021},
  publisher={ACM New York, NY, USA}
}

@article{leger2016blockmodels,
  title={Blockmodels: A R-package for estimating in Latent Block Model and Stochastic Block Model, with various probability functions, with or without covariates},
  author={Leger, Jean-Benoist},
  journal={arXiv preprint arXiv:1602.07587},
  year={2016}
}

@article{biernacki2002assessing,
  title={Assessing a mixture model for clustering with the integrated completed likelihood},
  author={Biernacki, Christophe and Celeux, Gilles and Govaert, G{\'e}rard},
  journal={IEEE transactions on pattern analysis and machine intelligence},
  volume={22},
  number={7},
  pages={719--725},
  year={2002},
  publisher={IEEE}
}

@book{menczer2020first,
  title={A first course in network science},
  author={Menczer, Filippo and Fortunato, Santo and Davis, Clayton A},
  year={2020},
  publisher={Cambridge University Press}
}

@book{newman2018networks,
  title={Networks},
  author={Newman, Mark},
  year={2018},
  publisher={Oxford university press}
}

@book{al2017python,
  title={Python for graph and network analysis},
  author={Al-Taie, Mohammed Zuhair and Kadry, Seifedine},
  year={2017},
  publisher={Springer}
}

@book{manning1999foundations,
  title={Foundations of statistical natural language processing},
  author={Manning, Christopher and Schutze, Hinrich},
  year={1999},
  publisher={MIT press}
}

@inproceedings{devlin2019bert,
  title={Bert: Pre-training of deep bidirectional transformers for language understanding},
  author={Devlin, Jacob and Chang, Ming-Wei and Lee, Kenton and Toutanova, Kristina},
  booktitle={Proceedings of the 2019 conference of the North American chapter of the association for computational linguistics: human language technologies, volume 1 (long and short papers)},
  pages={4171--4186},
  year={2019}
}

@article{vaswani2017attention,
  title={Attention is all you need},
  author={Vaswani, Ashish and Shazeer, Noam and Parmar, Niki and Uszkoreit, Jakob and Jones, Llion and Gomez, Aidan N and Kaiser, {\L}ukasz and Polosukhin, Illia},
  journal={Advances in neural information processing systems},
  volume={30},
  year={2017}
}

@article{luque2024caracterizacion,
  title={Caracterizaci{\'o}n del discurso de posesi{\'o}n presidencial e identificaci{\'o}n de comunidades pol{\'\i}ticas en Colombia: Aproximaci{\'o}n emp{\'\i}rica desde el an{\'a}lisis de redes sociales},
  author={Luque, Carolina and Agudelo, Isabella and Leal, Kevin and Sosa, Juan},
  journal={Redes: revista hispana para el an{\'a}lisis de redes sociales},
  volume={35},
  number={1},
  pages={128--150},
  year={2024}
}

@article{fowler2006,
  author    = {Fowler, James H.},
  title     = {Legislative cosponsorship networks in the US House and Senate},
  journal   = {Social Networks},
  volume    = {28},
  number    = {4},
  pages     = {454--465},
  year      = {2006},
}

@article{abbe2023community,
  title={Community detection and stochastic block models: recent developments},
  author={Abbe, Emmanuel},
  year={2018},
  journal={Journal of Machine Learning Research},
  volume={18},
  number={177},
  pages={1--86}
}

@book{silge2017text,
  title        = {Text Mining with R: A Tidy Approach},
  author       = {Silge, Julia and Robinson, David},
  year         = {2017},
  publisher    = {O'Reilly Media},
  address      = {Sebastopol, CA, USA},
  isbn         = {978-1-491-98165-8},
  url          = {https://www.tidytextmining.com/},
  note         = {First Edition, June 2017}
}

@book{kolaczyk2020statistical,
  title        = {Statistical Analysis of Network Data with R},
  author       = {Kolaczyk, Eric D. and Cs{\'a}rdi, G{\'a}bor},
  year         = {2020},
  edition      = {2nd},
  publisher    = {Springer Nature Switzerland AG},
  series       = {UseR!},
  address      = {Cham, Switzerland},
  isbn         = {978-3-030-44128-9},
  doi          = {10.1007/978-3-030-44129-6},
  url          = {https://doi.org/10.1007/978-3-030-44129-6},
  note         = {Second edition}
}

@article{sosa2025bayesian,
  title        = {Bayesian Sociality Models: A Scalable and Flexible Alternative for Network Analysis},
  author       = {Sosa, Juan and Mart{\'i}nez, Carlos},
  year         = {2025},
  journal      = {arXiv preprint arXiv:2503.14697},
  url          = {https://arxiv.org/abs/2503.14697},
  institution  = {Universidad Nacional de Colombia},
  note         = {License: CC BY 4.0},
}

@article{grun2011topicmodels,
  title        = {topicmodels: An R Package for Fitting Topic Models},
  author       = {Gr{\"u}n, Bettina and Hornik, Kurt},
  journal      = {Journal of Statistical Software},
  volume       = {40},
  number       = {13},
  pages        = {1--30},
  year         = {2011},
  url          = {https://www.jstatsoft.org/v40/i13/},
  doi          = {10.18637/jss.v040.i13},
  publisher    = {Foundation for Open Access Statistics},
  address      = {Los Angeles, CA, USA}
}

@article{gonzalez2015identidad,
  title        = {Identidad nacional, bipartidismo y violencia en Colombia: los desafíos de la multiculturalidad consagrada por la Constitución de 1991},
  author       = {González, Fernán E.},
  journal      = {Centro de Investigación y Educación Popular (CINEP)},
  year         = {2015},
  address      = {Bogotá, Colombia},
  url          = {mailto:fernan@cinep.org.co},
  note         = {Recibido: 15 septiembre 2014 / Revisado: 18 febrero 2015 / Aceptado: 6 mayo 2015 / Publicado: 15 junio 2015}
}

@article{alonso2014ensamblajes,
  title        = {Ensamblajes institucionales y guerras civiles en la Colombia del siglo XIX},
  author       = {Alonso Espinal, Manuel Alberto},
  journal      = {Revista Co-herencia},
  volume       = {11},
  number       = {21},
  pages        = {169--190},
  year         = {2014},
  address      = {Medellín, Colombia},
  issn         = {1794-5887},
  note         = {Recibido: 8 septiembre 2014 / Aprobado: 26 octubre 2014},
  url          = {mailto:manuel.alonso@udea.edu.co}
}

@article{chenou_restrepo2023naciondividida,
  title        = {Una nación dividida: análisis del discurso político en redes sociales antes del plebiscito del acuerdo de paz con las FARC},
  author       = {Chenou, Jean-Marie and Restrepo, Elvira María},
  journal      = {Análisis Político},
  volume       = {36},
  number       = {106},
  pages        = {e111038},
  year         = {2023},
  address      = {Bogotá, Colombia},
  issn         = {0121-4705},
  doi          = {10.15446/anpol.v36n106.111038},
  url          = {https://doi.org/10.15446/anpol.v36n106.111038},
  note         = {Epub: 30 septiembre 2023},
}

@article{kim2025collaboration,
  title        = {Collaboration Dynamics in Legislative Co-Sponsorship Networks: Evidence from Korea},
  author       = {Kim, Sunjin and Ryu, Doojin and Song, Chang Geun},
  journal      = {Economics},
  year         = {2025},
  doi          = {10.1515/econ-2025-0142},
  url          = {https://doi.org/10.1515/econ-2025-0142},
  note         = {Received: 1 September 2024; Accepted: 25 February 2025},
  jel_codes    = {D02, D72, P48}
}

@article{robles2025dime,
  title        = {Dime con quién andas y te diré quién eres: análisis estructural de las redes de los senadores de Colombia de los periodos 2010-2014 y 2014-2018},
  author       = {Robles Dávila, Esteban and Manfredi, Luciana C. and Sayago Gómez, Juan Tomás and Franco Jurado, Juan Manuel},
  journal      = {Cuadernos de Economía},
  volume       = {44},
  number       = {94},
  pages        = {307--330},
  year         = {2025},
  doi          = {10.15446/cuad.econ.v44n94.106787},
  url          = {https://doi.org/10.15446/cuad.econ.v44n94.106787},
  note         = {Recibido: 16 enero 2023; Ajustado: 7 septiembre 2023; Aprobado: 15 noviembre 2023},
  address      = {Cali, Colombia},
  issn = {0121-4772}
}

@inproceedings{hachaj2018bigrams,
  title        = {What Can Be Learned from Bigrams Analysis of Messages in Social Network?},
  author       = {Hachaj, Tomasz and Ogiela, Marek R.},
  booktitle    = {2018 11th International Congress on Image and Signal Processing, BioMedical Engineering and Informatics (CISP-BMEI)},
  year         = {2018},
  address      = {Beijing, China},
  publisher    = {IEEE},
  doi          = {10.1109/CISP-BMEI.2018.8633045},
  url          = {https://doi.org/10.1109/CISP-BMEI.2018.8633045}
}

@article{galvez2018copalabras,
  title        = {Análisis de co-palabras aplicado a los artículos muy citados en Biblioteconomía y Ciencias de la Información (2007-2017)},
  author       = {Gálvez, Carmen},
  journal      = {Transinformação},
  volume       = {30},
  number       = {3},
  pages        = {277--286},
  year         = {2018},
  publisher    = {Pontifícia Universidade Católica de Campinas},
  doi          = {10.1590/2318-08892018000300001},
  url          = {https://doi.org/10.1590/2318-08892018000300001},
  note         = {Recibido: 13 noviembre 2017; Revisado: 17 enero 2018; Aprobado: 1 marzo 2018}
}

@misc{aicher2014learning,
  title         = {Learning Latent Block Structure in Weighted Networks},
  author        = {Aicher, Christopher and Jacobs, Abigail Z. and Clauset, Aaron},
  year          = {2014},
  eprint        = {1404.0431},
  archivePrefix = {arXiv},
  primaryClass  = {stat.ML},
  version       = {2},
  url           = {https://arxiv.org/abs/1404.0431},
  note          = {arXiv:1404.0431v2 [stat.ML], 3 Jun 2014}
}

@article{blei2003lda,
  title        = {Latent Dirichlet Allocation},
  author       = {Blei, David M. and Ng, Andrew Y. and Jordan, Michael I.},
  journal      = {Journal of Machine Learning Research},
  volume       = {3},
  pages        = {993--1022},
  year         = {2003},
  editor       = {Lafferty, John},
  url          = {https://www.jmlr.org/papers/v3/blei03a.html},
  note         = {Submitted: 2/02; Published: 1/03}
}

\appendix

\section{Caracterización de redes bipartitas}

\begin{table}[H]
\centering
\footnotesize

\textit{Panel (a). Métricas globales para las redes de proyección.}

\medskip
\begin{tabular}{llp{2cm}p{2cm}p{2cm}p{2cm}}
\toprule
Gobierno & Tipo de red & Diám. & G. Prom. & G. Máx. & Den. \\
\midrule
Santos & Individuos--Leyes & 19 &  6.28 &  96 & 0.00 \\
Santos & Partidos--Leyes   &  6 &  3.21 & 237 & 0.00 \\
Duque  & Individuos--Leyes & 14 & 11.47 & 171 & 0.01 \\
Duque  & Partidos--Leyes   &  6 &  4.54 & 647 & 0.00 \\
Petro  & Individuos--Leyes & 11 & 15.23 & 140 & 0.01 \\
Petro  & Partidos--Leyes   &  6 &  7.05 & 531 & 0.01 \\
\bottomrule
\end{tabular}

\bigskip
\textit{Panel (b). Nodos con mayor grado y mayor \textit{closeness}.}

\medskip
\begin{tabular}{llp{4.5cm}p{4.5cm}}
\toprule
Gobierno & Tipo de red & Nodo mayor grado & Nodo mayor \textit{closeness} \\
\midrule
Santos & Individuos--Leyes & Carlos Guevara Villabón & 120 \\
Santos & Partidos--Leyes   & Partido Liberal          & 864 \\
Duque  & Individuos--Leyes & Fabián Díaz Plata        & 1008 \\
Duque  & Partidos--Leyes   & Partido Liberal          & 1220 \\
Petro  & Individuos--Leyes & Gabriel Parrado Durán    & 2821 \\
Petro  & Partidos--Leyes   & Partido Liberal          & 3532 \\
\bottomrule
\end{tabular}

\bigskip
\textit{Panel (c). Nodos con mayor \textit{betweenness} y mayor \textit{eigenvector}.}

\medskip
\begin{tabular}{llp{4.5cm}p{4.5cm}}
\toprule
Gobierno & Tipo de red & Nodo mayor \textit{betweenness} & Nodo mayor \textit{eigenvector} \\
\midrule
Santos & Individuos--Leyes & 575 & Esperanza Pinzón Jiménez \\
Santos & Partidos--Leyes   & Otros & Partido Liberal \\
Duque  & Individuos--Leyes & Fabián Díaz Plata & César Lorduy Maldonado \\
Duque  & Partidos--Leyes   & Partido Liberal & Partido Liberal \\
Petro  & Individuos--Leyes & 3912 & Gabriel Parrado Durán \\
Petro  & Partidos--Leyes   & Partido Conservador Colombiano & Partido Liberal \\
\bottomrule
\end{tabular}

\caption{{\footnotesize Resumen de métricas globales y nodos más importantes por gobierno y red bipartita. En el panel (a), Diám. denota el diámetro de la red, G.\ Prom.\ denota el grado promedio, G.\ Máx.\ denota el grado máximo y Den.\ denota la densidad. En los paneles (b) y (c), cuando el nodo corresponde a un proyecto de ley, el número reportado corresponde al identificador del proyecto en la base de datos.}}
\label{tab:metricas_redes_bipartitas}

\end{table}


\section{Caracterización de proyecciones}

\begin{table}[H]
\centering
\footnotesize

\textit{Panel (a). Métricas globales para las redes proyectadas.}

\medskip
\begin{tabular}{llp{1.6cm}p{1.6cm}p{1.6cm}p{1.6cm}p{1.6cm}}
\toprule
Gobierno & Tipo de red & Diám. & G. Prom. & G. Máx. & Den. & Tran. \\
\midrule
Santos & Partidos        & 2 &  14.12 &  16 & 0.88 & 0.92 \\
Santos & Representantes  & 9 &  55.72 & 164 & 0.18 & 0.69 \\
Duque  & Partidos        & 2 &  18.57 &  20 & 0.93 & 0.94 \\
Duque  & Representantes  & 7 &  78.67 & 196 & 0.22 & 0.79 \\
Petro  & Partidos        & 2 &  30.87 &  38 & 0.81 & 0.96 \\
Petro  & Representantes  & 5 & 119.80 & 216 & 0.41 & 0.86 \\
\bottomrule
\end{tabular}

\bigskip
\textit{Panel (b). Nodos con mayor grado y mayor fuerza.}

\medskip
\begin{tabular}{llp{4.5cm}p{4.5cm}}
\toprule
Gobierno & Tipo de red & Nodo mayor grado & Nodo mayor fuerza \\
\midrule
Santos & Partidos       & Partido Conservador, Centro Democrático, Partido de la U, Alianza Verde & Partido Liberal \\
Santos & Representantes & Víctor Javier Correa Vélez & Víctor Javier Correa Vélez \\
Duque  & Partidos       & Diversos partidos & Centro Democrático \\
Duque  & Representantes & Jennifer Arias Falla & Eloy Quintero Romero \\
Petro  & Partidos       & Alianza Verde, Pacto Histórico, Comunes, MAIS & Coalición Pacto Histórico \\
Petro  & Representantes & Eduard Sarmiento Hidalgo & Carolina Giraldo Botero \\
\bottomrule
\end{tabular}

\bigskip
\textit{Panel (c). Nodos con mayor \textit{closeness}, mayor \textit{betweenness} y mayor \textit{eigenvector}.}

\medskip
\begin{tabular}{llp{3cm}p{3cm}p{3cm}}
\toprule
Gobierno & Tipo de red & Nodo mayor \textit{closeness} & Nodo mayor \textit{betweenness} & Nodo mayor \textit{eigenvector} \\
\midrule
Santos & Partidos       & Partido Conservador Colombiano & Partido Conservador Colombiano & Partido de la U \\
Santos & Representantes & Gonzálo Araújo Muñoz & Enrique Gil Botero & Víctor Correa Vélez \\
Duque  & Partidos       & C.\ Pacto Histórico & C.\ Pacto Histórico & Partido de la U \\
Duque  & Representantes & Luis Murillo Urrutia & Alicia Arango Olmos & Jennifer Arias Falla \\
Petro  & Partidos       & Partido Alianza Verde & Partido Alianza Verde & Partido Alianza Verde \\
Petro  & Representantes & Jose Antonio Ocampo & David Racero Mayorca & Gabriel Parrado Durán \\
\bottomrule
\end{tabular}

\caption{{\footnotesize Resumen de métricas globales y nodos más importantes en las redes bipartitas por gobierno. En el panel (a), Diám.\ denota el diámetro, G.\ Prom.\ denota el grado promedio, G.\ Máx.\ denota el grado máximo, Den.\ denota la densidad y Tran.\ denota la transitividad.}}
\label{tab:metricas_proyeccion_gobiernos}
\end{table}


\section{Visualización redes bipartitas} 

\section{Visualización de proyecciones}

\section{Visualización de redes semánticas: gobierno Santos}

\section{Visualización de redes semánticas: gobierno Duque}

\section{Visualización de redes semánticas: gobierno Petro}

\section{Matrices de probabilidades de interacción}

\end{document}